\documentclass[11pt]{article}
\pdfoutput=1
\usepackage[dvipsnames]{xcolor}
	\definecolor{light-gray}{gray}{0.9}
\usepackage{jheppub}
\usepackage{amssymb,amsfonts,amsmath,amsthm,lineno}
\usepackage{enumerate}
\usepackage{mathrsfs}
\usepackage{bbold}
\usepackage{tikz}
\usetikzlibrary{shapes.geometric, arrows}
\usepackage{hyperref}
\usepackage{slashed}
\usepackage{pgfplots}
\pgfplotsset{compat=1.15}
\usepgfplotslibrary{fillbetween}
\makeatletter
\newcommand{\Vast}{\bBigg@{4.75}}
\makeatother
\usepackage{empheq}

\usepackage{empheq}
\usepackage{amsmath}

\newcommand{\be}{\begin{equation}}
\newcommand{\ee}{\end{equation}}
\newcommand{\bea}{\begin{eqnarray}}
\newcommand{\eea}{\end{eqnarray}}

\newcommand{\CB}{\mathcal{B}}
\newcommand{\CC}{\mathcal{C}}

\newcommand{\CE}{\mathcal{E}}
\newcommand{\CF}{\mathcal{F}}

\newcommand{\CJ}{\mathcal{J}}

\newcommand{\CL}{\mathcal{L}}
\newcommand{\CN}{\mathcal{N}}
\newcommand{\CM}{\mathcal{M}}
\newcommand{\CO}{\mathcal{O}}

\newcommand{\CW}{\mathcal{W}}

\newcommand{\lr}{\left (}
\newcommand{\rr}{\right )}
\newcommand{\ls}{\left [}
\newcommand{\rs}{\right ]}
\newcommand{\lc}{\left \{}
\newcommand{\rc}{\right \}}

\newcommand\qt\tau

\newcommand{\p}{\partial}

\renewcommand{\tilde}[1]{\widetilde{#1}}
\newcommand{\tr}{\text{tr}}

\makeatletter
\renewcommand{\@seccntformat}[1]{\csname the#1\endcsname.\,\,}
\makeatother
\allowdisplaybreaks

\DeclareMathOperator{\sgn}{sgn}

\let \savenumberline \numberline
\def \numberline#1{\savenumberline{#1.}}

\makeatletter
\def\@fpheader{\relax}
\makeatother

\def\bea{\begin{eqnarray}}
\def\eea{\end{eqnarray}}

\usetikzlibrary{shapes,arrows,chains}
\usetikzlibrary{decorations.markings}
\usetikzlibrary{decorations.pathmorphing}
\usetikzlibrary{shapes.multipart}
\tikzset{snake it/.style={decorate, decoration=snake}}

\title{\ \vspace{1.6cm} \\
\scalebox{0.97}{Anisotropic Compactification of Nonrelativistic M-Theory}
}

\author[1]{Stephen Ebert}
\author[2]{and Ziqi Yan}

\affiliation[1]{\medskip
Mani L. Bhaumik Institute for Theoretical Physics\\Department of Physics and Astronomy\\ University of California, Los Angeles, CA 90095, USA\medskip}

\affiliation[2]{Nordita, KTH Royal Institute of Technology and Stockholm University \\
Hannes Alfv\'{e}ns v\"{a}g 12, SE-106 91 Stockholm, Sweden}

\emailAdd{stephenebert@physics.ucla.edu}
\emailAdd{ziqi.yan@su.se}

\abstract{We study a decoupling limit of M-theory where the three-form gauge potential becomes critical. This limit leads to nonrelativistic M-theory coupled to a non-Lorentzian spacetime geometry. Nonrelativistic M-theory is U-dual to M-theory in the discrete light cone quantization, a non-perturbative approach related to the Matrix theory description of M-theory. We focus on the compactification of nonrelativistic M-theory over a two-torus that exhibits anisotropic behaviors due to the foliation structure of the spacetime geometry. We develop a frame covariant formalism of the toroidal geometry, which provides a geometrical interpretation of the recently discovered polynomial realization of SL($2\,,\mathbb{Z}$) duality in nonrelativistic type IIB superstring theory. We will show that the nonrelativistic IIB string background fields transform as polynomials of an effective Galilean ``boost velocity" on the two-torus. As an application, we construct an action principle describing a single M5-brane in nonrelativistic M-theory and study its compactification over the anisotropic two-torus. This procedure leads to a D3-brane action in nonrelativistic IIB string theory that makes the SL($2\,,\mathbb{Z}$) invariance manifest in the polynomial realization.
}

\begin{document} 
\maketitle
\vfill\eject

\section{Introduction}

As one of the many heritages from the second superstring revolution, M-theory provides powerful insights for probing various non-perturbative aspects of superstring theories. Considerable knowledge of this conjectured unification has become accessible based on Matrix theory and the AdS/CFT correspondence, even though a comprehensive formulation of M-theory remains mysterious. In Matrix theory, M-theory compactified on a lightlike circle is described by a Matrix quantum mechanical system of D0-particles \cite{Banks:1996vh, deWit:1988wri}. The study of Matrix theory provides valuable guidance for formulating the full M-theory. 

In the literature, M-theory on a lightlike compactification is called the Discrete Light Cone Quantization (DLCQ) of M-theory \cite{Susskind:1997cw}. DLCQ is an important non-perturbative approach in both quantum field theory \cite{Kogut:1979wt} and string/M-theory \cite{Susskind:1997cw, Seiberg:1997ad, Sen:1997we}. Decoding the exotic physics in DLCQ M-theory plays a central role in the Matrix theory approach to M-theory. Intriguingly, there exists a U-dual description of DLCQ M-theory that we refer to as \emph{nonrelativistic M-theory} \cite{Gomis:2000bd, Garcia:2002fa}, where \emph{no} lightlike compactification is present anymore. Instead, nonrelativistic M-theory exhibits salient non-Lorentzian behaviors: the eleven-dimensional spacetime develops a foliation structure with a three-dimensional longitudinal sector and an eight-dimensional transverse sector. These two sectors are related through (membrane) Galilean boosts. This geometry is referred to as the \emph{membrane Newton-Cartan geometry}, which naturally generalizes Newton-Cartan geometry associated with the covariantization of Newtonian gravity to the higher-dimensional foliation structure \cite{Blair:2021waq, Ebert:2021mfu} (also see \cite{Brugues:2006yd, Pereniguez:2019eoq, Bergshoeff:2020xhv, Bergshoeff:2023rkk} for more general $p$-brane Newton-Cartan geometries).   

The U-duality relating DLCQ and nonrelativistic M-theory \cite{uduality} is an M-theory uplift of the T-duality relation between DLCQ and nonrelativistic string theory \cite{Bergshoeff:2018yvt, Gomis:2000bd, Danielsson:2000gi}.\,\footnote{Note that the relation to DLCQ M-theory and Matrix theory has been commented on in \cite{Danielsson:2000gi}.} Nonrelativistic string theory is unitary, ultra-violet complete, and has a Galilean invariant string spectrum \cite{Klebanov:2000pp, Gomis:2000bd, Danielsson:2000gi} (see \cite{Oling:2022fft} for a review). This theory couples to the background \emph{string Newton-Cartan geometry} \cite{Andringa:2012uz}, which contains a two-dimensional longitudinal sector and an eight-dimensional transverse sector, related to each other via (string) Galilean boosts. Under T-duality, a spacelike longitudinal compactification in nonrelativistic string theory maps to the lightlike circle in DLCQ string theory. Via this T-dual relation, nonrelativistic string theory provides a first principles definition of DLCQ string theory.

In this paper, motivated to understand the relationship between nonrelativistic M-theory and type IIB superstring theory, we analyze the toroidal compactification of nonrelativistic M-theory. This study may also serve as a preparation for establishing the U-duality relation between DLCQ and nonrelativistic string theory, which requires compactifying nonrelativistic M-theory over a three-torus and will be discussed separately in \cite{uduality}. This three-torus has to be ``anisotropic," with two cycles in the longitudinal sector and the third cycle in the transverse sector of the membrane Newton-Cartan geometry. Dimensionally reducing nonrelativistic M-theory along one of the longitudinal cycles leads to nonrelativistic IIA string theory. Then, T-dualizing along the other longitudinal circle produces the DLCQ of IIB string theory. Finally, we also have to T-dualize the original transverse circle to acquire the DLCQ of IIA string theory, which uplifts to DLCQ M-theory. Generally, we refer to such a compactification over a compact manifold that splits between the longitudinal and transverse sectors in nonrelativistic M-theory as an \emph{anisotropic compactification}.\,\footnote{In terms of this terminology, the conventional toroidal compactification would be referred to as an ``isotropic compactification," implying that different cycles are treated equally on the same footing.} 

To focus on the novelty brought by the anisotropy of this compactification, we perform in this paper a systematic study of compactifying nonrelativistic M-theory over a simpler anisotropic two-torus,\,\footnote{It is shown in \cite{wfmt, longpaper} that the topology of this anisotropic torus is the same as a pinched torus.} where novel structures already arise. Compactifying M-theory over a two-torus leads to type IIB superstring theory in the relativistic framework.\,\footnote{The seeming mismatch between dimensions has been elucidated in \cite{Schwarz:1995dk}: Compactifying M-theory over a two-torus can be equivalently viewed as compactifying type IIA superstring theory over a spatial circle of radius $R$\,, which is T-dual to type IIB superstring theory over the dual circle of radius $\alpha' / R$\,, with $\alpha'$ the Regge slope. Therefore, shrinking the torus corresponds to taking a small $R$\,, which decompactifies the dual circle in the T-dual IIB theory.} In the anisotropic compactification of nonrelativistic M-theory, we require one cycle of the two-torus to be longitudinal and the other transverse. Depending on which cycle on the anisotropic two-torus is taken to be the nonrelativistic M-theory circle, we arrive at \emph{different} IIB string theories that are S-dual to each other:  if the M-theory circle is longitudinal, we find nonrelativistic IIB string theory that arises from a critical Kalb-Ramond field limit of relativistic IIB string theory; if the M-theory circle is transverse, we find instead a different limit of relativistic IIB string theory associated with a critical Ramond-Ramond (RR) two-form field. These corners associated with the critical Kalb-Ramond and RR two-forms are closely related to Matrix string theory \cite{Banks:1996my, Motl:1997th, Dijkgraaf:1997vv, Danielsson:2000gi}. The T-duality connection to general decoupling limits associated with different critical RR potentials has been studied in \cite{Gopakumar:2000ep, Harmark:2000ff, Danielsson:2000gi} and is further explored in \cite{uduality, wfmt}, which reveal a duality web unifying a zoo of decoupling limits of string/M-theory, including Matrix (gauge) theories. 

It is famously known that the S-duality in type IIB superstring theory extends to the SL($2\,,\mathbb{Z}$) duality \cite{Schwarz:1995du, Schwarz:1995jq, Aspinwall:1995fw}, which acquires a simple geometric interpretation in M-theory as the isometry group of the two-torus over which M-theory compactifies. Shown in \cite{Bergshoeff:2022iss, Bergshoeff:2023ogz}, the SL($2\,,\mathbb{Z}$) duality can also be realized in nonrelativistic IIB string theory. To be more precise, we define the SL($2\,,\mathbb{Z}$) transformation matrix to be 
\be \label{eq:sl2zg}
    \Lambda = 
    \begin{pmatrix}
        \alpha &\,\,\,\, \beta \\[2pt]
        \gamma &\,\,\,\, \delta 
    \end{pmatrix}\,,
        \qquad%
    \alpha \, \delta - \beta \, \gamma = 1\,,
        \qquad%
    \alpha\,, \beta\,, \gamma\,, \delta \in \mathbb{Z}\,.
\ee
We define the RR zero-form in nonrelativistic IIB theory as $C^{(0)}$\,. When $\gamma \, C^{(0)} + \delta = 0$\,, nonrelativistic string theory gets mapped to the critical RR two-form limit of IIB string theory, and the $\mathbb{Z}^{}_2$ part of this set of transformations is the S-duality that we have discussed in the previous paragraph. Instead, when $\gamma \, C^{(0)} + \delta \neq 0$\,, in terms of a set of properly chosen variables, the background fields transform under the SL($2\,,\mathbb{Z}$) duality as a polynomial of a quantity $\kappa$ formed by the group parameters and the modulus of the two-torus (see Eq.~\eqref{eq:trnsfeaaomega}). We refer to this realization of the global SL($2\,,\mathbb{Z}$) symmetry in nonrelativistic IIB string theory as a \emph{polynomial realization} \cite{Bergshoeff:2023ogz}. This novel mathematical structure bears an intriguing connection to the classical invariant theory of binary forms (homogeneous polynomials in two variables) in algebra and provides a useful formalism for classifying SL($2\,,\mathbb{Z}$) invariants in nonrelativistic IIB string theory. For example, in \cite{Bergshoeff:2023ogz}, the complete set of field strengths describing nonrelativistic IIB supergravity, together with their SL($2\,,\mathbb{Z}$) transformation rules, are encoded by a quadratic and quartic binary form. 
Our study of anisotropic compactification of nonrelativistic M-theory in this paper will provide a natural geometric interpretation of the polynomial realization of SL($2\,,\mathbb{Z}$) duality. We will find that the parameter $\kappa$\,, in terms of which the background fields transform polynomially, acquires a physical meaning in M-theory as an effective Galilean boost velocity on the two-torus. See section~\ref{sec:bvbut} for details. 

As an application of the formalism developed in this paper, we will consider an anisotropic toroidal compactification of a single M5-brane in nonrelativistic M-theory. The action of the nonrelativistic M5-brane will be derived in Eq.~\eqref{eq:nrm5a} for the first time,\,\footnote{The construction of nonrelativistic M5-brane has been attempted in the an older version of \cite{Roychowdhury:2022kdh}, where an expansion in the presence of a critical three-form gauge field is considered. However, the expansion in \cite{Roychowdhury:2022kdh} is invalid as it contains the inverse of an identically zero determinant. This determinant is over the longitudinal metric, which is of rank three, being pulled back to the six-dimensional M5-brane worldvolume. A newer version of \cite{Roychowdhury:2022kdh} takes a different starting point, where another limit of M-theory is introduced. This limit involves an extra rescaling of the M5-brane tension in addition to the reparametrization of the spacetime metric that we consider, and is thus \emph{not} the nonrelativistic membrane limit.}
which is essential for furthering the understanding of nonrelativistic M-theory. This advance also provides a concrete starting point for revisiting open membrane theory \cite{Gopakumar:2000ep,Bergshoeff:2000ai}. Finally, compactifying the M5-brane action leads to a manifestly SL($2\,,\mathbb{Z}$) invariant formalism of the D3-brane in nonrelativistic IIB superstring theory.\,\footnote{See \cite{Berman:1998va, Berman:1998sz} for an M5-brane on a two-torus and its relation to D3-brane in relativistic M-theory.} We will show that the associated D3-brane action~\eqref{eq:d33} makes the polynomial realization manifest and significantly simplifies the rather complicated SL($2\,,\mathbb{Z}$)-invariant D3-brane action derived in \cite{Bergshoeff:2022iss}. 

This paper is organized as follows. In section~\ref{sec:benmt}, we review nonrelativistic string theory and its uplift to nonrelativistic M-theory. We then define the anisotropic toroidal compactification of nonrelativistic M-theory and explain its dual relation to DLCQ M-theory. In section~\ref{sec:atbsld}, we develop the zweibein formalism to describe a toroidal geometry, using which we provide a geometric interpretation of the polynomial realization of SL($2\,, \mathbb{Z}$) in nonrelativistic type IIB superstring theory. In section~\ref{eq:mbat}, we apply this formalism to compactify a single M5-brane in nonrelativistic M-theory. First, we review the Pasti-Sorokin-Tonin (PST) formalism \cite{Pasti:1997gx, Pasti:1995tn,Pasti:1996vs} of M5-brane and its compactification over a torus in relativistic M-theory \cite{Berman:1998va}. Then, we construct the M5-brane action~\eqref{eq:nrm5a} in nonrelativistic M-theory and study its toroidal compactification. We conclude the paper and give outlooks in section~\ref{sec:concl}. 
Appendix~\ref{sec:prid} discusses the Iwasawa decomposition of SL($2\,, \mathbb{R}$) in connection to the zweibein formalism of the two-torus. Appendix~\ref{app:quadM5} contains an expansion of the nonrelativistic M5-brane action to the quadratic order with respect to the worldvolume $U(1)$ gauge field strength. Appendix~\ref{app:D3} includes a derivation of the dual D3-brane action in section~\ref{sec:dtbantbst}. 

\section{Basic Elements of Nonrelativistic M-Theory} \label{sec:benmt}

We start by reviewing nonrelativistic string theory and its uplift to nonrelativistic M-theory. Then, we introduce the concept of anisotropic compactification in nonrelativistic M-theory and illustrate the dual relation between nonrelativistic and DLCQ M-theory.  

\subsection{The M-Theory Uplift of Nonrelativistic String Theory} \label{eq:mtunst}

Nonrelativistic string theory arises as a stringy limit of relativistic string theory, which requires reparametrizing the target space background fields in terms of a constant parameter $\omega$ followed by an infinite $\omega$ limit. In the following, we use the hatted letters to denote quantities in relativistic string theory and unhatted ones for nonrelativistic string theory. We will focus on the bosonic sector of type II superstring theories, where the relativistic closed-string background fields include the metric $\hat{G}_\text{MN}$\,, Kalb-Ramond field $\hat{B}_\text{MN}$\,, dilaton field $\hat{\Phi}$\,, and RR potentials $\hat{C}^{(q)}$\,, with $\text{M} = 0 \,, \,\cdots\,,\,9$\,. Note that the superscript ``$(q)$" implies that the quantity is a differential $q$-form. We will work in the Einstein frame, where the metric $\hat{G}^{}_\text{MN}$ is related to the string-frame metric $\hat{G}^\text{\,string}_\text{MN}$ via
\be
    \hat{G}^{}_\text{MN} = e^{-\hat{\Phi}/2} \, \hat{G}^\text{\,string}_\text{MN}\,.
\ee 
It is useful to introduce the vielbein fields $\hat{E}^{}_\text{M}{}^A$ and $\hat{E}^{}_\text{M}{}^{A'}$, such that
\be
    \hat{G}^{}_\text{MN} = \hat{E}^{}_\text{M}{}^A \, \hat{E}^{}_\text{N}{}^B \, \eta^{}_{AB} + \hat{E}^{}_\text{M}{}^{A'} \, \hat{E}^{}_\text{N}{}^{A'}\,, 
\ee
where we artificially split the frame index into the longitudinal sector labeled by a two-dimensional Minkowski index $A = 0\,, \, 1$ and the transverse sector by an eight-dimensional index $A'$\,. We then rescale the associated vielbein fields as
\begin{align} \label{eq:dte}
    \hat{E}^{}_\text{M}{}^A = \omega^{3/4} \, \tau^{}_\text{M}{}^A\,,
        \qquad%
    \hat{E}^{}_\text{M}{}^{A'} = \omega^{-1/4} \, E^{}_\text{M}{}^{A'}\,.
\end{align}
In the $\omega \rightarrow \infty$ limit, the target space develops an induced codimension-two foliation structure \cite{Andringa:2012uz}.
The longitudinal and transverse geometries are encoded by the vielbein fields $\tau^{}_\text{M}{}^A$ and $E^{}_\text{M}{}^{A'}$\,, respectively. The spacetime Poincar\'{e} group breaks down into the spacetime string Galilean group, where the local, finite string Galilean boost parametrized by the Lie group parameter $\lambda^{}_{AA'}$ is
\be \label{eq:sgb}
    \delta_\text{G} \tau^{}_\text{M}{}^A = 0\,,
        \qquad%
    \delta_\text{G} E^{}_\text{M}{}^{A'} = - \lambda_A{}^{A'} \, \tau^{}_\text{M}{}^A\,,
\ee
which relates the transverse sector to the longitudinal sector. 

Moreover, we reparametrize the remaining background fields as \cite{Bergshoeff:2019pij, Ebert:2021mfu}
\begin{subequations} \label{eq:bcprep}
\begin{align}
    \hat{B}^{(2)} & = - \omega^2 \, e^{\Phi/2} \, \ell^{(2)} + B^{(2)}\,, 
        &%
    \hat{\Phi} & = \Phi + \ln \omega\,, \\[4pt]
    \hat{C}^{(q)} & = \omega^2 \, e^{\Phi/2} \, \ell^{(2)} \wedge C^{(q-2)} + C^{(q)}\,,
\end{align}
\end{subequations}
where $B^{(2)}$, $\Phi$\,, and $C^{(q)}$ are respectively the Kalb-Ramond, dilaton, and RR fields in nonrelativistic string theory. Note that $q$ is odd for type IIA string theory and is even for type IIB string theory. We also defined 
\be \label{eq:framel}
    \ell^{(2)} = \frac{1}{2} \, \tau^A \wedge \tau^B \, \epsilon^{}_{AB}\,.
\ee
Here, the Levi-Civita symbol $\epsilon^{}_{AB}$ with frame indices is defined via $\epsilon^{}_{01} = 1$\,.
The spacetime geometry in the $\omega \rightarrow \infty$ limit is the \emph{(torsional) string Newton-Cartan geometry} \cite{Bergshoeff:2021bmc, Bidussi:2021ujm}, which is required to be supplement with intrinsic torsional constraints due to supersymmetry \cite{Bergshoeff:2021tfn} and quantum consistency \cite{Gomis:2019zyu, Yan:2021lbe}. In the resulting parametrization of the string Newton-Cartan geometry, $B^{(2)}$ and $C^{(q)}$ transform non-trivially under the local string Galilean boost \eqref{eq:sgb}, with
\begin{subequations} \label{eq:sgbbc}
\begin{align} 
    \delta^{}_\text{G} B^{(2)} & = \epsilon^{}_{AB} \, \lambda^{B}{}_{A'} \, e^{\Phi/2} \, \tau^A \wedge E^{A'}\,, \\[4pt]
    \delta^{}_\text{G} C^{(q)} & = - \epsilon^{}_{AB} \, \lambda^{B}{}_{A'} \, e^{\Phi/2} \, \tau^A \wedge E^{A'} \wedge C^{(q-2)}\,.
\end{align}
\end{subequations}
Moreover, the one-form gauge potential $\hat{A}^{(1)}$ coupled to relativistic open strings remains the same under the above reparametrization, \emph{i.e.}, $\hat{A}^{(1)} = A^{(1)}$\,. In relativistic IIB superstring theory, this one-form gauge potential can be extended to the Born-Infeld vector $\hat{A}^{(1)} = \bigl( \hat{A}^\text{\scalebox{0.8}{B}}\,, \hat{A}^\text{\scalebox{0.8}{C}} \bigr)^\intercal$\,, where $\hat{A}^\text{\scalebox{0.8}{B}}$ and $\hat{A}^{\text{\scalebox{0.8}{C}}}$ are associated with the Kalb-Ramond field $\hat{B}^{(2)}$ and RR two-form $\hat{C}^{(2)}$, respectively. Accordingly,
\be \label{eq:haa}
    \hat{A}^\text{\scalebox{0.8}{B}} = A^\text{\scalebox{0.8}{B}}\,,
        \qquad%
    \hat{A}^\text{\scalebox{0.8}{C}} = A^\text{\scalebox{0.8}{C}}\,,
\ee
define the Born-Infeld vector in nonrelativistic IIB theory. Since the Born-Infeld vector is unaffected by the nonrelativistic string limit, we will omit the hat for $U(1)$ gauge potentials in relativistic string theory and write them respectively as $A^\text{\scalebox{0.8}{B}}$ and $A^\text{\scalebox{0.8}{C}}$\,.

In the $\omega \rightarrow \infty$ limit of the relativistic string action, the Kalb-Ramond field is fine-tuned to cancel the string tension, which leads to the defining worldsheet theory for nonrelativistic string theory \cite{Gomis:2000bd}. The beta functions associated with the open string background fields were derived in \cite{Gomis:2020fui}, from which the effective D$p$-brane actions in nonrelativistic string theory are constructed. In particular, performing an S-duality transformation of the $U(1)$ gauge field on a single D2-brane in nonrelativistic string theory gives rise to the M2-brane action in nonrelativistic M-theory \cite{Ebert:2021mfu}.
Intriguingly, the resulting M2-brane couples to spacetime equipped with a codimension-three foliation structure.\,\footnote{See related work \cite{Kluson:2019uza} on the M2-brane in nonrelativistic M-theory. Also see \cite{Roychowdhury:2022est} for a nonrelativistic expansion of the M2-brane in relativistic M-theory.} 

The above duality relation determines the limiting prescription for deriving nonrelativistic M-theory from relativistic M-theory, which we detail below. The closed membrane background fields in relativistic M-theory include a metric $\hat{\mathbb{G}}^{}_{\mathbb{M}\mathbb{N}}$\,, $\mathbb{M} = 0\,, \, 1\,, \cdots\,, \, 10$\,, a three-form gauge potential $\hat{\mathbb{C}}^{(3)}$\,, and a six-form gauge potential $\hat{\mathbb{C}}^{(6)}$\,. As in the string case, it is also helpful to introduce the vielbein formalism for the M-theory metric $\hat{\mathbb{G}}^{}_{\mathbb{M}\mathbb{N}}$\,, with
\be \label{eq:mtgr}
    \hat{\mathbb{G}}^{}_{\mathbb{M}\mathbb{N}} = \hat{\mathbb{E}}^{}_{\,\mathbb{M}}{}^u \, \hat{\mathbb{E}}^{}_\mathbb{N}{}^v \, \eta^{}_{uv} + \hat{\mathbb{E}}^{}_{\,\mathbb{M}}{}^{A'} \, \hat{\mathbb{E}}^{}_\mathbb{N}{}^{A'}\,,
\ee
where the longitudinal $u$ index extends the two-dimensional index $A$ in nonrelativistic string theory to be three-dimensional. 
The appropriate rescalings of the vielbein fields are
\be \label{eq:resmth}
    \hat{\mathbb{E}}^{}_{\mathbb{M}}{}^u = \omega^{2/3} \, \gamma^{}_{\mathbb{M}}{}^u\,,
        \qquad%
    \hat{\mathbb{E}}^{}_{\mathbb{M}}{}^{A'} = \omega^{-1/3} \, \mathbb{E}^{}_{\mathbb{M}}{}^{A'}\,.
\ee
After performing the $\omega \rightarrow \infty$ limit, $\gamma^{}_{\mathbb{M}}{}^u$ ($\mathbb{E}^{}_\mathbb{M}{}^{A'}$) encodes the geometry in the three-dimensional (eight-dimensional) longitudinal (transverse) sector. These sectors are related to each other via a local, finite membrane Galilean boost parametrized by the Lie group parameter $\lambda^{}_{uA'}$\,,
\be \label{eq:mgbge}
    \delta^{}_\text{G} \gamma^{}_{\mathbb{M}}{}^u = 0 \,,
        \qquad%
    \delta^{}_\text{G} \mathbb{E}^{}_{\mathbb{M}}{}^{A'} = - \lambda^{}_u{}^{A'} \, \gamma^{}_{\mathbb{M}}{}^u\,.
\ee
Moreover, the higher-form gauge potentials $\hat{\mathbb{C}}^{(3)}$ and $\hat{\mathbb{C}}^{(6)}$ from eleven-dimensional supergravity \cite{Candiello:1993di,Aharony:1996wp} are reparametrized as 
\begin{subequations} \label{eq:hc36rp}
\begin{align}
    \hat{\mathbb{C}}^{(3)} & = - \omega^2 \, \Gamma^{(3)} + \mathbb{C}^{(3)}\,, \\[4pt]
    \hat{\mathbb{C}}^{(6)} & = \omega^2 \, \Gamma^{(3)} \wedge \mathbb{C}^{(3)} + \mathbb{C}^{(6)}\,,
\end{align}
\end{subequations}
where
\be \label{eq:defgt}
    \Gamma^{(3)} = \frac{1}{3!} \, \gamma^u \wedge \gamma^v \wedge \gamma^w \, \epsilon_{uvw}\,.
\ee
The spacetime geometry in the $\omega \rightarrow \infty$ limit is the \emph{membrane Newton-Cartan geometry} \cite{Blair:2021waq, Ebert:2021mfu}. In terms of the above prescriptions, the higher-form gauge potentials transform under the membrane Galilean boost as follows:
\begin{subequations} \label{eq:mgbc36}
\begin{align} 
    \mathbb{C}^{(3)} & \rightarrow \mathbb{C}^{(3)} - \frac{1}{2} \, \epsilon^{}_{uvw} \, \lambda^u{}^{}_{\!A'} \, \gamma^v \wedge \gamma^w \wedge \mathbb{E}^{A'}\,, \\[4pt]
    \mathbb{C}^{(6)} & \rightarrow \mathbb{C}^{(6)} + \epsilon^{}_{uvw} \, \lambda^u{}^{}_{\!A'} \, \gamma^v \wedge \gamma^w \wedge \mathbb{E}^{A'} \wedge \mathbb{C}^{(3)}\,.
\end{align}
\end{subequations}
We refer to this $\omega \rightarrow \infty$ limit of relativistic M-theory as the nonrelativistic membrane limit. Just like the string case, the two-form gauge potential $\mathbb{A}^{(2)}$ coupled to open membranes is unaffected by the limit. 

Upon compactifying nonrelativistic M-theory over a longitudinal spatial circle, it reduces to nonrelativistic type IIA superstring theory coupled to a ten-dimensional string Newton-Cartan geometry and equips to a codimension-two foliation structure. The Kaluza-Klein reduction map from nonrelativistic M-theory to IIA superstring theory is given by 
\begin{subequations} \label{eq:rm}
\be 
    \gamma^{}_\mathbb{M}{}^u = 
    e^{2\Phi/3}
    \begin{pmatrix}
        e^{-\Phi/2} \, \tau^{}_\text{M}{}^A &\,\, C^{}_\text{M}\\[4pt]
        0 &\,\, 1
    \end{pmatrix}\,,
        \qquad%
    \mathbb{E}^{}_\mathbb{M}{}^{A'} =
    e^{\Phi/6}
    \begin{pmatrix}
        E^{}_\text{M}{}^{A'} &\,\, 0 \\[4pt]
        0 &\,\, 0
    \end{pmatrix}\,.
\ee
Moreover, 
\begin{align}
    \mathbb{C}^{}_\text{MNL} & = C^{}_\text{MNL}\,,  
        &%
    \mathbb{C}^{}_{\text{M}^{}_0 \cdots \text{M}^{}_4,10} & = 2 \, \CC^{}_{\text{M}^{}_0 \cdots \text{M}^{}_4}\,, \\[4pt]
    \mathbb{C}^{}_{\text{MN},10} & = B^{}_\text{MN}\,,
        &%
    \mathbb{A}^{}_{\text{M},10} & = A^{}_\text{M}\,,
\end{align}
\end{subequations}
where
\be
    \CC^{(5)} = C^{(5)}_{} + \frac{1}{2} \, C^{(3)}_{} \wedge B^{(2)}_{}\,.
\ee
Note that the above reduction rules take the same form as the Kaluza-Klein reduction rules from relativistic M-theory to IIA theory. Moreover, the component $\mathbb{C}^{}_{\text{M}^{}_0 \cdots \text{M}^{}_5}$ gives rise to the magnetic dual of the $B$-field in ten dimensions.

\begin{figure}[t!]
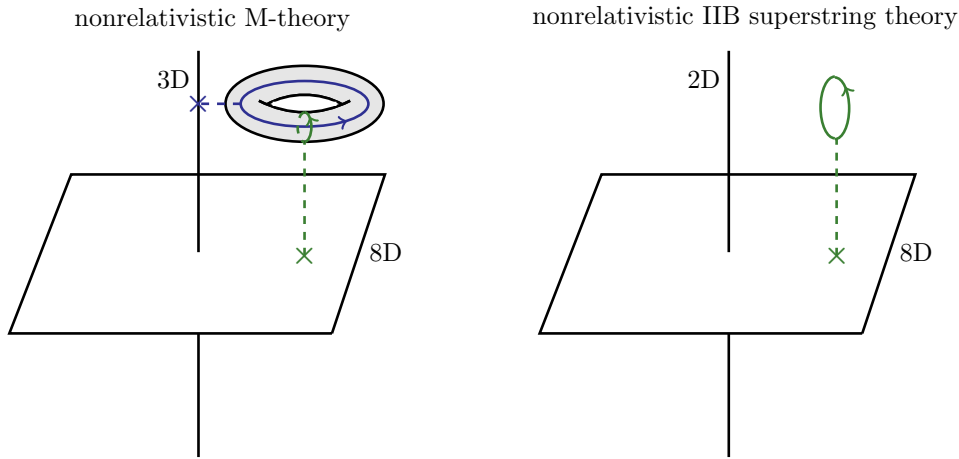

\begin{center}
\tikzpicture[scale=0.47, line width=0.35mm]
\draw (3.375+1,-4.5) -- (-3.625-0.125-1,-4.5) -- (-2-1,0) -- (5-0.125+1,0) -- (3.375+1,-4.5);

\draw (1,4.4) node{\scalebox{0.9}{nonrelativistic M-theory}};

\draw (0.6,-8) -- (0.6,-4.5);
\draw (0.6,-2.2)--(0.6,3.5);

\draw (-0.1,2.7) node{\scalebox{0.9}{3D}};
\draw (4.9+1,-2.2) node{\scalebox{0.9}{8D}};

\draw [OliveGreen,dashed] (3.6,-2.3)--(3.6,1);
\draw[OliveGreen] (3.6,-2.3) node{$\boldsymbol{\times}$};

\draw[OliveGreen, decoration={markings, mark=at position 0.8 with {\arrow{>}}}, postaction={decorate}] (18.1+0.5,1.5-0.5) arc (-90:90:0.45cm*0.9 and 0.97*0.9cm);
\draw[OliveGreen] (18.1+0.5,1.5+1.9*0.7+0.04+0.4-0.5) arc (90:270:0.45cm*0.9 and 0.97*0.9cm);

%\draw (1,-9) node {string Newton-Cartan};

\draw (15+3.375+1,-4.5) -- (15-3.625-0.125-1,-4.5) -- (15-2-1,0) -- (15+5-0.125+1,0) -- (15+3.375+1,-4.5);

\draw (15+0.6,-8) -- (15+0.6,-4.5);
\draw (15+0.6,-2.2)--(15+0.6,3.5);

\draw [OliveGreen,dashed] (15+2.4+0.5+0.75,-2.3)--(15+2.4+0.5+0.75,1);
\draw[OliveGreen] (15+2.4+0.5+0.75,-2.3) node{$\boldsymbol{\times}$};

%\draw[] (15+1,-9) node {membrane Newton-Cartan};

\draw (15-0.1,2.7) node{\scalebox{0.9}{2D}};
\draw (15+4.9+1,-2.2) node{\scalebox{0.9}{8D}};

%%%%%%%%%%%%

\scope[xshift=3.6cm,yshift=2cm,scale=.43]
\fill[light-gray] (0,0) ellipse (5.2 and 2.5);

\begin{scope}
  \clip (0,0) ellipse (5.2 cm and 2.5 cm);
  \path[draw,name path= A](-3,0.2) .. controls (-1,-0.8) and (1,-0.8) .. (3,0.2);
  \clip (0,0) ellipse (5.2 cm and 1.5 cm);
  \path[draw,name path=B] (-2.5,-0.05) .. controls (-1,0.8) and (1,0.8) .. (2.5,-0.05);
\fill [white, intersection segments={of=A and B, sequence={A1--B1}}];
\end{scope}

\draw[] (0,0) ellipse (5.2 cm and 2.5cm);
\draw[] (-3,0.2) .. controls (-1,-0.8) and (1,-0.8) .. (3,0.2);
\draw[] (-2.5,-0.05) .. controls (-1,0.8) and (1,0.8) .. (2.5,-0.05);

\draw[Blue,  decoration={markings, mark=at position 0.9 with {\arrow{>}}},
        postaction={decorate}](0,0) ellipse  (4.2 cm and 1.5cm);

\draw[OliveGreen,  decoration={markings, mark=at position 0.8 with {\arrow{>}}},
        postaction={decorate}] (0,-2.5) arc (-90:90:0.45cm and 0.97cm);
\draw[ OliveGreen,dashed] (0,-0.575) arc (90:270:0.45cm and 0.97cm);

\draw[ultra thick, White] (-2.1,-.0035) -- (2.1,-.0035);

\draw (29, 5.7) node{\scalebox{0.9}{nonrelativistic IIB superstring theory}};

%\draw[OliveGreen] (0,-7) node{$\times$};
%\draw [OliveGreen,dashed] (0,-6.8)--(0,-2.4);
\draw[Blue] (-7,0) node{$\boldsymbol{\times}$};
\draw [Blue,dashed] (-6.8,0)--(-4.2,0);
\endscope
\endtikzpicture
\caption{We show the relation between nonrelativistic M-theory over an anisotropic two-torus and nonrelativistic IIB superstring theory. The figure on the left displays nonrelativistic M-theory coupled to an eleven-dimensional membrane Newton-Cartan geometry containing a three-dimensional longitudinal sector and an eight-dimensional transverse sector. The figure on the right depicts nonrelativistic IIB superstring theory coupling to a ten-dimensional string Newton-Cartan geometry with a two-dimensional longitudinal sector and an eight-dimensional transverse sector. Here, nonrelativistic M-theory compactified over an anisotropic torus with the blue cycle in the longitudinal sector and the green cycle in the transverse sector. While the blue cycle is the M-theory circle, the green cycle is T-dualized to the transverse circle in nonrelativistic IIB superstring theory. 
\label{fig:foliation}}
\end{center}
\end{figure}

\subsection{Anisotropic Two-Torus and \texorpdfstring{SL($2\,,\mathbb{Z}$)}{SLZ} Duality} \label{sec:acsld}

In this subsection, the global SL($2\,,\mathbb{Z}$) duality in type IIB superstring theory gains a geometric interpretation as the isometry group on the two-torus over which M-theory compactifies. In nonrelativistic M-theory, the analog of this toroidal compactification has an intriguing interplay with the codimension-three foliation structure. 

We already discussed in section~\ref{eq:mtunst} that compactifying nonrelativistic M-theory coupled to membrane Newton-Cartan geometry over a longitudinal spatial direction gives rise to nonrelativistic IIA superstring theory coupling to string Newton-Cartan geometry. To consider the SL($2\,,\,\mathbb{Z}$) duality in nonrelativistic IIB superstring theory, we further compactify nonrelativistic IIA superstring theory over a second spatial circle. Performing a T-duality along this second spatial circle gives rise to IIB superstring theory. If this extra spatial circle is in the longitudinal sector, the two-torus is isotropic, and the resulting description is the DLCQ of relativistic IIB string theory. Instead, if this second circle is in the transverse sector, the two-torus is anisotropic, and the resulting theory is nonrelativistic IIB string theory. This latter configuration with a transverse compactification is what we are interested in. Therefore, compactifying nonrelativistic M-theory over an anisotropic torus leads to nonrelativistic IIB string theory. 

\begin{figure}[t!]
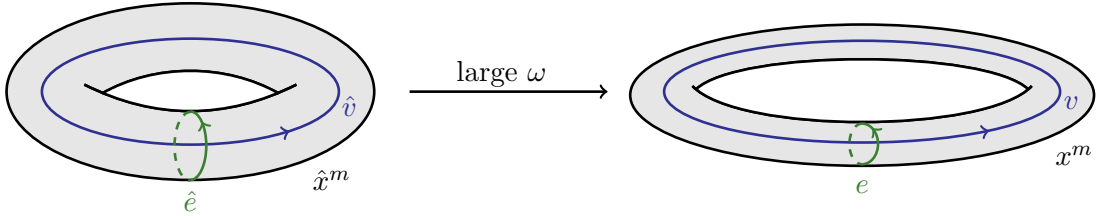

\begin{center}
\tikzpicture[scale=0.47, line width=0.35mm]

\fill[light-gray] (0,0) ellipse (5.2 and 2.5);

\begin{scope}
  \clip (0,0) ellipse (5.2 cm and 2.5 cm);
  \path[draw,name path= A](-3,0.2) .. controls (-1,-0.8) and (1,-0.8) .. (3,0.2);
  \clip (0,0) ellipse (5.2 cm and 1.5 cm);
  \path[draw,name path=B] (-2.5,-0.05) .. controls (-1,0.8) and (1,0.8) .. (2.5,-0.05);
\fill [white,
          intersection segments={
            of=A and B,
            sequence={A1--B1}
          }];
\end{scope}

\draw[] (0,0) ellipse  (5.2 cm and 2.5cm);
\draw[] (-3,0.2) .. controls (-1,-0.8) and (1,-0.8) .. (3,0.2);
\draw[] (-2.5,-0.05) .. controls (-1,0.8) and (1,0.8) .. (2.5,-0.05);

\draw[ Blue,  decoration={markings, mark=at position 0.9 with {\arrow{>}}},
        postaction={decorate} ](0,0) ellipse  (4.2 cm and 1.5cm);

\draw[OliveGreen,  decoration={markings, mark=at position 0.8 with {\arrow{>}}},
        postaction={decorate}] (0,-2.5) arc (-90:90:0.45cm and 0.97cm);
\draw[ OliveGreen,dashed] (0,-0.575) arc (90:270:0.45cm and 0.97cm);
\draw[Blue] (4.5,-0.4) node{$\hat{v}$};
\draw[OliveGreen] (0,-3.1) node{$\hat{e}$};
\draw[->] (6.2,0) -- (11.8,0);
\draw (8.8,.5) node{large $\omega$};
\draw (4,-2.5) node{$\hat{x}^m$};
\draw (25,-1.8) node{$x^m$};
\draw[ultra thick, White] (-2.21,-.0035) -- (2.21,-.0035);

\scope[xshift=19cm,yshift=0cm,scale=1]
\fill[light-gray] (0,-0.1) ellipse (8.2*.8 and 2.5*.8);
\begin{scope}
  \clip (0,0) ellipse (8.2*.8 cm and 2.5*.8 cm);
  \path[draw,name path= A] (-6*.8,0.2*.8) .. controls (-3.5,-1.2) and (3.5,-1.2) .. (6*.8,0.2*.8);
  \clip (0,0) ellipse (8.2*.8 cm and 1.5*.8 cm);
  \path[draw,name path=B] (-5.75*.8-0.1,0.08*.8) .. controls (-3.5,1.2) and (3.5,1.2) .. (5.75*.8+0.1,0.08*.8);
\fill [white,
          intersection segments={
            of=A and B,
            sequence={A1--B1}
          }];
\end{scope}

\draw[] (0,-0.1) ellipse  (8.2*0.8 cm and 2.5*.8cm);
\draw[] (-4.8,0.16) .. controls (-3.5,-1.2) %(-0.8,-1.2) 
and (3.5,-1.2)
%(1*.8,-1.5*.8) 
.. (6*.8,0.2*.8);
\draw[] (-5.75*.8-0.1,0.08*.8) .. controls (-3.5,1.2) %(-2*.8,1.4*.8) 
and (3.5,1.2) %(2*.8,1.4*.8) 
.. (5.75*.8+0.1,0.08*.8);

\draw[Blue,  decoration={markings, mark=at position .9 with {\arrow{>}}},
        postaction={decorate}](0,0) ellipse  (7*.8cm and 1.8*.8cm);

\draw[OliveGreen,  decoration={markings, mark=at position 1*.8 with {\arrow{>}}},
        postaction={decorate}] (0,-2.57*.8) arc (-90:90:0.45cm and 0.72*.8cm);
\draw[ OliveGreen,dashed] (0,-1.13*.8) arc (90:270:0.45*.8cm and 0.72*.8 cm);
\draw[Blue] (7.4*.8,-0.4*.8) node{$v$};
\draw[OliveGreen] (0,-2.7) node{$e$};
\draw[ultra thick, White] (-5.65*.8,.087*.8) -- (5.65*.8,.087*.8);
\endscope
\endtikzpicture
\caption{Reparametrization of a two-torus. The blue cycle lies along a longitudinal direction, while the green cycle lies along a transverse direction. In the $\omega \rightarrow\infty$ limit, the longitudinal cycle becomes infinitely large while the transverse cycle shrinks to zero size. This limit implies that the metric on the two-torus becomes singular, and one has to describe the toroidal geometry using vielbein fields. \label{fig:torus}}
\end{center}
\end{figure}

The above observation implies that the SL($2\,,\,\mathbb{Z}$) duality in nonrelativistic IIB theory originates from compactifying nonrelativistic M-theory over an anisotropic torus that straddles between the longitudinal and transverse sectors in the target space membrane Newton-Cartan geometry. It is illustrative to consider the two-torus in the flat limit with $\gamma^{}_\mathbb{M}{}^u = \delta_\mathbb{M}^u$ and $\mathbb{E}^{}_\mathbb{M}{}^{A'} = \delta_\mathbb{M}^{A'}$\,. In this flat case, the anisotropic torus arises from rescaling the coordinates on a Riemannian torus. Define the target space coordinates in relativistic M-theory to be $\hat{X}^\mathbb{M} = (\hat{x}^\mu\,, \, \hat{x}^m)$\,, where $\hat{x}^m$\,, $m=9\,,\,10$ are coordinates on the internal torus and $\hat{x}^\mu$ are spacetime coordinates in relativistic type IIB superstring theory. The rescalings of the vielbein fields in Eq.~\eqref{eq:resmth} imply 
\be
    \hat{x}^{0\,,\, 1\,,\, 9} = \omega^{2/3} \, x^{0\,,\, 1\,,\, 9}
\ee
in the three-dimensional longitudinal sector and
\be
    \hat{x}^{2\,,\, \cdots\,,\, 8\,, \, 10} = \omega^{-1/3} \, x^{2\,,\, \cdots\,,\, 8\,, \, 10}
\ee
in the eight-dimensional transverse sector of the membrane Newton-Cartan geometry. Therefore, on the compactified torus, at large $\omega$\,, the longitudinal cycle is enlarged while the transverse cycle shrinks. In the $\omega\rightarrow\infty$ limit, the toroidal geometry becomes singular: one seems to find a ring-like object of an infinite radius. Nevertheless, this singularly limiting behavior does not indicate any sickness of the geometry but only implies that it is inappropriate to use a metric description anymore. In fact, in the $\omega \rightarrow \infty$ limit, the toroidal geometry becomes non-Riemannian and should be described using vielbein fields. See figure~\ref{fig:torus} for an illustration. In the next section, we quantify this non-Riemannian geometry on the anisotropic torus.

We have seen that compactifying nonrelativistic M-theory over an anisotropic two-torus leads to nonrelativistic IIB string theory. This relation implicitly assumes that we regard the longitudinal cycle on the two-torus to be the M-theory circle. However, it is physically equivalent if we instead take the transverse cycle on the two-torus be the M-theory circle. In this latter case, we find the limit of relativistic IIB superstring theory where the RR two-form fine-tunes to cancel the D1-brane tension. This critical RR two-form limit of the D1-brane is closely related to Matrix string theory \cite{Danielsson:2000gi} (also see \cite{uduality} for further discussions). This limit implies that nonrelativistic IIB superstring theory is S-dual to the critical RR two-form limit of relativistic IIB superstring theory. As we will see later in section~\ref{sec:onebranelimit}, there is a fraction of the SL($2\,, \mathbb{Z}$) group transformations that map these two corners. 

\subsection{Anisotropic Three-Torus and Matrix Theory}

\begin{figure}[t!]
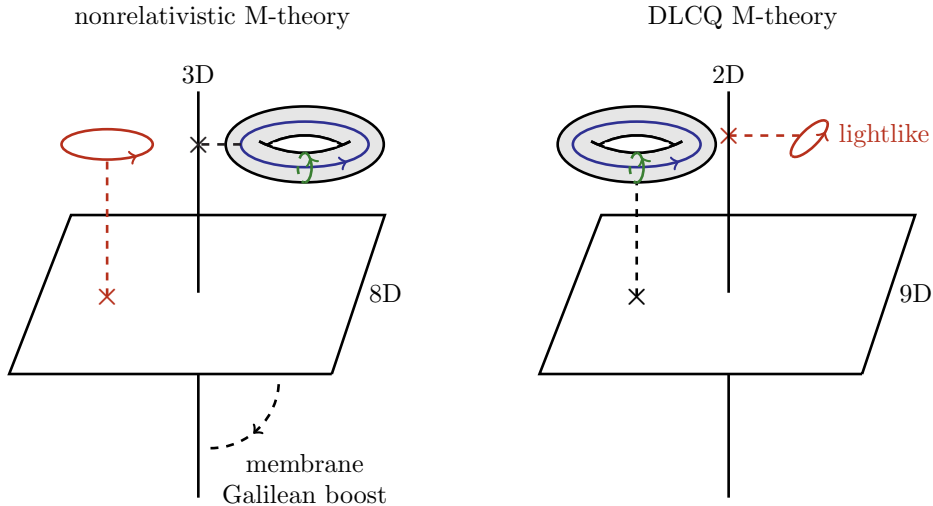

\begin{center}

\tikzpicture[scale=0.47, line width=0.35mm]

\draw (3.375+1,-4.5) -- (-3.625-0.125-1,-4.5) -- (-2-1,0) -- (5-0.125+1,0) -- (3.375+1,-4.5);

\draw (0.6,-8) -- (0.6,-4.5);
\draw (0.6,-2.2)--(0.6,3.5);
\draw (1,5.6) node{\scalebox{0.9}{nonrelativistic M-theory}};
\draw (0.6,4) node{\scalebox{0.9}{3D}};
\draw (4.9+1,-2.2) node{\scalebox{0.9}{8D}};
\draw (20,2.25) node{\scalebox{0.9}{\textcolor{BrickRed}{lightlike}}};
\draw [BrickRed,dashed] (15.6,2+.25)--(15.6+2,2+.25);
\draw[BrickRed] (15.6cm,2+.25) node{$\boldsymbol{\times}$};
\draw[BrickRed, decoration={markings, mark=at position 0.8 with {\arrow{>}}}, postaction={decorate},cm={cos(45) ,-sin(45) ,sin(45) ,cos(45) ,(14.6 cm, 2.7 cm)}] (2.4+0.3,1.25) arc (-90:90:0.35cm*0.7 and 0.99*0.7cm);
\draw[BrickRed,cm={cos(45) ,-sin(45) ,sin(45) ,cos(45) ,(14.6 cm, 2.7 cm)}] (2.4+0.3,1.25+1.9*0.7+0.05) arc (90:270:0.35cm*0.7 and 0.99*0.7cm);

\draw (15+3.375+1,-4.5) -- (15-3.625-0.125-1,-4.5) -- (15-2-1,0) -- (15+5-0.125+1,0) -- (15+3.375+1,-4.5);

\draw (15+0.6,-8) -- (15+0.6,-4.5);
\draw (15+0.6,-2.2)--(15+0.6,3.5);

\draw (15+1,5.6) node{\scalebox{0.9}{DLCQ M-theory}};
\draw (15+0.6,4) node{\scalebox{0.9}{2D}};
\draw (15+4.9+1,-2.2) node{\scalebox{0.9}{9D}};
\scope[xshift=13 cm,yshift=2cm,scale=.43]

\fill[light-gray] (0,0) ellipse (5.2 and 2.5);

\begin{scope}
  \clip (0,0) ellipse (5.2 cm and 2.5 cm);
  \path[draw,name path= A](-3,0.2) .. controls (-1,-0.8) and (1,-0.8) .. (3,0.2);
  \clip (0,0) ellipse (5.2 cm and 1.5 cm);
  \path[draw,name path=B] (-2.5,-0.05) .. controls (-1,0.8) and (1,0.8) .. (2.5,-0.05);
\fill [white, intersection segments={of=A and B, sequence={A1--B1}}];
\end{scope}

\draw[] (0,0) ellipse (5.2 cm and 2.5cm);
\draw[] (-3,0.2) .. controls (-1,-0.8) and (1,-0.8) .. (3,0.2);
\draw[] (-2.5,-0.05) .. controls (-1,0.8) and (1,0.8) .. (2.5,-0.05);

\draw[Blue,  decoration={markings, mark=at position 0.9 with {\arrow{>}}},
        postaction={decorate}](0,0) ellipse  (4.2 cm and 1.5cm);

\draw[OliveGreen,  decoration={markings, mark=at position 0.8 with {\arrow{>}}},
        postaction={decorate}] (0,-2.5) arc (-90:90:0.45cm and 0.97cm);
\draw[ OliveGreen,dashed] (0,-0.575) arc (90:270:0.45cm and 0.97cm);

\draw[ultra thick, White] (-2.1,-.0035) -- (2.1,-.0035);

\draw[black] (0,-10) node{$\boldsymbol{\times}$};
\draw [black,dashed] (0,-10)--(0,-2.6);
\endscope

%%%%%%%%%%%% SECOND TORUS

\scope[xshift=3.61cm,yshift=2cm,scale=.43]

\fill[light-gray] (0,0) ellipse (5.2 and 2.5);

\begin{scope}
  \clip (0,0) ellipse (5.2 cm and 2.5 cm);
  \path[draw,name path= A](-3,0.2) .. controls (-1,-0.8) and (1,-0.8) .. (3,0.2);
  \clip (0,0) ellipse (5.2 cm and 1.5 cm);
  \path[draw,name path=B] (-2.5,-0.05) .. controls (-1,0.8) and (1,0.8) .. (2.5,-0.05);
\fill [white, intersection segments={of=A and B, sequence={A1--B1}}];
\end{scope}

\draw[] (0,0) ellipse (5.2 cm and 2.5cm);
\draw[] (-3,0.2) .. controls (-1,-0.8) and (1,-0.8) .. (3,0.2);
\draw[] (-2.5,-0.05) .. controls (-1,0.8) and (1,0.8) .. (2.5,-0.05);

\draw[Blue,  decoration={markings, mark=at position 0.9 with {\arrow{>}}},
        postaction={decorate}](0,0) ellipse  (4.2 cm and 1.5cm);

\draw[OliveGreen,  decoration={markings, mark=at position 0.8 with {\arrow{>}}},
        postaction={decorate}] (0,-2.5) arc (-90:90:0.45cm and 0.97cm);
\draw[ OliveGreen,dashed] (0,-0.575) arc (90:270:0.45cm and 0.97cm);

\draw[ultra thick, White] (-2.1,-.0035) -- (2.1,-.0035);

\draw[Black] (-7,0) node{$\boldsymbol{\times}$};
\draw [Black,dashed] (-6.8,0)--(-4.2,0);
\draw[BrickRed,  decoration={markings, mark=at position 0.9 with {\arrow{>}}},
        postaction={decorate}](-13,0) ellipse  (3 cm and 1cm);
\draw[BrickRed] (-13,-10) node{$\boldsymbol{\times}$};
\draw [BrickRed,dashed] (-13,-10)--(-13,-1);
\draw[decoration={markings, mark=at position 0.5 with {\arrow{<}}},
        postaction={decorate}, dashed] (-6.2,-20) arc (-90:0:4.5cm and 4.5cm);
\draw (0,-21) node{\scalebox{0.9}{membrane}};
\draw (0,-23) node{\scalebox{0.9}{Galilean boost}};
\endscope

\endtikzpicture
\caption{We illustrate the anisotropic compactification over a three-torus and duality between DLCQ and nonrelativistic M-theory. While the spacetime geometry in DLCQ M-theory is Riemannian, the membrane Newton-Cartan geometry in nonrelativistic M-theory is non-Riemannian and consequently does not admit any global metric.    
\label{fig:DLCQnrmt}}
\end{center}
\end{figure}

Until now, we have been focusing on the anisotropic compactification over a two-torus. In this subsection, we will instead consider an anisotropic compactification of nonrelativistic M-theory over a three-torus, which is mandatory for constructing the dual relation between nonrelativistic and DLCQ M-theory. We illustrate this duality relation below and sketch how Matrix theory arises in this context. See \cite{uduality} for the associated U-duality relation in eleven dimensions, where concrete relationships to Matrix (gauge) theories are studied. 

We start with nonrelativistic M-theory. For simplicity, we focus on flat spacetime. As described in \S\ref{eq:mtunst}, the spacetime directions are factorized into a three-dimensional longitudinal sector containing $x^{0,\,1,\,9}$ and an eight-dimensional transverse sector containing $x^{2,\,\cdots,\,8,\,10}$\,. We consider a compactification of the target space over a three-torus, with two cycles in the longitudinal $x^{1,\,9}$ directions and the remaining cycle in the transverse $x^{10}$ direction. Performing a dimensional reduction along the $x^{9}$ circle, we find nonrelativistic IIA superstring theory. Next, performing a longitudinal T-duality transformation along the $x^1$ circle gives rise to the DLCQ of relativistic IIB superstring theory, where the original spacelike $x^1$ circle in nonrelativistic IIB superstring theory now maps to a dual circle that is lightlike. Finally, we perform a transverse T-duality transformation along the $x^{10}$ circle, which gives rise to the DLCQ of relativistic IIA superstring theory. Uplifting this resulting DLCQ IIA theory to relativistic M-theory, we recover DLCQ M-theory. See figure~\ref{fig:DLCQnrmt} for an illustration. The above prescriptions essentially form a U-duality relation between nonrelativistic and DLCQ M-theory, where the latter corresponds to Matrix theory \cite{uduality}.

To see how matrix theory arises from the M2-branes in nonrelativistic M-theory, at least in a schematic way, we start with the bosonic action describing a single M2-brane in nonrelativistic M-theory \cite{Gomis:2004pw, Kluson:2019uza}, 
\be \label{eq:nrm2}
    S^{}_\text{M2} \sim - \frac{1}{2} \int d^3 \sigma \, \sqrt{-\gamma} \, \gamma^{\alpha\beta} \, \p_\alpha X^\mathbb{M} \, \p_\beta X^\mathbb{N} \, \mathbb{E}^{}_{\mathbb{M}\mathbb{N}} - \int \mathbb{C}^{(3)}\,.
\ee
The worldvolume fields $X^\mathbb{M}$ define  the embedding coordinates.
We also defined the pullback $\gamma^{}_{\alpha\beta} = \p^{}_\alpha X^\mathbb{M} \, \p^{}_\beta X^\mathbb{N} \, \gamma^{}_{\mathbb{M}\mathbb{N}}$\,, with $\gamma^{\alpha\beta}$ the inverse of $\gamma^{}_{\alpha\beta}$ and $\gamma = \det \gamma^{}_{\alpha\beta}$\,. Moreover, the Chern-Simons term $\mathbb{C}^{(3)}$ denotes the pullback of the three-form gauge potential to the worldvolume. We require that the M2-brane is localized in the $x^{9}$ circle. Perform a duality transformation of the Nambu-Goldstone boson that perturbs the shape of the M2-brane in $x^{9}$ gives rise to the following D2-brane action in nonrelativistic string theory \cite{Ebert:2021mfu}: 
\be \label{eq:nrsd2}
    S^{}_\text{D2} \sim - \int d^3\sigma \, e^{-\Phi} \sqrt{-\det 
    \begin{pmatrix}
        0 &\quad \tau^{}_\beta{}^0 + \tau^{}_\beta{}^1 \\[4
        pt]
        {\tau}^{}_\alpha{}^0 - {\tau}^{}_\alpha{}^1 &\quad E^{}_{\alpha\beta} + \CF^{}_{\alpha\beta}
    \end{pmatrix}} - \int \Bigl( C^{(3)} + C^{(1)} \wedge \CF \Bigr)\,.
\ee
Note that the collective coordinate $x^{9}$ is dualized to be the $U(1)$ gauge potential $A^{}_\alpha$ on the D2-brane in Eq.~\eqref{eq:nrsd2}. Note that the size of the longitudinal circle along $x^{10}$ controls the size of the string coupling $e^{\langle \Phi \rangle}$\,. Here, $\CF^{(2)} = B^{(2)} + dA^{(1)}$\,, and the background fields are pulled back from spacetime to the worldvolume. The reduction map, which establishes the correspondence between the components in Eq.\eqref{eq:nrsd2} and their counterparts in Eq.\eqref{eq:nrm2}, is provided in Eq.~\eqref{eq:rm}.  
Next, we perform a longitudinal T-duality transformation along the $x^1$ circle. This T-duality leads to a dual D1-string action in the DLCQ of relativistic IIB string theory, where there is a lightlike compactification along $x^+$ that is T-dual to $x^1$\,. The dual D1-string is localized in the lightlike $x^+$ circle.
Finally, performing a T-duality transformation along the transverse $x^{10}$ circle, we are led to the action describing a single D0-brane in the DLCQ of relativistic IIA superstring theory, which in flat spacetime reads:
\be \label{eq:dza}
    S \sim  \int dt \, \sqrt{ \dot{x}^- \, \dot{x}^+ + \dot{x}^{A'} \dot{x}^{A'}}\,. 
\ee
Here, the lightlike direction $x^+$ is compactified\,. N\"{a}ively, the Hamiltonian associated with the action~\eqref{eq:dza} is zero, which is due to the existence of the worldline diffeomorphism. To compute the spectrum of the system, we impose the gauge fixing condition $x^- = t$\,. Furthermore, we note that the momenta $p^{}_+$ and $p^{}_{A'}$, which are conjugate to the lightlike coordinate $x^+$ and the transverse coordinates $x^{A'}$ are, respectively,
\be \label{eq:p+}
    p^{}_+ = \frac{1}{2 \sqrt{\dot{x}^+ + \dot{x}^{A'} \, \dot{x}^{A'}}}\,,
        \qquad%
    p^{}_{A'} = 2 \, p^{}_+ \, \dot{x}^{A'}.
\ee
The Hamiltonian is 
\be \label{eq:hmt}
    H \sim \int dt \, \frac{p^{}_{A'} \, p^{}_{A'} - 1}{4 \, p_+}\,,
\ee
which one uplifts to the Kaluza-Klein modes in DLCQ M-theory. When considering a stack of D0-branes, the Hamiltonian \eqref{eq:hmt} is generalized to describe Matrix theory. Ultimately, understanding Matrix theory using nonrelativistic M-theory requires considering a stack of nonrelativistic M2-branes and possibly some limiting version of ABJM superconformal field theory \cite{Aharony:2008ug}.

\section{Anisotropic Torus and Branched \texorpdfstring{SL($2\,,\mathbb{Z}$)}{SLZ} Duality} \label{sec:atbsld}

We now formulate the compactification of nonrelativistic M-theory over an anisotropic two-torus. We have shown that nonrelativistic M-theory arises as an $\omega \rightarrow \infty$ limit of relativistic M-theory with the reparametrization of the target space vielbein fields in Eq.~\eqref{eq:resmth}. Moreover, we require that the torus lies in the isometry longitudinal $x^9$ and transverse $x^{10}$ direction. The reparametrization \eqref{eq:resmth} of the vielbein fields becomes
\begin{subequations} \label{eq:euapr}
\begin{align} 
    \hat{\mathbb{E}}^u & = \omega^{2/3} \, \gamma^u\,,
        \qquad\quad\,\,\,%
    u = 0\,, 1\,, 9\,; \\[4pt]
    \hat{\mathbb{E}}^{A'} & = \omega^{-1/3} \, \mathbb{E}^{A'}\,,
        \qquad%
    A' = 2\,, \cdots\,, 8\,, 10\,.
\end{align}
\end{subequations}
Before sending $\omega$ to infinity, there is a well-defined metric describing the Riemannian geometry of the torus. However, the metric description becomes invalid after performing the $\omega \rightarrow \infty$ limit. 
In the following, we will first develop the zweibein formalism of the toroidal geometry in the Riemannian case. This zweibein formalism will help facilitate with the $\omega \rightarrow \infty$ limit. We will then use this formalism to provide the M-theory interpretation of the SL($2\,,\mathbb{Z}$) duality in nonrelativistic IIB string theory. This procedure will naturally lead to the polynomial realization of SL($2\,,\mathbb{Z}$) discovered in \cite{Bergshoeff:2023ogz}, which we will review in section~\ref{sec:mtoprsl}.

\subsection{Zweibein Formalism of Torus Geometry} \label{sec:zftg}

We first consider the compactification of relativistic M-theory over a two-torus. The eleven-dimensional target space in M-theory equips to a metric background field $\mathbb{G}_{\mathbb{I}\mathbb{J}}$ such that $\mathbb{I}\,, \mathbb{J} = 0\,, \, 1\,, \, \cdots, 10$\,, for which we take the following dimension-reduction ansatz:
\be
    \hat{\mathbb{G}}^{}_{\mathbb{I}\mathbb{J}} = 
    \begin{pmatrix}
        \hat{G}_{\mu\nu} &\qquad 0 \\[4pt]
        0 &\qquad \hat{g}^{}_{mn}
    \end{pmatrix}\,,
        \qquad%
    \mu = 0\,, 1\,, \cdots\,, 8\,,
        \qquad%
    m = 9\,, 10\,.
\ee
We have excluded any Kaluza-Klein modes for simplicity. Here, $\hat{G}_{\mu\nu}$ is the metric in the target space of ten-dimensional superstring theory compactified over $S^1$\,, and $\hat{g}^{}_{mn}$ is the metric on the torus, with 
\be \label{eq:torusmetric}
    \hat{g}^{}_{mn} = \frac{\Gamma}{\hat{\tau}^{}_2}
    \begin{pmatrix}
        1 &\qquad - \hat{\tau}^{}_1 \\[4pt]
        - \hat{\tau}^{}_1 &\qquad \hat{\tau}^2_1 + \hat{\tau}^2_2
    \end{pmatrix}\,,
\ee
where $\hat{\tau} = \hat{\tau}^{}_1 + i \, \hat{\tau}^{}_2$ is the torus modulus. The second fundamental form of the torus is
\be \label{eq:fff}
    ds^2 = \frac{\Gamma}{\hat{\tau}^{}_2} \, \Bigl| d\hat{x}^{9} - \hat{\tau} \, d\hat{x}^{10} \Bigr|^2\,,
\ee
where $\Gamma$ is the surface area of the torus. For now, we set $\Gamma = 1$ for convenience. Later, when we consider the dimensional reduction to IIB superstring theory, the dependence on $\Gamma$ will be recovered, as the limit $\Gamma \rightarrow 0$ is required. Note that $\hat{x}^m$ satisfies the periodic boundary condition
\be \label{eq:pbc}
    \hat{x}^{9\,,\,10} \sim \hat{x}^{9\,,\,10} + 1\,.
\ee
The isometry group on the two-torus is SL($2\,, \mathbb{Z}$)\,, which acts on the coordinates $\hat{x}^m$ and the modulus $\hat{\tau}$ as
\be \label{eq:sl2zrel}
    \hat{x}^m
    \rightarrow
    \Lambda^m{}_n \, \hat{x}^n\,,
        \qquad%
    \hat{\tau} \rightarrow 
    \frac{\alpha \, \hat{\tau} + \beta}{\gamma \, \hat{\tau} + \delta}\,.
\ee
See Eq.~\eqref{eq:sl2zg} for the definition of $\Lambda$\,. Note that the group parameters must be integers to obey the periodic boundary conditions in Eq.~\eqref{eq:pbc}. The unimodularity condition $\alpha \, \delta - \beta \, \gamma = 1$ in Eq.~\eqref{eq:sl2zg} is imposed such that $ds^2$ remains invariant under Eq.~\eqref{eq:sl2zrel}. After the toroidal compactification, in IIB superstring theory, the modulus becomes
\be \label{eq:c0pt}
    \hat{\tau} = \hat{C}^{(0)} + i \, e^{- \hat{\Phi}}
\ee
where $\hat{\tau}^{}_1 = \hat{C}^{(0)}$ is the RR zero-form and $\hat{\tau}^{}_2 = e^{-\hat{\Phi}}$ is the inverse string coupling. 

To facilitate the nonrelativistic membrane limit, performed via rescaling the vielbein fields in spacetime as in Eq.~\eqref{eq:euapr}, we introduce a zweibein formalism of the toroidal geometry by rewriting the torus metric \eqref{eq:torusmetric} as
\be
    \hat{g}^{}_{mn} = \hat{e}^{}_m{}^a \, \hat{e}^{}_n{}^b \, \delta^{}_{ab} = \hat{v}_m \, \hat{v}_n + \hat{e}_m \, \hat{e}_n\,,
\ee
where we defined
\be
    \hat{e}_m{}^a
    = 
    \begin{pmatrix}
        \hat{v}^{}_m \\[4pt]
        \hat{e}^{}_m
    \end{pmatrix},
        \qquad%
     \hat{v}_m \equiv \hat{e}_m{}^9 = 
        \frac{1}{\sqrt{\hat{\tau}^{}_2}}
    \begin{pmatrix}
        1 \\[4pt]
        - \hat{\tau}^{}_1
    \end{pmatrix},
        \qquad%
    \hat{e}_m \equiv \hat{e}_m{}^{10} =
    \sqrt{\hat{\tau}^{}_2}
    \begin{pmatrix}
        0 \\[4pt]
        1
    \end{pmatrix}.
\ee
The SL($2\,,\mathbb{Z}$) transformations of $\hat{e}^{}_m{}^a$ are not particularly illuminating at first sight. However, it is intriguing to consider the SL($2\,,\mathbb{Z}$) transformation of the quantities 
\be \label{eq:defve}
    \hat{v} = \hat{v}_{m} \, d\hat{x}^m\,,
        \qquad%
    \hat{e} = \hat{e}_{m} \, d\hat{x}^m\,.
\ee
Using the transformations in Eq.~\eqref{eq:sl2zrel}, we find
\be \label{eq:trnsfeaa}
    \begin{pmatrix}
        \hat{v} \\[4pt]
        \hat{e}
    \end{pmatrix} 
        \rightarrow% 
    \frac{\sgn \bigl( \gamma \, \hat{\tau}^{}_1 + \delta \bigr)}{\sqrt{1 + \hat{\kappa}^2}} 
    \begin{pmatrix}
        1 &\quad -\hat{\kappa} \\[4pt]
        \hat{\kappa} &\quad 1
    \end{pmatrix}
    \begin{pmatrix}
        \hat{v} \\[4pt]
        \hat{e}
    \end{pmatrix},
        \qquad%
    \hat{\kappa} =
    \frac{\gamma \, \hat{\tau}^{}_2}{\gamma \, \hat{\tau}^{}_1 + \delta}\,.
\ee
Note that $\hat{\kappa}$ depends both on the parameters of the Lie group $\gamma$ and $\delta$ and the modulus $\hat{\tau} = \hat{\tau}^{}_1 + i \, \hat{\tau}^{}_2$\,.
We further define 
\be \label{eq:rdhk}
    \theta = \arctan \hat{\kappa} + \frac{\pi}{2} \bigl[ \sgn \bigl( \gamma \, \hat{\tau}^{}_1 + \delta \bigr) - 1 \bigr]\,,
\ee
which implies that
\be \label{eq:ktt}
    \hat{\kappa} = \tan \theta\,.
\ee
Here, $\arctan \hat{\kappa} \in \bigl( - \pi / 2\,, \pi / 2 \bigr)$ denotes the principal value. Note that the quantity $\hat{\kappa}$ depends on both the group parameters and modulus $\hat{\tau}^{}$\,. 
Applying Eqs.~\eqref{eq:rdhk} and \eqref{eq:ktt} to Eq.~\eqref{eq:trnsfeaa}, we find
\be \label{eq:trnsfe}
    \begin{pmatrix}
        \hat{v} \\[4pt]
        \hat{e}
    \end{pmatrix} 
        \rightarrow%  
    \begin{pmatrix}
        \cos\theta &\quad -\sin\theta \\[4pt]
        \sin\theta &\quad \cos\theta
    \end{pmatrix}
    \begin{pmatrix}
        \hat{v} \\[4pt]
        \hat{e}
    \end{pmatrix}\,.
\ee
Geometrically, if the zweibein $\hat{e}^{}_m{}^a$ is arbitrary, then the transformation \eqref{eq:trnsfe} can be interpreted as a local rotation on the torus. What we have done above is essentially a mapping of the global SL$(2\,,\mathbb{Z})$ group within a local SL$(2\,,\mathbb{R})$ group. For the quantity $\hat{v}$ and $\hat{e}$ defined in Eq.~\eqref{eq:defve}, the curved indices $m$'s are contracted, which implies that the local diffeomorphisms on the two-torus are automatically satisfied. Therefore, the only remaining gauge symmetry is the local SO(2) rotation in Eq.~\eqref{eq:trnsfe}. Furthermore, it follows that the vector $(\hat{v}^m \, \p_m\,, \, \hat{e}^m \, \p_m)^\intercal$ transforms as
\be \label{eq:trnsfe20}
    \begin{pmatrix}
        \hat{v}^m \\[4pt]
        \hat{e}^m
    \end{pmatrix} \frac{\p}{\p\hat{x}^m} 
        \rightarrow%  
    \begin{pmatrix}
        \cos\theta &\quad \sin\theta \\[4pt]
        - \sin\theta &\quad \cos\theta
    \end{pmatrix}
    \begin{pmatrix}
        \hat{v}^m \\[4pt]
        \hat{e}^m
    \end{pmatrix} \frac{\p}{\p\hat{x}^m}\,,
\ee
where the inverse vielbein fields are
\be \label{eq:invviel}
    \hat{v}^m =
        \sqrt{\hat{\tau}_2}
    \begin{pmatrix}
        1 \\[4pt]
        0
    \end{pmatrix}\,,
        \qquad%
    \hat{e}^m = 
        \frac{1}{\sqrt{\hat{\tau}_2}}
    \begin{pmatrix}
        \hat{\tau}_1 \\[4pt]
        1
    \end{pmatrix}\,,
\ee
such that the following orthogonality conditions hold: 
\be
    \hat{v}^m \, \hat{v}_m = \hat{e}^m \, \hat{e}_m = 1\,,
        \qquad%
    \hat{v}^m \, \hat{e}_m = \hat{e}^m \, \hat{v}_m = 0\,,
        \qquad%
    \hat{v}^n \, \hat{v}_m + \hat{e}^n \, \hat{e}_m = \delta_m^n\,.
\ee
Note that the branching factor $\sgn \bigl( \gamma \, \hat{\tau}^{}_1 + \delta \bigr)$ in Eq.~\eqref{eq:trnsfeaa} does not have any physical significance here. The reason is that the zweibein formalism developed here is an alternative rewriting of the metric formalism on the two-torus. In the metric formalism, the branching factor cancels. However, after taking the membrane limit later in section~\ref{sec:bvbut}, the metric formalism becomes invalid, and the anisotropic toroidal geometry is only accessible via the vielbein fields. In this latter case of anisotropic compactification, the branching factor becomes physical and is responsible for the branched SL($2\,,\mathbb{Z}$) duality in nonrelativistic IIB superstring theory \cite{Bergshoeff:2023ogz}.   

As a simple example for illustrating how the above zweibein formalism generates the SL($2\,,\mathbb{Z}$) transformations of all the background fields in type IIB superstring theory, we consider the gauge theory to arise from the system with open M2-branes ending on M5-branes. The open membrane gauge potential is a two-form field ${\mathbb{A}}^{(2)}$\,. We also define the three-form field strength $\mathbb{F}^{(3)} = d\mathbb{A}^{(2)}$. Compactifying over a two-torus, the vector mode of ${\mathbb{A}}^{(2)}$ gives rise to the Born-Infeld vector $({A}^\text{\scalebox{0.8}{B}}\,,\,{A}^\text{\scalebox{0.8}{C}})^\intercal$ and their field strengths ${F}^\text{\scalebox{0.8}{B},\,\scalebox{0.8}{C}} = d{A}^\text{\scalebox{0.8}{B},\,\scalebox{0.8}{C}}$,
\be \label{eq:habc}
    \mathbb{A}^{(2)}_m = 
    \begin{pmatrix}
        {\mathbb{A}}^{}_{\mu \,9} \\[4pt]
        {\mathbb{A}}^{}_{\mu 10}
    \end{pmatrix}
    =
    \begin{pmatrix}
        {A}^\text{\scalebox{0.8}{B}}_\mu \\[4pt]
        {A}^\text{\scalebox{0.8}{C}}_\mu
    \end{pmatrix}\,,
        \qquad%
    \mathbb{F}^{(3)}_m =
    \begin{pmatrix}
        {\mathbb{F}}^{}_{\!\mu\nu \,9} \\[4pt]
        {\mathbb{F}}^{}_{\!\mu\nu 10}
   \end{pmatrix}
   =
   \begin{pmatrix}
        {F}^\text{\scalebox{0.8}{B}}_{\mu\nu} \\[4pt]
        {F}^\text{\scalebox{0.8}{C}}_{\mu\nu}
   \end{pmatrix}\,.
\ee
Here, ${A}^\text{\scalebox{0.8}{B}}$ and ${A}^\text{\scalebox{0.8}{C}}$ are gauge potentials associated with the Kalb-Ramond field 
$\hat{B}^{(2)}$ and the RR two-form 
$\hat{C}^{(2)}$\,, respectively.  
We also fix the gauge such that ${\mathbb{A}}_{mn} = 0$\,. 
In M-theory, ${\mathbb{F}}^{(3)}_m$ acts as $\p_m$ under the SL($2\,, \mathbb{Z}$) transformation on the compactified torus. According to Eq.~\eqref{eq:trnsfe20}, we find
\be \label{eq:ftrsnf}
    \begin{pmatrix}
        \hat{\CF}^\text{\scalebox{0.8}{B}} \\[4pt]
        \hat{\CF}^\text{\scalebox{0.8}{C}}
    \end{pmatrix} 
        \rightarrow%  
    \begin{pmatrix}
        \cos\theta &\quad \sin\theta \\[4pt]
        - \sin\theta &\quad \cos\theta
    \end{pmatrix}
    \begin{pmatrix}
        \hat{\CF}^\text{\scalebox{0.8}{B}} \\[4pt]
        \hat{\CF}^\text{\scalebox{0.8}{C}}
    \end{pmatrix} \,,
\ee
where
\be \label{eq:fbcdef}
    \begin{pmatrix}
        \hat{\CF}^\text{\scalebox{0.8}{B}} \\[4pt]
        \hat{\CF}^\text{\scalebox{0.8}{C}}
    \end{pmatrix}
    =
    \begin{pmatrix}
        \hat{v}^m \, {\mathbb{F}}^{(3)}_m \\[4pt]
        \hat{e}^m \, \hat{\mathbb{F}}^{(3)}_m
    \end{pmatrix}
    =
    \begin{pmatrix}
        \hat{\tau}^{1/2}_2 \, {F}^\text{\scalebox{0.8}{B}} \\[4pt]
        \hat{\tau}^{-1/2}_2 \, \bigl( {F}^\text{\scalebox{0.8}{C}} + \hat{\tau}^{}_1 \,  {F}^\text{\scalebox{0.8}{B}} \bigr)
    \end{pmatrix}\,.
\ee
In terms of the complex field strength $\hat{\CW} = \hat{\CF}^\text{\scalebox{0.8}{C}} + i \, \hat{\CF}^\text{\scalebox{0.8}{B}}$\,, 
the SL($2\,, \mathbb{Z}$) transformations in Eq.~\eqref{eq:ftrsnf} become
$\hat{\CW} \rightarrow e^{i\theta} \, \hat{\CW}$\,.
Therefore, the quantity $\hat{\CW}^* \, \hat{\CW} = \hat{\tau}^{-1}_2 \, \bigl| \hat{\tau} \, {F}^\text{\scalebox{0.8}{B}} + {F}^\text{\scalebox{0.8}{C}} \bigr|^2$ is manifestly SL($2\,, \mathbb{Z}$) invariant. 
This observation is nothing new from the metric formalism of the toroidal geometry as 
$\hat{\CW}^* \, \hat{\CW} = \hat{g}^{mn} \, \mathbb{F}_m \, \mathbb{F}_n$\,,
where $\hat{g}^{mn}$ is the inverse of the toroidal metric $\hat{g}_{mn}$ in Eq.~\eqref{eq:torusmetric}. 

Similarly, the two-form fields $\hat{B}^{(2)}$ and $\hat{C}^{(2)}$ arise from dimensionally reducing the three-form gauge potential $\mathbb{C}^{(3)}$ in M-theory, with
$\hat{\mathbb{C}}^{(3)}_m = \bigl( \hat{B}^{(2)},\,\hat{C}^{(2)} \bigr)^\intercal$\,.
Analogous to Eqs.~\eqref{eq:ftrsnf} and \eqref{eq:fbcdef}, we have
\be \label{eq:mcbctrnsf}
    \begin{pmatrix}
        \hat{\CB}^{(2)} \\[4pt]
        \hat{\CC}^{(2)}
    \end{pmatrix} 
        \rightarrow%  
    \begin{pmatrix}
        \cos\theta &\quad \sin\theta \\[4pt]
        - \sin\theta &\quad \cos\theta
    \end{pmatrix}
    \begin{pmatrix}
        \hat{\CB}^{(2)} \\[4pt]
        \hat{\CC}^{(2)}
    \end{pmatrix} \,,
\ee
where
\be \label{eq:hmcbc}
    \begin{pmatrix}
        \hat{\CB}^{(2)} \\[4pt]
        \hat{\CC}^{(2)}
    \end{pmatrix} 
    =
    \begin{pmatrix}
        \hat{v}^m \, \hat{\mathbb{C}}^{(3)}_m \\[4pt]
        \hat{e}^m \, \hat{\mathbb{C}}^{(3)}_m
    \end{pmatrix}
    =
    \begin{pmatrix}
        \hat{\tau}^{1/2}_2 \, \hat{B}^{(2)} \\[4pt]
        \hat{\tau}^{-1/2}_2 \, \bigl( \hat{C}^{(2)} + \hat{\tau}^{}_1 \,  \hat{B}^{(2)} \bigr)
    \end{pmatrix}.
\ee
We discover a new basis for the background fields, as shown in Eq.~\eqref{eq:hmcbc}, where all the spacetime indices are projected onto frame indices by the vielbein fields on the torus.

\subsection{Anisotropy from Blowing Up the Torus} \label{sec:bvbut}

We are ready to perform the membrane limit introduced in section~\ref{eq:mtunst}. The reparameterizations in Eq.~\eqref{eq:resmth} of the vielbein fields in the target space membrane Newton-Cartan geometry imply that, on the compactified torus, 
\be \label{eq:hverep}
    \hat{v} = \omega^{2/3} \, v\,,
        \qquad%
    \hat{e} = \omega^{-1/3} \, e\,,
\ee
where $v = v^{}_m \, dx^m$, $e = e^{}_m \, dx^m$ and
\be
    v^{}_m = 
        \frac{1}{\sqrt{\tau^{}_2}}
    \begin{pmatrix}
        1 \\[4pt]
        - \tau^{}_1
    \end{pmatrix},
        \qquad%
    e^{}_m =
    \sqrt{\tau^{}_2}
    \begin{pmatrix}
        0 \\[4pt]
        1
    \end{pmatrix}\,.
\ee
Moreover, compare Eqs.~\eqref{eq:bcprep} and \eqref{eq:c0pt}, we find
\be \label{eq:hattt}
    \hat{\tau}^{}_1 = \tau^{}_1\,,
        \qquad%
    \hat{\tau}^{}_2 = \omega^{-1} \, \tau^{}_2\,,
\ee
where $\tau = \tau^{}_1 + i \, \tau^{}_2$ will be the modulus of the anisotropic torus after performing the $\omega \rightarrow \infty$ limit. Matching Eqs.~\eqref{eq:hverep} and \eqref{eq:hattt} requires
\be \label{eq:rescalexm}
    \hat{x}^m = \omega^{1/6} \, x^m\,. 
\ee
In terms of the prescriptions \eqref{eq:hverep} $\sim$ \eqref{eq:rescalexm}, we find that the SL($2\,,\mathbb{Z}$) transformations of $\hat{v}$ and $\hat{e}$ in Eq.~\eqref{eq:trnsfeaa} now become
\be \label{eq:trnsfeaaomega}
    \begin{pmatrix}
        v \\[4pt]
        e
    \end{pmatrix} 
        \rightarrow% 
    \frac{\sgn \bigl( \gamma \, \tau^{}_1 + \delta \bigr)}{\sqrt{1 + \kappa^2 / \omega^2}} 
    \begin{pmatrix}
        1 &\quad -\kappa/\omega^2 \\[4pt]
        \kappa &\quad 1
    \end{pmatrix}
    \begin{pmatrix}
        v \\[4pt]
        e
    \end{pmatrix}\,,
        \qquad%
    \kappa =
    \frac{\gamma \, \tau^{}_2}{\gamma \, \tau^{}_1 + \delta}\,.
\ee

In the case where $\gamma \, \tau^{}_1 + \delta \neq 0$\,, we are allowed to perform the limit $\omega \rightarrow \infty$ directly, which leads to the following SL($2\,,\mathbb{Z}$) transformation of vielbein fields $v$ and $e$\,:
\be \label{eq:vegr}
    \begin{pmatrix}
        v \\[4pt]
        e
    \end{pmatrix}
        \rightarrow  
    \sgn \bigl( \gamma \, \tau^{}_1 + \delta \bigr)
    \begin{pmatrix}
        1 &\quad 0 \\[4pt]
        \kappa &\quad 1
    \end{pmatrix}
    \begin{pmatrix}
        v \\[4pt]
        e
    \end{pmatrix}\,.
\ee
The torus after the $\omega \rightarrow \infty$ limit is an \emph{anisotropic torus}, which does not admit a global metric. The anisotropic torus geometry is encoded by the longitudinal vielbein $v$ and transverse vielbein $e$ in the background membrane Newton-Cartan geometry.
Regarding the vielbein fields to be arbitrary, we will refer to the transformation \eqref{eq:vegr} as a local \emph{Galilean rotation}. 
Similarly, for the inverse vielbein fields $v^m$ and $e^m$ defined via the orthogonality conditions,
\be
    v^m \, v_m = e^m \, e_m = 1\,,
        \qquad%
    v^m \, e_m = e^m \, v_m = 0\,,
        \qquad%
    v^n \, v_m + e^n \, e_m = \delta_m^n\,.
\ee
we have
\be \label{eq:trnsfe2}
    \begin{pmatrix}
        v^m \\[4pt]
        e^m
    \end{pmatrix}
    \frac{\p}{\p x^m}
        \rightarrow%  
    \begin{pmatrix}
        1 &\quad -\kappa \\[4pt]
        0 &\quad 1
    \end{pmatrix}
    \begin{pmatrix}
        v^m \\[4pt]
        e^m
    \end{pmatrix}
    \frac{\p}{\p x^m}\,.
\ee
We will discuss the case where $\gamma \, \tau^{}_1 + \delta = 0$ later in section~\ref{sec:onebranelimit}.

In order to elucidate the physical meaning of the aforementioned limiting procedure, certain rewriting in the relativistic case before sending $\omega$ to infinity is required. We perform the following steps:
\begin{enumerate}[1.]

\item

First, note that $\hat{\tau}^{}_2$ in Eq.~\eqref{eq:sl2zrel} transforms as
\be \label{eq:ht2trnsf}
    \hat{\tau}^{}_2 \rightarrow \frac{1}{1 + \hat{\kappa}^2} \frac{\hat{\tau}^{}_2}{( \gamma \, \hat{\tau}^{}_1 + \delta )^2}\,.
\ee
The branching factor $\sgn (\gamma \, \hat{\tau}_1 + \delta)$ in Eq.~\eqref{eq:trnsfeaa} arises as
\be \label{eq:shttsgn}
    \sqrt{\hat{\tau}^{}_2} \rightarrow \frac{\sgn \bigl( \gamma \, \hat{\tau}^{}_1 + \delta \bigr)}{\sqrt{1 + \hat{\kappa}^2}} \frac{\sqrt{\hat{\tau}^{}_2}}{\gamma \, \hat{\tau}^{}_1 + \delta}\,.
\ee
The sign function in Eq.~\eqref{eq:shttsgn} is the origin of the branching factor in Eq.~\eqref{eq:trnsfeaa}. 

\item Next, we show that this branching factor can be relocated into the choice of branching in the dilaton field in type IIB superstring theory. 
This IIB theory arises from compactifying relativistic M-theory over the torus, where $\hat{\tau}^{}_2 = e^{-\hat{\Phi}}$ with $\hat{\Phi}$ the dilaton field and $\hat{\tau}^{}_1 = \hat{C}^{(0)}$ the RR zero-form. The transformation $\hat{\tau}^{}_2$ in Eq.~\eqref{eq:ht2trnsf} implies
\be
    \hat{\Phi} \rightarrow \hat{\Phi} + \ln \bigl( 1 + \hat{\kappa}^2 \bigr) + 2 \, \ln \bigl| \gamma \, \hat{C}^{(0)} + \delta \bigr|\,.
\ee
Modify the SL($2\,,\mathbb{Z}$) transformation of the dilaton field to be \cite{Bergshoeff:2023ogz},
\be \label{eq:hphitrnsf}
    \hat{\Phi} \rightarrow \hat{\Phi} + \ln \bigl( 1 + \hat{\kappa}^2 \bigr) + 2 \, \ln \bigl( \gamma \, \hat{C}^{(0)} + \delta \bigr)\,,
\ee
we find that Eq.~\eqref{eq:shttsgn} now becomes
\be
    \sqrt{\hat{\tau}^{}_2} \rightarrow \frac{1}{\sqrt{1 + \hat{\kappa}^2}} \frac{\sqrt{\hat{\tau}^{}_2}}{\gamma \, \hat{\tau}^{}_1 + \delta}\,.
\ee
As a result, Eq.~\eqref{eq:trnsfeaa} now simplifies to be
\be \label{eq:trnsfeaa2}
    \begin{pmatrix}
        \hat{v} \\[4pt]
        \hat{e}
    \end{pmatrix} 
        \rightarrow% 
    \frac{1}{\sqrt{1 + \hat{\kappa}^2}} 
    \begin{pmatrix}
        1 &\quad -\hat{\kappa} \\[4pt]
        \hat{\kappa} &\quad 1
    \end{pmatrix}
    \begin{pmatrix}
        \hat{v} \\[4pt]
        \hat{e}
    \end{pmatrix},
\ee
where the branching factor is removed. 
The new transformation \eqref{eq:hphitrnsf} implies that $\hat{\Phi}$ is complexified and will generally gain a shift of $2 \pi i$ when $\gamma \, \hat{C}^{(1)} + \delta < 0$\,, which, however, does not affect the positivity of the string coupling $\hat{g}^{}_s = e^{\langle \hat{\Phi} \rangle}$\,.

\item Using Eqs.~\eqref{eq:trnsfeaa} and \eqref{eq:hattt}, we find $\hat{\kappa} = \kappa / \omega$\,.
Finally, we take the redefinition $\omega = i \, c$ such that Eq.~\eqref{eq:trnsfeaa2} becomes 
\be \label{eq:lb}
    \begin{pmatrix}
        v \\[4pt]
        e
    \end{pmatrix} 
        \rightarrow% 
    \frac{1}{\sqrt{1 - \kappa^2 / c^2}} 
    \begin{pmatrix}
        1 &\quad \kappa/c^2 \\[4pt]
        \kappa &\quad 1
    \end{pmatrix}
    \begin{pmatrix}
        v \\[4pt]
        e
    \end{pmatrix}\,,
\ee
where we applied the reparametrizations in Eq.~\eqref{eq:hverep}. The above procedure effectively performs a Wick rotation of the longitudinal direction on the two-torus. As a result, the expression in Eq.~\eqref{eq:lb} is in form the same as a local Lorentz boosts on the two-torus: the mysterious quantity $\kappa$ now gains a physical interpretation as an effective Lorentz boost velocity in the transverse direction, while $c$\,, if taken to be real-valued, plays the role of an effective speed of light. 

\end{enumerate}

After the above preparation, it is now straightforward to understand the physical meaning of Eq.~\eqref{eq:vegr} in nonrelativistic M-theory. Taking the infinite speed of light limit $c \rightarrow \infty$ in Eq.~\eqref{eq:lb}, we find the induced transformation
\be \label{eq:lb2}
    \begin{pmatrix}
        v \\[4pt]
        e
    \end{pmatrix} 
        \rightarrow% 
    \begin{pmatrix}
        1 &\quad 0 \\[4pt]
        \kappa &\quad 1
    \end{pmatrix}
    \begin{pmatrix}
        v \\[4pt]
        e
    \end{pmatrix}
\ee
that replaces Eq.~\eqref{eq:vegr}. The Galilean rotation Eq.~\eqref{eq:vegr} on the anisotropic torus now becomes an effective ``Galilean boost" in Eq.~\eqref{eq:lb2}. Effectively, on a local patch of the toroidal manifold, $v$ can be thought of as the absolute time $t$ in Newton-Cartan geometry, while $e$ represents a spatial direction $x$ along which a Galilean boost is performed. Then, Eq.~\eqref{eq:lb2} becomes 
\be
    t \rightarrow t\,,
        \qquad%
    x \rightarrow x + \kappa \, t\,,
\ee
which is genuinely speaking a local Galilean boost transformation, with $\kappa$ the boost velocity. Of course, the Galilean boost between a spatial direction and an absolute time direction is fundamentally different from the Galilean rotation between two spatial directions on the anisotropic torus, where the latter is what we ultimately care about. We will not keep emphasizing this distinction throughout the paper but loosely refer to $\kappa$ as the boost velocity. This slight abuse of terminology has the benefit of bringing up the more familiar intuition from Newton-Cartan geometry. 
Finally, we note that the SL($2\,,\mathbb{Z}$) transformation of the dilaton field in the resulting nonrelativistic IIB string theory after the anisotropic toroidal reduction is
\be
    \Phi \rightarrow \Phi + 2 \, \ln \bigl( \gamma \, C^{(0)} + \delta \bigr)\,,
\ee
which matches the $c \rightarrow \infty$ limit of the dilaton transformation \eqref{eq:hphitrnsf} in relativistic IIB string theory. 

The procedure of complexifying $\omega = i \, c$ has an interesting effect on the nonrelativistic string limit defined by Eqs.~\eqref{eq:dte} and \eqref{eq:bcprep}, which we transcribe as below:
\begin{subequations}
\begin{align}
    \hat{B}^{(2)} & = - \omega^2 \, e^{\Phi/2} \, \ell^{(2)} + B^{(2)}\,,
        &
    \hat{G}^{}_\text{MN} & = \omega^{3/2} \, \tau^{}_\text{MN} + \omega^{-1/2} \, E^{}_\text{MN}\,, \\[4pt]
    \hat{C}^{(q)} & = \omega^2 \, e^{\Phi/2} \, \ell^{(2)} \wedge C^{(q-2)} + C^{(q)}\,,
        &
    \hat{\Phi} & = \Phi + \ln \omega\,. 
\end{align}
\end{subequations}
In terms of $\omega = i \, c$\,, and redefining $\hat{G}^{}_\text{MN}$ and $\hat{\Phi}$ such that the T-duality invariant quantity $e^{\hat{\Phi}} \, \hat{G}$ in the Einstein frame is unchanged, we find the following reparametrizations:
\begin{subequations}
\begin{align}
    \hat{B}^{(2)} & = c^2 \, e^{\Phi/2} \, \ell^{(2)} + B^{(2)}\,,
        &
    \hat{G}^{}_\text{MN} & = c^{3/2} \, \tau^{}_\text{MN} + c^{-1/2} \, E^{}_\text{MN}\,, \\[4pt]
    \hat{C}^{(q)} & = - c^2 \, e^{\Phi/2} \, \ell^{(2)} \wedge C^{(q-2)} + C^{(q)}\,,
        &
    \hat{\Phi} & = \Phi + \ln c\,. 
\end{align}
\end{subequations}
Shown in \cite{Bergshoeff:2022iss}, these two limiting prescriptions lead to different sectors in nonrelativistic IIB string theory, which are related to each other via SL($2\,,\mathbb{Z}$) transformations satisfying $\gamma \, C^{(0)} + \delta < 0$\,. On the other hand, when $\gamma \, C^{(0)} + \delta > 0$\,, each of these sectors is mapped to itself.

\subsection{M-Theory Origin of Polynomial Realization of \texorpdfstring{SL($2\,,\mathbb{Z}$)}{SLZ}} \label{sec:mtoprsl}

We are ready to apply the machinery developed to the anisotropic toroidal compactification of nonrelativistic M-theory. This compactification will give rise to a natural set of variables in nonrelativistic IIB superstring theory and provide a geometrical interpretation of the exotic polynomial realization of SL($2\,,\mathbb{R}$) in nonrelativistic IIB supergravity \cite{Bergshoeff:2023ogz}. 
We start with a geometrical derivation of the SL($2\,,\mathbb{Z}$) transformations of various fundamental fields in nonrelativistic IIB superstring theory, namely, the vielbein, dilaton, Kalb-Ramond, and RR fields. In the Einstein frame, the vielbein fields are automatically invariant under the SL($2\,,\mathbb{Z}$) transformations. Therefore, we only need to focus on how the dilaton and higher-form gauge fields transform under the action of SL($2\,,\mathbb{R}$). 

We first derive the SL($2\,,\mathbb{Z}$) transformations of the dilaton $\Phi$ and zero-form $C^{(0)}$ in nonrelativistic IIB string theory. From Eq.~\eqref{eq:lb2}\,, we learned that $v = v_m \, dx^m$ is invariant under the local Galilean boost on the anisotropic torus. Since the toroidal coordinates $x^m$ transform linearly under the SL($2\,,\mathbb{Z}$) isometry, $x^m \rightarrow \Lambda^m{}_n \, x^n$\,, we find
\be \label{eq:vmlv}
    v_m \rightarrow \Lambda_m{}^n \, v_n\,,  
        \qquad%
    v_m = \frac{1}{\sqrt{\tau^{}_2}}
    \begin{pmatrix}
        1 \\[4pt]
        - \tau^{}_1
    \end{pmatrix},
        \qquad%
    \Lambda_m{}^n 
    =
    \begin{pmatrix}
        \delta &\,\, - \gamma \\[4pt]
        - \beta &\,\, \alpha
    \end{pmatrix}.
\ee
Similarly, $e^m \, \p_m$ is also invariant under the local Galilean boost. Therefore,
\be \label{eq:emle}
    e^m \rightarrow \Lambda^m{}_n \, e^n\,,
        \qquad%
    e^m = \frac{1}{\sqrt{\tau_2}}
    \begin{pmatrix}
        \tau^{}_1 \\[4pt]
        1
    \end{pmatrix},
        \qquad%
    \Lambda^m{}_n 
    =
    \begin{pmatrix}
        \alpha &\quad \beta \\[4pt]
        \gamma &\quad \delta
    \end{pmatrix}.
\ee
Since $\tau = C^{(0)} + i \, e^{-\Phi}$\,, both Eqs.~\eqref{eq:vmlv} and \eqref{eq:emle} imply the following SL($2\,, \mathbb{Z}$) transformations:
\be \label{eq:phic0}
    \Phi \rightarrow \Phi + 2 \, \ln \bigl( \gamma \, C^{(0)} + \delta \bigr)\,,
        \qquad%
    C^{(0)} \rightarrow \frac{\alpha \, C^{(0)} + \beta}{\gamma \, C^{(0)} + \delta}\,.
\ee

Next, we focus on the SL($2\,,\mathbb{Z}$) transformations of the two-forms $B^{(2)}$ and $C^{(2)}$\,. In terms of the reparametrizations in Eq.~\eqref{eq:bcprep}, we find the following induced reparametrizations of $\hat{\CB}^{(2)}$ and $\hat{\CC}^{(2)}$ defined in Eq.~\eqref{eq:hmcbc}:
\be \label{eq:mcb2c2exp}
    \hat{\CB}^{(2)} = - \omega^{3/2} \, \ell^{(2)} + \omega^{-1/2} \, \CB^{(2)}\,,
        \qquad%
    \hat{\CC}^{(2)} = \omega^{1/2} \, \CC^{(2)}\,.
\ee
Here, analogous to Eq.~\eqref{eq:hmcbc}, $\CB^{(2)}$ and $\CC^{(2)}$ arise from projecting the three-form gauge field $\mathbb{C}^{(3)}$ in nonrelativistic M-theory, with 
\be \label{eq:hmcbcnh}
    \begin{pmatrix}
        {\CB}^{(2)} \\[4pt]
        {\CC}^{(2)}
    \end{pmatrix} 
    =
    \begin{pmatrix}
        v^m \, {\mathbb{C}}^{(3)}_m \\[4pt]
        e^m \, {\mathbb{C}}^{(3)}_m
    \end{pmatrix}
    =
    \begin{pmatrix}
        e^{-\Phi/2} \, {B}^{(2)} \\[4pt]
        e^{\Phi/2} \, \bigl( {C}^{(2)} + C^{(0)} \,  {B}^{(2)} \bigr)
    \end{pmatrix}.
\ee

The quantities $\hat{\CB}^{(2)}$ ($\hat{\CC}^{(2)}$) transform in the same way as $\hat{v}$ ($\hat{e}$). Therefore, Eqs.~\eqref{eq:trnsfeaa} and \eqref{eq:mcb2c2exp} imply
\be \label{eq:mcbctrnsf2}
    \begin{pmatrix}
        - \omega^2 \, \ell^{(2)} + {\CB}^{(2)} \\[4pt]
        {\CC}^{(2)}
    \end{pmatrix} 
        \rightarrow%  
    \frac{1}{\sqrt{1+\kappa^2/\omega^2}}
    \begin{pmatrix}
        1 &\quad -\kappa \\[4pt]
        \kappa/\omega^2 &\quad 1
    \end{pmatrix}
    \begin{pmatrix}
        - \omega^2 \, \ell^{(2)} + \CB^{(2)} \\[4pt]
        \CC^{(2)}
    \end{pmatrix} \,,
\ee
In the $\omega \rightarrow \infty$ limit, we obtain
\be \label{eq:polb2c2}
    \CB^{(2)} \rightarrow \CB^{(2)} - \kappa \, \CC^{(2)} + \tfrac{1}{2} \, \kappa^2 \, \ell^{(2)}\,, 
        \qquad%
    \CC^{(2)} \rightarrow \CC^{(2)} - \kappa \, \ell^{(2)}\,,
\ee
which shows explicitly that $\CB^{(2)}$ and $\CC^{(2)}$ transform as polynomials in $\kappa$\,. 

Finally, we turn to the SL($2\,,\mathbb{Z}$) transformation of the RR four-form $C^{(4)}$ in nonrelativistic string theory. In the relativistic IIB superstring theory, the four-form field 
\be \label{eq:hc4}
    \hat{\CC}^{(4)} \equiv \hat{C}^{(4)} + \frac{1}{2} \, \hat{B}^{(2)} \wedge \hat{C}^{(2)}
\ee
forms an SL($2\,,\mathbb{Z}$) singlet. Uplifting to M-theory, the four-form in Eq.~\eqref{eq:hc4} arises from the six-form gauge potential $\mathbb{C}^{(6)}$\,, \emph{i.e.},
\be \label{eq:hcc4}
    \hat{\CC}^{(4)} = \frac{1}{2} \, \hat{\mathbb{C}}^{(6)}_{mn} \, \hat{v}^m \, \hat{e}^m\,,
\ee
where the toroidal indices $m$ and $n$ in $\hat{\mathbb{C}}^{(6)}_{mn}$ are projected to be the frame indices using the inverse vielbein fields $v^m$ and $e^m$\,. The quantity in Eq.~\eqref{eq:hcc4} is manifestly invariant under the local rotation transformation \eqref{eq:trnsfe} on the torus. Plugging Eq.~\eqref{eq:bcprep} into Eq.~\eqref{eq:hc4}, we find
\be \label{eq:hcfcf}
    \hat{\CC}^{(4)} = \frac{1}{2} \, \omega^2 \, \CC^{(2)} \wedge \ell^{(2)} + \CC^{(4)}\,,
        \qquad%
    \CC^{(4)} = C^{(4)} + \frac{1}{2} \, B^{(2)} \wedge C^{(2)}\,. 
\ee
From Eq.~\eqref{eq:mcbctrnsf2}, we read off the SL($2\,,\mathbb{Z}$) transformation
\be
    \hat{\CC}^{(2)} \rightarrow \frac{1}{\sqrt{1+\kappa^2/\omega^2}} \lr - \kappa \, \ell^{(2)} + \frac{\kappa}{\omega^2} \, {\CB}^{(2)} + {\CC}^{(2)} \rr.
\ee
Together with the SL($2\,,\mathbb{Z}$) transformation $\hat{\CC}^{(4)} \rightarrow \hat{\CC}^{(4)}$ and Eq.~\eqref{eq:hcfcf}, the $\omega \rightarrow \infty$ limit gives
\be \label{eq:polc4}
    \CC^{(4)} \rightarrow \CC^{(4)} - \frac{1}{2} \, \kappa \, \CB^{(2)} \wedge \ell^{(2)} + \frac{1}{4} \, \kappa^2 \, \CC^{(2)} \wedge \ell^{(2)}\,.
\ee
We summarize what we have derived so far. while \eqref{eq:phic0} of $\Phi$ and $C^{(0)}$ form an SL($2\,,\mathbb{Z}$) doublet,  
\be \label{eq:czft}
    e^{\Phi/2} 
    \begin{pmatrix}
        C^{(0)} \\[4pt]
        1
    \end{pmatrix}
    \rightarrow 
    \begin{pmatrix}
        \alpha &\quad \beta \\[4pt]
        \gamma &\quad \delta
    \end{pmatrix}
    \begin{pmatrix}
        C^{(0)} \\[4pt]
        1
    \end{pmatrix},
\ee
which has the geometric interpretation as the inverse transverse vielbein denoted as $e^m$ on the torus. Note Eq.~\eqref{eq:czft} is simply a rewriting of Eq.~\eqref{eq:emle} in IIB string variables. 
The SL($2\,,\mathbb{Z}$) transformations \eqref{eq:polb2c2} and \eqref{eq:polc4} of the higher-form potential fields are polynomials in $\kappa$\,, which form two three-dimensional polynomial realizations of SL($2\,,\mathbb{Z}$),
\begin{align} \label{eq:sttf}
    \mathbf{S}^{(2)}_3 =
    \begin{pmatrix}
        \ell^{(2)} \\[4pt]
        \CC^{(2)} \\[4pt]
        \CB^{(2)}
    \end{pmatrix}\,,
        \qquad%
    \mathbf{S}^{(4)}_3 =
    \begin{pmatrix}
        \CC^{(2)} \wedge \ell^{(2)} \\[4pt]
        \CB^{(2)} \wedge \ell^{(2)}  \\[4pt]
        2 \, \CC^{(4)}
    \end{pmatrix}\,.
\end{align}
Explicitly, the SL($2\,,\mathbb{Z}$) transformations of the two three-vectors in Eq.~\eqref{eq:sttf} take the form $\mathbf{S}^{}_3 \rightarrow \mathbf{U}^{}_3 \, \mathbf{S}^{}_3$\,, with
\be
    \mathbf{U}^{}_3 = 
    \begin{pmatrix}
        1 &\,\, 0 &\,\, 0 \\[4pt]
        -\kappa &\,\, 1 &\,\, 0 \\[4pt]
        \frac{1}{2} \, \kappa^2 &\,\, - \kappa &\,\, 1
    \end{pmatrix}\,.
\ee
These polynomial realizations of SL($2\,,\mathbb{Z}$) have the geometric interpretation as local Galilean boosts on the compactified torus, with $\kappa$ the boost velocity.  

\subsection{S-Duality and Critical Ramond-Ramond Two-Form} \label{sec:onebranelimit}

The discussions in sections~\ref{sec:bvbut} and \ref{sec:mtoprsl} only apply to $\gamma \, \tau^{}_1 + \delta \neq 0$\,. When $\gamma \, \tau^{}_1 + \delta = 0$\,, we have to revisit Eq.~\eqref{eq:sl2zrel} before taking the $\omega \rightarrow \infty$ limit. Using Eq.~\eqref{eq:hattt}, we find
\be \label{eq:tpo}
    \hat{\tau}'_1 = \frac{\alpha}{\gamma}\,,
        \qquad%
    \hat{\tau}'_2 = \frac{\omega}{\gamma^2 \, \tau^{}_2}\,.
\ee
Under the condition $\gamma \, \tau^{}_1 + \delta = 0$\,, the SL($2\,,\mathbb{Z}$) transformation \eqref{eq:trnsfe20} becomes
\be \label{eq:trnsfeaa3}
    \begin{pmatrix}
        \hat{v}'{}^m \\[4pt]
        \hat{e}'{}^m
    \end{pmatrix} \frac{\p}{\p\hat{x}^m} 
        \rightarrow%  
    \begin{pmatrix}
        0 &\quad 1 \\[4pt]
        - 1 &\quad 0
    \end{pmatrix}
    \begin{pmatrix}
        \hat{v}^m \\[4pt]
        \hat{e}^m
    \end{pmatrix} \frac{\p}{\p\hat{x}^m}\,.
\ee
We have introduced the primed notation to denote the variables after the SL($2\,,\mathbb{Z}$) transformation.
We already learned around Eq.~\eqref{eq:mcbctrnsf} that $(\hat{\CB}^{(2)},\, \hat{\CC}^{(2)})^\intercal$ transform in the same way as in Eq.~\eqref{eq:trnsfeaa3}. Therefore,
\be \label{eq:bpc}
    \hat{\CB}^{\prime(2)} = \hat{\CC}^{(2)}\,,
        \qquad%
    \hat{\CC}^{\prime(2)} = - \hat{\CB}^{(2)}\,.
\ee
Plugging Eq.~\eqref{eq:mcb2c2exp} into Eq.~\eqref{eq:bpc}, we find 
\be \label{eq:mcb2c2exp220}
    \hat{\CB}^{\prime(2)} = \omega^{1/2} \, \CC^{(2)}\,,
        \qquad%
    \hat{\CC}^{\prime(2)} = \omega^{3/2} \, \ell^{(2)} - \omega^{-1/2} \, \CB^{(2)}\,.
\ee
From Eqs.~\eqref{eq:tpo} and \eqref{eq:mcb2c2exp220}, we find the dual theory described by the primed notation is defined by the following reparametrizations of the background fields in relativistic IIB string theory:
\begin{subequations} \label{eq:mcb2c2exp22}
\begin{align} 
    \hat{\tau}'_1 & = \tau'_1\,,
        &
    \hat{\CB}^{\prime(2)} & = \omega^{1/2} \, \CB^{\prime(2)}\,, \\[4pt]
    \hat{\tau}'_2 & = \omega \, \tau'_2\,,
        &
    \hat{\CC}^{\prime(2)} & = \omega^{3/2} \, \ell^{\prime(2)} + \omega^{-1/2} \, \CC^{\prime(2)}\,.
\end{align}
The reparametrizations of the vielbein fields remain the same as in Eq.~\eqref{eq:dte}, with
\be
    \hat{E}^{\prime A} = \omega^{3/4} \, \tau^{\prime A}\,,
        \qquad%
    \hat{E}^{\prime A'} = \omega^{-1/4} \, E^{\prime A'}\,.
\ee
\end{subequations}
These reparametrizations of the relativistic background fields are different from Eqs.~\eqref{eq:hattt} and~\eqref{eq:mcb2c2exp} associated with the nonrelativistic string limit: instead of tuning the Kalb-Ramond field $\hat{\CB}^{\prime(2)}$ to its critical value to cancel the fundamental string tension, we are now in the dual frame where the RR two-form $\hat{\CC}^{\prime(2)}$ is fine-tuned to cancel the D1-string tension. This execution is the critical RR two-form limit of relativistic IIB string theory. 

To make the above observation manifest, we apply the reparametrizations in Eq.~\eqref{eq:mcb2c2exp22} to the SL($2\,,\mathbb{Z}$)-invariant action describing the $(p\,,q)$-string in relativistic IIB string theory, which is a bound state of $p$ fundamental strings and $q$ D1-strings. For simplicity's sake, we drop the primes in Eq.~\eqref{eq:mcb2c2exp22} and in the rest of this subsection. Note that the notation here without the primes is for the critical RR two-form limit, and they should not be confused with the notation in nonrelativistic string theory. In terms of the new variables in Eq.~\eqref{eq:hmcbc}, the $(p\,,q)$-string action can be written as
\be \label{eq:hss}
    \hat{S}^{}_\text{string} = - T \int d^2 \sigma \, \sqrt{- \bigl( \hat{P}^2 + \hat{Q}^2 \bigr) \, \det \hat{G}^{}_{\alpha\beta}} + T \int \Bigl( \hat{P} \, \hat{\CB}^{(2)} + \hat{Q} \, \hat{\CC}^{(2)} \Bigr)\,,
\ee
where
\be
    \hat{P} = w^m \, \hat{v}^{}_m\,,
        \qquad%
    \hat{Q} = w^m \, \hat{e}^{}_m\,,
        \qquad%
    w^m = 
    \begin{pmatrix}
        p \\[4pt]
        q
    \end{pmatrix}\,.
\ee
Note that $w^m$ transforms as an SL($2\,,\mathbb{Z}$) doublet. The action~\eqref{eq:hss} is manifestly SL$(2\,,\mathbb{Z})$ invariant. Note that $\hat{\tau}^{}_1 = \tau^{}_1$\,. 
Plugging Eq.~\eqref{eq:mcb2c2exp22} into \eqref{eq:hss}, we find the one-brane $\omega \rightarrow \infty$ limit of Eq.~\eqref{eq:hss} is
\be \label{eq:ssaoi}
    S^{}_\text{string} = \frac{T \, Q}{2} \int d^2 \sigma \, \sqrt{-\tau} \, \tau^{\alpha\beta} \, E^{}_{\alpha\beta} + T \, Q \int \Bigl( \CC^{(2)} - \chi^{-1} \, \CB^{(2)} - \tfrac{1}{2} \, \chi^{-2} \, \ell^{(2)} \Bigr)\,,
\ee
where 
\be
    Q = w^m \, e^{}_m = q \, \sqrt{\tau^{}_2} > 0\,,
        \qquad%
    \chi = - \frac{q \, \tau^{}_2}{p - q \, \tau^{}_1}\,.
\ee
It is clear from Eq.~\eqref{eq:ssaoi} that the $(p\,,q)$-states in the critical RR two-form limit have to contain D1-strings. In contrast, because there is no state left when $q = 0$\,, there is no independent fundamental string state anymore. Under the SL($2\,,\mathbb{Z}$) transformations, the action~\eqref{eq:ssaoi} is invariant if $\gamma = 0$\,, but is mapped to the $(p\,,q)$-string action in nonrelativistic string theory if $\gamma \neq 0$\,. See \cite{uduality} for further discussions on the complete SL($2\,,\mathbb{Z}$) transformations relating the nonrelativistic string and critical RR two-form limit of type IIB superstring theory and its connection to Matrix string theory.

\section{Application: M5-Brane on Anisotropic Torus} \label{eq:mbat}

We apply the general theory of anisotropic compactification to study the anisotropic compactification of the M5-brane, \emph{i.e.}, the magnetic dual of the M2-brane in nonrelativistic M-theory. We start with an application from the zweibein formalism for the toroidal compactification introduced in section~\ref{sec:zftg} to relativistic M-theory. We will review compactifying relativistic M5-brane over a torus and revisit its relation to the SL($2\,,\mathbb{Z}$)-invariant D3-brane action \cite{Berman:1998va}. In terms of the variables whose indices are projected by the zweibeine fields as in section~\ref{sec:zftg}, we construct a convenient way of writing the manifestly SL($2\,,\mathbb{Z}$)-invariant D$p$-branes. We will then study the nonrelativistic membrane limit of the M5-brane in nonrelativistic M-theory and its anisotropic toroidal compactification. We will relate the compactified M5-brane action to the manifestly SL($2\,,\mathbb{Z}$)-invariant D3-brane in nonrelativistic IIB string theory, which provides an M-theory explanation for the intricate branched SL($2\,,\mathbb{Z}$) structure found in \cite{Bergshoeff:2023ogz}.

\subsection{Relativistic M5-Brane over a Torus}

Before considering the nonrelativistic membrane limit, we first focus on the relativistic M5-brane, whose formalism is sufficiently involved, and it is worthwhile to sort out the convention before we move on to the nonrelativistic M5-brane. 

\subsubsection{The PST Formalism of M5-Brane} \label{sec:pstfmb}

We start with reviewing the Pasti-Sorokin-Tonin (PST) formalism of a single M5-brane with general covariance in relativistic M-theory \cite{Pasti:1997gx, Pasti:1995tn,Pasti:1996vs}.\,\footnote{Also see \cite{Perry:1996mk,Schwarz:1997mc} that captures the five-dimensional covariance instead of the general covariance of the six-dimensional M5-brane worldvolume. Historically, these works led up to the discovery of the PST action.} We will focus on the bosonic contents, but note that this bosonic sector is part of a supersymmetric theory that enjoys kappa symmetry \cite{Bandos:1997ui}. Just like how an open string is coupled to a one-form gauge field $A^{(1)}$\,, an open M2-brane couples to an antisymmetric two-form gauge potential $\mathbb{A}^{(2)}$\,. The gauge potential $\mathbb{A}^{(2)}$ is also coupled to the M5-brane when open M2-branes are ending on it, which is similar to how a D-brane couples to a vector gauge potential when there are open strings ending on it. In the free field limit, the field strength associated with $\mathbb{A}^{(2)}$ is self-dual on the six-dimensional worldvolume of the M5-brane. The PST formalism is a Born-Infeld-Dirac-like action describing a single bosonic M5-brane that realizes this self-duality condition at non-linear order, which is made possible after introducing an auxiliary worldvolume one-form field $a^{(1)}$\,.

Consider an embedding of the six-dimensional worldvolume of the M5-brane with coordinates $\sigma^\mu$\,, $\mu = 0\,, \, 1\,, \cdots\,, 5$ within eleven-dimensional spacetime. The PST action is\,\footnote{See \cite{Bergshoeff:1998vx} for the Hamiltonian formalism of the relativistic M5-brane.}
\begin{align} \label{eq:pstaction}
\begin{split}
    \hat{S}^{}_\text{M5} = - T^{}_\text{M5} & \int d^6 \sigma \, \sqrt{- \det \Bigl( \hat{\mathbb{G}}^{}_{\mu\nu} + i \, \hat{\mathbb{\Theta}}^{}_{\mu\nu} \Bigr)} \\[6pt]
    - \frac{T^{}_\text{M5}}{4} & \int d^6 \sigma \, \sqrt{-\hat{\mathbb{G}}} \, \hat{\mathbb{\Theta}}^{\mu\nu} \, \hat{\mathbb{H}}^{}_{\mu\nu\rho} \, \hat{\mathbb{N}}^\rho - \frac{T^{}_\text{M5}}{2} \int \lr \hat{\mathbb{C}}^{(6)} + {\mathbb{F}}^{(3)} \wedge \hat{\mathbb{C}}^{(3)} \rr\,,
\end{split}
\end{align}
where $\hat{\mathbb{N}}^\mu = \hat{\mathbb{G}}^{\mu\nu} \, \hat{\mathbb{N}}^{}_\nu$\,,
\be
    \hat{\mathbb{N}}^{(1)} = \frac{a^{(1)}}{\sqrt{a^{}_{\mu} \, \hat{\mathbb{G}}^{\mu \nu} \, a^{}_{\nu}}}\,,
        \qquad%
    \mathbb{F}^{(3)} = d {\mathbb{A}}^{(2)}\,,
        \qquad%
    \hat{\mathbb{H}}^{(3)} = \hat{\mathbb{C}}^{(3)} + \mathbb{F}^{(3)}\,,
\ee
and
\be
    \hat{\mathbb{\Theta}}^{\mu\nu} = \frac{1}{3! \, \sqrt{-\hat{\mathbb{G}}}} \, \epsilon^{\mu\nu\rho\sigma\lambda\kappa} \, \hat{\mathbb{H}}^{}_{\rho\sigma\lambda} \, \hat{\mathbb{N}}^{}_\kappa\,.
\ee
Here $\hat{\mathbb{G}}^{}_{\mu\nu}$\,, $\hat{\mathbb{H}}^{}_{\mu\nu\rho}$\,, $\hat{\mathbb{C}}^{(3)}$\,, and $\hat{\mathbb{C}}^{(6)}$ have been pulled back from the target space to the worldvolume, \emph{e.g.}, $\hat{\mathbb{G}}^{}_{\mu\nu} = \p^{}_\mu x^\mathbb{M} \, \p^{}_\nu x^\mathbb{N} \, \hat{\mathbb{G}}^{}_{\mathbb{M}\mathbb{N}}$\,.
Locally, we require $a^{(1)}$ to be exact, \emph{i.e.},
\be
    a^{(1)} = da^{(0)}\,.
\ee
In the flat limit with $\hat{\mathbb{G}}^{}_{\mu\nu} = \eta^{}_{\mu\nu}$ and $\hat{\mathbb{C}}^{(3)} = \hat{\mathbb{C}}^{(6)} = 0$\,, we find the part of the PST action \eqref{eq:pstaction} that is quadratic in $\mathbb{F}^{(3)}$ 
\begin{align}\label{eq:self-dual}
\begin{split}
    \hat{S}^\text{quad.}_\text{M5}
    & = - T^{}_\text{M5} \! \int \! d^6 \sigma \, \biggl\{ \frac{1}{24} \, \mathbb{F}^{\mu\nu\rho} \, \mathbb{F}^{}_{\mu\nu\rho} - \frac{\p^{}_\rho a \, \p^\sigma a}{8 \, \p^{}_\lambda a \, \p^\lambda a} \, \Bigl[ \mathbb{F}^{\mu\nu\rho} \! - \! \bigl( \star \mathbb{F} \bigr)^{\mu\nu\rho} \Bigr] \, \Bigl[ \mathbb{F}_{\mu\nu\sigma} \! - \! \bigl( \star \mathbb{F} \bigr)_{\mu\nu\sigma} \Bigr] \biggr\}\,. 
\end{split}
\end{align}
This action describes a free field $\mathbb{A}^{(2)}$ satisfying the self-dual condition $\mathbb{F}^{(3)} = \star \mathbb{F}^{(3)}$\,. Such a constrained two-form potential is said to be \emph{chiral}.

Besides the worldvolume diffeomorphisms, the PST action also enjoys rich gauge symmetries, which we classify below:
\begin{enumerate}

\item

\emph{One-form gauge symmetry of the chiral two-form} only acts non-trivially on $\mathbb{A}^{(2)}$, with
\be
    \delta^{}_\xi \mathbb{A}^{(2)} = d \xi^{(1)}\,.
\ee

\item

\emph{The PST symmetries} parametrized by a zero-form $\varphi$ and a one-form $\chi^{(1)}$ involve the scalar $a$ and act nontrivially on the chiral two-form $\mathbb{A}^{(2)}$, with
\be \label{eq:pstsym}
    \delta^{}_\text{PST} a = \varphi\,,
    \qquad%
    \delta^{}_\text{PST} \mathbb{A}^{(2)} = \frac{1}{2} \, \biggl( \frac{\varphi \,\CW^{(2)}}{\p^{}_\mathbb{\mu} a \, \p^\mathbb{\mu} a} + da^{(0)} \wedge \chi^{(1)} \biggr)\,,
\ee
where
\be
    \CW^{\mu\nu} = \hat{\mathbb{\Theta}}^{\mu\nu\rho} \, \p^{}_\rho a + 2 \, \frac{\sqrt{\p^{}_\lambda a \, \p^\lambda a}}{\sqrt{- \hat{\mathbb{G}}}} \frac{\delta}{\delta \hat{\mathbb{\Theta}}^{}_{\mu\nu}} \sqrt{- \det \bigl( \hat{\mathbb{G}}^{}_{\rho\sigma} + i \, \hat{\mathbb{\Theta}}^{}_{\rho\sigma}\bigr)}\,.
\ee
The PST symmetry parametrized by $\varphi$ says that the scalar mode $a$ is pure gauge.

\item

\emph{Higher-form gauge symmetries} that involve the three- and six-form gauge potentials,
\be
    \delta \hat{\mathbb{C}}^{(3)} = d\xi^{(2)}\,,
        \qquad%
    \delta \mathbb{A}^{(2)} = - \xi^{(2)}\,,
        \qquad%
    \delta \hat{\mathbb{C}}^{(6)} = d\xi^{(5)} + d\xi^{(2)} \wedge \hat{\mathbb{C}}^{(3)}\,.
\ee

\end{enumerate}
For example, one may use the PST symmetries to impose gauge fixing $\p^{}_\mu a = \delta_\mu^5$ and $\mathbb{A}^{}_{\mu5} = 0$\,. This gauge choice recovers the action principle considered in \cite{Perry:1996mk}.

\subsubsection{Double-Dimensional Reduction over a Torus} \label{sec:ddrt}

In \cite{Berman:1998va}, the double-dimensional reduction of M5-brane on a two-torus was studied, where shown that the resulting action on the four-dimensional worldvolume describes the D3-brane and its SL($2\,,\mathbb{Z}$) duals under special gauge choices. It is expected that a more general gauge choice leads to a manifestly SL($2\,,\mathbb{Z}$)-invariant D3-brane action as in \cite{Bergshoeff:2006gs}. We now review the double-dimensional reduction of M5-brane in \cite{Berman:1998va} and present a calculation that makes the SL($2\,,\mathbb{Z}$) invariance manifest. We will recast this calculation in the zweibein formalism on the torus developed in section~\ref{sec:zftg}. A solid understanding of this construction in the relativistic framework will provide useful guidance for our later discussion on the anisotropic compactification of the M5-brane in nonrelativistic M-theory.

We review \cite{Berman:1998va} in the following. We will start with introducing the reduction ansatz for the metric field in item~(\hyperref[item:a]{a}). Then, we will discuss about the gauge fixing in item~(\hyperref[item:b]{b}). To make contact with the D3-brane action directly, we must perform an electromagnetic duality transformation of the dimensionally reduced chiral field on the four-dimensional worldvolume as in item~(\hyperref[item:c]{c}), with the dual field being the Nambu-Goldstone boson that perturbs the shape of the D3-brane. Finally, we provide a list of mappings between the ingredients on the M5- and D3-brane in item~(\hyperref[item:d]{d}), which will eventually lead us to the manifestly SL($2\,,\mathbb{Z}$)-invariant D3-brane action in item~(\hyperref[item:e]{e}).

\vspace{3mm}

\noindent~\textbf{(a)~Metric reduction.\label{item:a}} We consider a dimensional reduction over a two-torus whose cycles extend in $x^9$ and $x^{10}$\,, and adopt the following ansatz for the target space metric:
\be \label{eq:rdmgg}
    \hat{\mathbb{G}}^{}_{\mathbb{M}\mathbb{N}} =
    \begin{pmatrix}
        \hat{G}^{}_\text{MN} &\quad 0 \\[4pt]
        0 &\quad \hat{g}^{}_{mn}
    \end{pmatrix},
        \qquad%
    \hat{g}^{}_{mn} = \frac{\Gamma}{\hat{\tau}^{}_2}
    \begin{pmatrix}
        1 &\qquad - \hat{\tau}^{}_1 \\[4pt]
        - \hat{\tau}^{}_1 &\qquad \hat{\tau}^2_1 + \hat{\tau}^2_2
    \end{pmatrix},
\ee
where $\hat{g}^{}_{mn}$ is the metric on the torus as given in Eq.~\eqref{eq:torusmetric} and $\Gamma$ is the surface area. Moreover, $\hat{G}^{}_\text{MN}$ is the metric in the Einstein frame, $\text{M}\,, \text{N} = 0\,, 1\,, \cdots\,, 8$ and $m\,, n = 9\,, 10$\,. Now, we consider a double-dimensional reduction of the M5-brane over the torus extending in the $x^9$ and $x^{10}$ directions in the target space. We require that the M5-brane wrap around this two-torus such that $x^9 = \sigma^4$ and $x^{10} = \sigma^5$\,. The pullback metric on the worldvolume also factorizes, 
\be
	\hat{\mathbb{G}}^{}_{\mu\nu} = 
	\begin{pmatrix}
		\hat{G}^{}_{\alpha\beta} &\quad 0 \\[4pt]
		0 &\quad \hat{g}^{}_{\text{mn}}
	\end{pmatrix}.
\ee
Here, $\alpha = 0\,, 1 \,, 2\,, 3$\,, $\text{m}\,, \text{n} = 4\,, 5$\,, and
\be
	\hat{G}^{}_{\alpha\beta} = \p^{}_\alpha x^\text{M} \, \p^{}_\beta x^\text{N} \, \hat{G}^{}_{\text{MN}}\,, 
		\qquad
	\hat{g}^{}_\text{mn} = \p^{}_\text{m} x^m \, \p^{}_\text{n} x^n \, \hat{g}^{}_{mn}\,.
\ee
Recall that the inverse vielbein fields on the target space torus defined in Eq.~\eqref{eq:invviel}, whose pullbacks to the worldvolume torus give the following frame fields:
\begin{align} \label{eq:emaw}
	\hat{v}^\text{m}  =  \sqrt{\frac{\hat{\tau}_2}{\Gamma}}
    	\begin{pmatrix}
		1 \\[4pt]
        		0
    	\end{pmatrix},
		\qquad%
   	\hat{e}^\text{m} = 
        \frac{1}{\sqrt{\Gamma \, \hat{\tau}_2}}
    	\begin{pmatrix}
       	 	\hat{\tau}_1 \\[4pt]
        		1
    	\end{pmatrix},
\end{align}
with 
\begin{align}
    \hat{g}^\text{mn} = \hat{e}^\text{m}{}_\text{a} \, \hat{e}^\text{n}{}_\text{b} \, \delta^{\text{ab}},
		\qquad%
    \hat{e}^\text{m}{}_\text{a}
    =
    \begin{pmatrix}
        \hat{v}^\text{m} \\[4pt]
        \hat{e}^\text{m}
    \end{pmatrix}.
\end{align}
Here, we recover the dependence on the two-torus area $\Gamma$.  

By writing Eq.~\eqref{eq:rdmgg}, we have truncated the Kaluza-Klein excitations from the toroidal compactification \cite{Berman:1998va}. Including these excitations would give rise to ($p\,,q$)-string states in IIB string theory, which implies that the Kaluza-Klein vectors correspond to the reductions of the two-form Kalb-Ramond and RR potential over a circle. The origin of the ($p\,,q$)-string states can be explained by the following argument: compactifying M-theory over the torus gives rise to IIA string theory on a circle, T-dualizing which gives rise to IIB string theory. In IIA, the Kaluza-Klein states correspond to the D0-brane states, which become ($p\,,q$)-string states after the T-duality transformation. The Kaluza-Klein momentum corresponds to the winding number of these ($p\,,q$)-strings. Therefore, the compactification of the M5-brane over the torus is, in fact, a bound state of a D3-brane and wrapped ($p\,,q$)-strings. The reduction map in Eq.~\eqref{eq:rdmgg} allows us to focus only on the D3-brane dynamics.\,\footnote{We thank Johannes Lahnsteiner for valuable discussions on compactifications in supergravity.} 

\vspace{3mm}

\noindent \textbf{(b)~Fixing the PST gauge.\label{item:b}} We gauge fix the one-form $a^{(1)}$ to be an element in the first de Rham cohomology group on the torus, with
\be \label{eq:gfao}
	a^{(1)} = q \, d\sigma^4 - p \, d\sigma^5\,,
		\qquad%
	p\,, q \in \mathbb{Z}\,.
\ee
Under the action of the SL($2\,,\mathbb{Z}$) isometry group on the torus, we have
\be \label{eq:wmdef}
    \begin{pmatrix}
        p \\[4pt]
        q 
    \end{pmatrix} \rightarrow
    \begin{pmatrix}
        \alpha &\,\,\,\, \beta \\[4pt]
        \gamma &\,\,\,\, \delta
    \end{pmatrix} 
    \begin{pmatrix}
        p \\[4pt]
        q 
    \end{pmatrix}\,. 
\ee 
Note that $p$ and $q$ label the number of the fundamental and D1-strings in a $(p\,,q)$-string state, respectively. It follows that $a^{}_\alpha = 0$\,.
Using the PST symmetry parametrized by $\chi^{(1)}$ in Eq.~\eqref{eq:pstsym}, we gauge fix $\mathbb{A}^{(2)}_\text{mn} = 0$\,. It follows that $\mathbb{F}^{}_{\!\alpha \text{mn}} = 0$\,.
Using the above prescriptions, and projecting the ``m" index to the frame index ``a" using Eq.~\eqref{eq:emaw}, we find the following reduced action on the four-dimensional worldvolume $\CM_4$\,:
\begin{align} \label{eq:fda}
    S^{}_\text{d.d.} = & - T^{}_\text{M5} \, \Gamma \int_{\CM_4} \!\! d^4 \sigma \, \sqrt{- \det \! \ls \hat{G}^{}_{\!\alpha\beta} - i \, \epsilon^{\text{ab}} \, \mathbb{N}^{\phantom{(}}_\text{a} \, \Bigl( \star \hat{\mathbb{H}}^{(3)}_\text{b} \Bigr)_{\!\alpha\beta} - \Bigl( \star \hat{\mathbb{H}}^{(3)}_{} \Bigr)^{}_{\!\alpha} \Bigl( \star \hat{\mathbb{H}}^{(3)}_{} \Bigr)^{}_{\!\beta} \rs} \\[4pt]
    & - \frac{T^{}_\text{M5}}{2} \int_{\CM_4} \! \ls \hat{\mathbb{C}}^{(6)}_{45} + \lr \hat{\mathbb{C}}^{(3)} + 2 \, \mathbb{F}^{(3)} \rr \wedge \hat{\mathbb{C}}^{(3)}_{45} - \Gamma \, \epsilon^\text{ab} \, \Bigl( \hat{\mathbb{N}}^{}_\text{a} \, \hat{\mathbb{H}}^{(3)}_\text{b} \! \wedge \hat{\mathbb{H}}^{(3)}_\text{c} \, \hat{\mathbb{N}}^\text{c} - \mathbb{F}^{(3)}_\text{a} \! \wedge \hat{\mathbb{C}}^{(3)}_\text{b} \Bigr) \rs. \notag
\end{align}
It is understood that all the differential forms and the Hodge star are defined on $\CM_4$\,. \emph{E.g.},
$\hat{\mathbb{H}}^{(3)}_\text{a} = \frac{1}{2} \, \hat{\mathbb{H}}^{(3)}_{\alpha\beta\text{a}} \, d\sigma^\alpha \wedge d\sigma^\beta$
is a two-form on $\CM_4$\,. 

\vspace{3mm}

\noindent \textbf{(c)~Electromagnetic duality.\label{item:c}} The four-dimensional action \eqref{eq:fda} is \emph{not yet} the D3-brane in the target space compactified over a circle: We still need to dualize the two-form gauge potential $\mathbb{A}^{(2)}_{\alpha\beta}$ on $\CM^{}_4$\,. Implementing this duality transformation amounts to adding to the action \eqref{eq:fda} a generating functional, which leads to the parent action,
\be
    S^{}_\text{parent} = S^{}_\text{d.d.} - \frac{T^{}_\text{M5}}{3!} \int_{\CM_4} d^4 \sigma \, \tilde{\mathbb{F}}^{\alpha\beta\gamma} \, \bigl( \mathbb{F}^{(3)} - d\mathbb{A}^{(2)} \bigr)_{\alpha\beta\gamma}\,.
\ee
Integrating out $\tilde{\mathbb{F}}^{\alpha\beta\gamma}$ enforces $\mathbb{F}^{(3)} = d \mathbb{A}^{(2)}$ and leads us back to the original action \eqref{eq:fda}. Instead, we now integrate out $\mathbb{A}^{(2)}$\,, which imposes the constraint $\p^{}_{\alpha} \tilde{\mathbb{F}}^{\alpha\beta\gamma} = 0$\,. This constraint is solved locally by
\be
    \tilde{\mathbb{F}}^{\alpha\beta\gamma} = \epsilon^{\alpha\beta\gamma\delta} \, \pi^{}_\delta\,,
        \qquad%
    \pi^{(1)} = d \pi\,,
\ee
\emph{i.e.}, the two-form $\mathbb{A}^{(2)}$ is dual to a scalar mode $\pi$\,. Note that the Levi-Civita symbol $\epsilon^{\alpha\beta\gamma\delta}$ with curved worldvolume indices is defined via $\epsilon^{01234} = 1$\,. Upon performing integration over $\mathbb{F}^{(3)}$, now regarded as an independent field, it leads to the emergence of the dual action,
\begin{align} \label{eq:sdualm5}
    S^{}_\text{dual} = & - T^{}_\text{M5} \, \Gamma \int_{\CM_4} d^4\sigma \sqrt{-\det \! \ls \hat{G}^{}_{\alpha\beta} + \epsilon^{\text{ab}} \, \hat{\mathbb{N}}^{}_\text{a} \Bigl( \hat{\mathbb{H}}^{(3)}_\text{b} \Bigr)_{\alpha\beta} + \frac{\hat{\Pi}^{}_\alpha \, \hat{\Pi}^{}_\beta}{\Gamma^2} \rs} + T^{}_\text{M5} \int \hat{\mathbb{C}}^{(3)} \wedge \pi^{(1)} \notag \\[4pt]
    & - \frac{T^{}_\text{M5}}{2} \int_{\CM_4} \biggl[ \lr \hat{\mathbb{C}}^{(6)}_{45} - \hat{\mathbb{C}}^{(3)} \wedge \hat{\mathbb{C}}^{(3)}_{45} \rr - \Gamma \, \epsilon^\text{ab} \, \Bigl( \hat{\mathbb{N}}^{}_\text{a} \, \hat{\mathbb{H}}^{(3)}_\text{b} \wedge \hat{\mathbb{H}}^{(3)}_\text{c} \, \hat{\mathbb{N}}^\text{c} - \mathbb{F}^{(3)}_\text{a} \wedge \hat{\mathbb{C}}^{(3)}_\text{b} \Bigr) \bigg]\,,
\end{align}
where
\be
    \hat{\Pi}^{(1)} = \pi^{(1)} + \Gamma \,  \hat{\mathbb{C}}^{(1)}\,. 
\ee
The action \eqref{eq:sdualm5} is manifestly invariant under the SL($2\,,\mathbb{Z}$) duality, which acts on the frame index ``a" as an SO(2) rotation. The scalar $\pi$ is the Nambu-Goldstone boson from the spontaneous breaking of the isometry direction where the D3-brane is localized.

\vspace{3mm}

\noindent \textbf{(d)~Dictionary between M5 and D3 data.\label{item:d}} To make the relation between Eq.~\eqref{eq:sdualm5} and the D3-brane action manifest, we need to re-express the action written in terms of the stringy ingredients. From the string-theoretical perspective, we have the following dictionary between quantities on the M5- and D3-brane:
\begin{enumerate}[(d1)]

\item 

The modulus $\hat{\tau} = \hat{\tau}^{}_1 + i \, \hat{\tau}^{}_2$ on the torus corresponds to the string dilaton and RR zero-form, respectively, with $\hat{\tau}^{}_1 = \hat{C}^{(0)}$\,, $\hat{\tau}^{}_2 = e^{-\hat{\Phi}}$\,.

\item

The Born-Infeld vectors $A^\text{\scalebox{0.8}{B}}$ and $A^\text{\scalebox{0.8}{C}}$ on the D3-brane, which are associated with the Kalb-Ramond and RR two-form, are related to the two-form gauge potential coupled to open membranes via $\mathbb{A}^{(2)}_\text{m} = \bigl( A^\text{\scalebox{0.8}{B}}, A^\text{\scalebox{0.8}{C}} \bigr)^\intercal$\,. The associated field strengths $F^\text{\scalebox{0.8}{B}}$ and $F^\text{\scalebox{0.8}{C}}$ are given by $\mathbb{F}^{(3)}_\text{m} = \bigl( F^\text{\scalebox{0.8}{B}}, F^\text{\scalebox{0.8}{C}} \bigr)^\intercal$\,.

\item

The higher-form RR potentials $\hat{C}^{(2)}$ and $\hat{C}^{(4)}$ in IIB string theory and $\hat{C}^{(1)}$ and $\hat{C}^{(3)}$ in IIA string theory on a circle are associated with the higher-form gauge potentials $\hat{\mathbb{C}}^{(3)}$ and $\hat{\mathbb{C}}^{(6)}$ in M-theory, with
\begin{subequations}
\be
    \hat{\mathbb{C}}^{(6)}_{45} = \Gamma \, \Bigl( 2 \, \hat{\CC}^{(4)} + \hat{C}^{(3)} \wedge \hat{C}^{(1)} \Bigr)\,,
        \qquad%
    \hat{\CC}^{(4)} = \hat{C}^{(4)} + \frac{1}{2} \hat{B}^{(2)} \wedge \hat{C}^{(2)}\,,
\ee
and
\begin{align}
    \hat{\mathbb{C}}^{(3)}_{45} = \Gamma \, e^{-\hat{\Phi}/4} \, \hat{C}^{(1)}\,,
        \qquad%
    \hat{\mathbb{C}}^{(3)}_{\text{m}} =
    \begin{pmatrix}
		\hat{B}^{(2)} \\[4pt]
		\hat{C}^{(2)}
	\end{pmatrix},
        \qquad%
	\hat{\mathbb{C}}^{(3)} = e^{\hat{\Phi}/4} \, \hat{C}^{(3)}\,.
\end{align}
\end{subequations}
Here, $\hat{B}^{(2)}$ is the Kalb-Ramond field and $\hat{C}^{(q)}$ are the RR potentials. Note that the odd forms, $\hat{C}^{(1)}$ and $\hat{C}^{(3)}$\,, also appear since the D3-brane lives in an effectively nine-dimensional target space, as the tenth dimension compactifies over a circle.  

\item

The gauge-fixing condition \eqref{eq:gfao} of $a^{(1)}$ contains the integers $p$ and $q$\,, corresponding to the number of the fundamental strings and of the D1-strings in a $(p\,,q)$-string bound state, respectively. These $(p\,,q)$-strings are smeared over the D3-brane. 

\end{enumerate}

\vspace{3mm}

\noindent \textbf{(e)~Manifestly SL($2\,,\mathbb{Z}$)-invariant D3-brane action.\label{item:e}}
In terms of the string-theoretical ingredients in item~(\hyperref[item:d]{d}), we find the quantities in Eq.~\eqref{eq:sdualm5} take the following form:
\begin{subequations}
\begin{align} 
    \hat{\mathbb{N}}^{}_\text{a} & = - \frac{\sgn\bigl(p - q \, \hat{C}^{(0)}\bigr)}{\sqrt{1 + \hat{\chi}^2}} 
    \begin{pmatrix}
        \hat{\chi} \\[4pt]
        1
    \end{pmatrix}, 
        &%
    \hat{\chi} &= - \frac{q \, e^{-\hat{\Phi}}}{p - q \, \hat{C}^{(0)}}\,, \label{eq:nachi} \\[4pt]
    \hat{\mathbb{H}}^{(3)}_\text{a} 
        &=%
    \frac{1}{\sqrt{\Gamma}} \begin{pmatrix}
        \hat{\mathscr{F}}^{\text{\scalebox{0.8}{B}}} \\[4pt]
        \hat{\mathscr{F}}^{\text{\scalebox{0.8}{C}}}
    \end{pmatrix} 
        =%
    \frac{1}{\sqrt{\Gamma}} \begin{pmatrix}
        \hat{\CB}^{(2)} + \hat{\CF}^{\text{\scalebox{0.8}{B}}} \\[4pt]
        \hat{\CC}^{(2)} + \hat{\CF}^{\text{\scalebox{0.8}{C}}}
    \end{pmatrix}.
\end{align}
\end{subequations}
Here, $\hat{\CB}^{(2)}$ and  $\hat{\CC}^{(2)}$ are defined in Eq.~\eqref{eq:hmcbc} and $\hat{\CF}^{\text{\scalebox{0.8}{B}}}$ and $\hat{\CF}^{\text{\scalebox{0.8}{B}}}$ are defined in Eq.~\eqref{eq:fbcdef}. Using the above prescriptions, together with the rescalings,
\be
    \lc \hat{\CB}^{(2)}\,, \, \hat{\CC}^{(2)}\,, \, \hat{\CF}^\text{\scalebox{0.8}{B}}\,, \, \hat{\CF}^\text{\scalebox{0.8}{C}} \rc \rightarrow \sqrt{\Gamma} \lc \hat{\CB}^{(2)}\,, \, \hat{\CC}^{(2)}\,, \, \hat{\CF}^\text{\scalebox{0.8}{B}}\,, \, \hat{\CF}^\text{\scalebox{0.8}{C}} \rc\,,
\ee
we find Eq.~\eqref{eq:sdualm5} becomes
\begin{align} \label{eq:d3}
    \hat{S}^{}_\text{dual} = & - T^{}_\text{D3} \int_{\CM_4} d^4\sigma \sqrt{-\det \! \ls \hat{G}^{}_{\alpha\beta} + e^{-\frac{\hat{\Phi}}{2}} \lr R \, \p_\alpha \pi + \hat{C}^{}_\alpha \rr \lr R \, \p_\beta \pi + \hat{C}^{}_\beta \rr + \hat{\mathscr{F}}^{}_{\alpha\beta} \rs} \notag \\[4pt]
    & - T^{}_\text{D3} \int_{\CM_4} \ls \lr \hat{\CC}^{(4)} + R \, \pi^{(1)} \wedge \hat{C}^{(3)} \rr + \tfrac{1}{2} \lr \hat{\CF}^{\text{\scalebox{0.8}{B}}} \wedge \hat{\CC}^{(2)} - \hat{\CF}^\text{\scalebox{0.8}{C}} \wedge \hat{\CB}^{(2)} \rr \rs \notag \\[4pt]
    & - T^{}_\text{D3} \int_{\CM_4} \frac{\bigl( \hat{\chi} \, \hat{\mathscr{F}}^{\text{\scalebox{0.8}{B}}} + \hat{\mathscr{F}}^{\text{\scalebox{0.8}{C}}} \bigr) \wedge \hat{\mathscr{F}}^{(2)}}{2 \sqrt{1+\hat{\chi}^2}}\,,
\end{align}
where
\be \label{eq:defhmsf}
    \hat{\mathscr{F}}^{(2)} = \frac{\hat{\mathscr{F}}^\text{\scalebox{0.8}{B}} - \hat{\chi} \, \hat{\mathscr{F}}^\text{\scalebox{0.8}{C}}}{\sqrt{1+\hat{\chi}^2}}\,.
\ee
Moreover, $T^{}_\text{D3} = \Gamma \, T^{}_\text{M5}$ is the effective D3-brane tension, and $R = e^{\frac{\Phi}{4}} \, \Gamma^{-1}$ is the radius of the circle over which type IIB superstring theory compactifies. In the limit where the torus shrinks to zero, \emph{i.e.}, $\Gamma \rightarrow 0$\,, we have $R \rightarrow \infty$ and thus ten-dimensional IIB superstring theory. The action \eqref{eq:d3} now describes D3-brane in the non-compact ten-dimensional target space and takes the following form:
\begin{align} \label{eq:d3f}
    \hat{S}^{}_\text{D3} = & - T^{}_\text{D3} \int d^4\sigma \sqrt{-\det \! \lr \hat{G}^{}_{\alpha\beta} + \hat{\mathscr{F}}^{}_{\alpha\beta} \rr} \\[4pt]
    & - T^{}_\text{D3} \int \ls \hat{\CC}^{(4)} + \frac{\bigl( \hat{\chi} \, \hat{\mathscr{F}}^{\text{\scalebox{0.8}{B}}} + \hat{\mathscr{F}}^{\text{\scalebox{0.8}{C}}} \bigr) \wedge \hat{\mathscr{F}}^{(2)}}{2 \sqrt{1+\hat{\chi}^2}} + \frac{1}{2} \lr \hat{\CF}^{\text{\scalebox{0.8}{B}}} \wedge \hat{\CC}^{(2)} - \hat{\CF}^\text{\scalebox{0.8}{C}} \wedge \hat{\CB}^{(2)} \rr \rs, \notag
\end{align}
It is understood that all the ingredients are pullbacks from the ten-dimensional target space to $\CM^{}_4$\,. Since we focus on the D3-brane action, it is clear that the final expression \eqref{eq:d3f} can already be derived in a concrete way by setting $R = 1$ and $\hat{C}^{(1)} = \hat{C}^{(3)} = 0$\,. We will take advantage of these simplifications when we move on to the analogous calculation in nonrelativistic string/M-theory later in this section. 

The manifestly SL($2\,,\mathbb{Z}$)-invariant D3-brane action matches the SL($2\,, \mathbb{Z}$)-invariant action in \cite{Bergshoeff:2006gs}, where, in addition to the SL($2\,,\mathbb{Z}$) doublet $w^\text{m} = (p\,, q)^\intercal$ in Eq.~\eqref{eq:wmdef}, a second SL($2\,,\mathbb{Z}$) doublet $\tilde{w}^\text{m} = (\tilde{p}\,, \tilde{q}\,)^\intercal$ is introduced, which satisfies
\be \label{eq:tpqr}
    p \, \tilde{q} - q \, \tilde{p} = 1\,. 
\ee
In terms of $\tilde{w}^m$\,, the D3-brane action in \cite{Bergshoeff:2006gs} can be written as (also see \cite{Bergshoeff:2022pzk})
\begin{align} \label{eq:hspdt}
    \hat{S}'_\text{D3} = & -T^{}_\text{D3} \int_{\CM_4} d^4 \sigma \sqrt{-\det \! \lr \hat{G}^{}_{\alpha\beta} + \frac{w^\text{m} \hat{\CF}^{}_\text{m}}{\sqrt{w^\text{n} \, w^{}_\text{n}}} \rr} - T^{}_\text{D3} \int \Bigl[ \hat{\CC}^{(4)} - \frac{1}{2} \bigl( w^\text{m} \hat{\Sigma}^{}_\text{m} \bigr) \wedge \bigl( \tilde{w}^\text{n} \hat{\Sigma}^{}_\text{n} \bigr) \Bigr] \notag \\[4pt]
    & - T^{}_\text{D3} \int_{\CM_4} \ls \bigl( \tilde{w}^\text{m} \hat{\Sigma}^{}_\text{m} \bigr) \wedge \bigl( w^\text{n} \hat{\CF}^{}_\text{n} \bigr) - \frac{1}{2} \frac{\tilde{w}^\text{m} \, w^{}_\text{m}}{w^\text{n} \, w^{}_\text{n}} \bigl( w^\text{k} \hat{\CF}^{}_\text{k} \bigr) \wedge \bigl( w^\text{r} \hat{\CF}^{}_\text{r} \bigr) \rs\,,
\end{align}
where $w^\text{n} \, w^{}_\text{n} = w^\text{m} \, g^{}_\text{mn} \, w^\text{n}$\,, and
\be
    \hat{\Sigma}^{}_\text{m} =
    \begin{pmatrix}
        \hat{B}^{(2)} \\[4pt]
        \hat{C}^{(2)}
    \end{pmatrix},
        \qquad%
    \hat{\CF}^{}_\text{m} = 
    \begin{pmatrix}
        \hat{B}^{(2)} + dA^\text{\scalebox{0.8}{B}} \\[4pt]
        \hat{C}^{(2)} + dA^\text{\scalebox{0.8}{C}}
    \end{pmatrix}
\ee
Using Eq.~\eqref{eq:tpqr}, we find that $\hat{S}'_\text{D3} = \hat{S}^{}_\text{D3}$ up to a boundary term. Therefore, the action~\eqref{eq:d3f} provides an alternative formalism of Eq.~\eqref{eq:hspdt}, but now without introducing $\tilde{w}^\text{m}$\,. 

When $p=1$ and $q=0$\,, the action \eqref{eq:d3f} reduces to the conventional D3-brane action in the Einstein frame
\begin{align*}
    \hat{S}^{}_\text{D3} \rightarrow -T^{}_\text{D3} \int_{\CM_4} \ls \, d^4 \sigma \sqrt{-\det \lr \hat{G}^{}_{\alpha\beta} + e^{-\hat{\Phi}/2} \, \hat{\CF}_{\alpha\beta} \rr} + \hat{\CC}^{(4)} + \hat{C}^{(2)} \! \wedge \hat{\CF} + \tfrac{1}{2} \, \hat{C}^{(0)} \hat{\CF} \wedge \hat{\CF} \, \rs, 
\end{align*}
where $\hat{\CF} = \hat{B}^{(2)} + dA^\text{\scalebox{0.8}{B}}$\,. 

\subsection{Nonrelativistic M5-Brane over an Anisotropic Torus} \label{sec:nrmfbat}

We are now ready to consider the nonrelativistic membrane limit of the M5-brane action \eqref{eq:pstaction}. This membrane limit has been introduced in section~\ref{eq:mtunst}, where we have reparametrized the vielbein fields as in Eq.~\eqref{eq:resmth} and the gauge potentials $\hat{\mathbb{C}}^{(3)}$ and $\hat{\mathbb{C}}^{(6)}$ as in Eq.~\eqref{eq:hc36rp}, and the limit is defined by sending $\omega$ to infinity. 

\subsubsection{M5-Brane Action in Nonrelativistic M-Theory}

For simplicity, we consider a special configuration of M5-brane embedded in the curved eleven-dimensional spacetime, where the transverse sector divides into two sectors with $A' = ( u'\,, \, 4\,, \, \cdots\,, \, 8 )$ and $u' = (2\,, \, 3\,, \, 10)$\,. We require 
\be \label{eq:special}
    \mathbb{E}_\mu{}^{A'} = 
    \begin{pmatrix}
        \mathbb{E}_\mu{}^{u'} &\,\, 0
    \end{pmatrix}.
\ee
It follows from Eq.~\eqref{eq:resmth} that
\be
    \hat{\mathbb{G}}^{}_{\mu\nu} = \Omega^2 \, \gamma^{}_{\mu\nu} + \Omega^{-1} \, \mathbb{E}_{\mu\nu}\,,
\ee
where $\Omega = \omega^{2/3}$\,, $\gamma^{}_{\mu\nu} = \gamma^{}_\mu{}^{u} \, \gamma^{}_\mu{}^{v} \, \eta^{}_{uv}$\,, and $\mathbb{E}^{}_{\mu\nu} = \mathbb{E}^{}_\mu{}^{u'} \, \mathbb{E}^{}_\nu{}^{v'} \, \delta^{}_{u'v'}$\,. 
The index $u = 0\,, \, 1\,, \, 9$ is the longitudinal frame index. This restriction will greatly simplify the derivation of the nonrelativistic M5-brane action from taking the membrane limit of Eq.~\eqref{eq:pstaction}. 
To further facilitate the large $\Omega$ expansion, we note 
the large $\Omega$ expansion
\be
    \hat{\mathbb{\Theta}}^\mu{}_\nu = \bigl( \mathbb{\Theta}_0 \bigr){}^\mu{}_\nu + \Omega^{-3} \, \bigl( \mathbb{\Theta}_3 \bigr){}^\mu{}_\nu + O (\Omega^{-6})\,,
\ee
where
\be \label{eq:theta03}
    \bigl( \mathbb{\Theta}_0 \bigr){}^\mu{}_\nu = \mathbb{\Theta}^{\mu\rho} \, \gamma^{}_{\rho\nu} - \tilde{\Gamma}^{\mu\rho} \, \mathbb{E}^{}_{\rho\nu}\,,
        \qquad%
    \bigl( \mathbb{\Theta}_3 \bigr){}^\mu{}_\nu = \mathbb{\Theta}^{\mu\rho} \, \mathbb{E}^{}_{\rho\nu} - \tfrac{1}{2} \, \mathbb{N}^{}_u \, \mathbb{N}^u \, \bigl( \mathbb{\Theta}_0 \bigr)^\mu{}_\nu\,.
\ee
We have defined
and
\begin{subequations} \label{eq:trnsfe22}
\begin{align}
    \tilde{\Gamma}^{\mu \nu} & = \frac{1}{3! \, \mathbb{E}} \, \epsilon^{\mu\nu\rho\sigma\lambda\kappa} \mathbb{\Gamma}^{}_{\rho\sigma\lambda} \, \mathbb{N}^{}_\kappa\,, 
        &%
    \mathbb{N}^{(1)} & = \frac{a^{(1)}}{\sqrt{a_{u'} \, a^{u'}}}\,, \\[4pt]
    \mathbb{\Theta}^{\mu\nu} & = \frac{1}{3! \, \mathbb{E}} \, \epsilon^{\mu\nu\rho\sigma\lambda\kappa} \mathbb{H}^{}_{\rho\sigma\lambda} \, \mathbb{N}^{}_\kappa\,,
        &%
    \mathbb{H}^{(3)} & = \mathbb{C}^{(3)} + \mathbb{F}^{(3)}\,.
\end{align}
\end{subequations}
Recall that $\Gamma^{(3)} = \gamma^0 \wedge \gamma^1 \wedge \gamma^{9}$ is defined in Eq.~\eqref{eq:defgt}. Here,
\be \label{eq:defe}
    \mathbb{\mathbb{E}} = \det \Bigl( \gamma_\mu{}^u \quad \mathbb{E}_\mu{}^{u'} \Bigr)\,.
\ee
Using the above ingredients and the identity from the Cayley-Hamilton theorem
\be \label{eq:dgpt}
    \det \Bigl( \delta^\mu_\nu + i \, \hat{\mathbb{\Theta}}^{\mu}{}_{\nu} \Bigr) = 1 + \tfrac{1}{2} \, \tr \bigl( \hat{\mathbb{\Theta}}^2 \bigr) + \tfrac{1}{8} \! \ls \tr \bigl( \hat{\mathbb{\Theta}}^2 \bigr) \rs^2 \!\!\! - \tfrac{1}{4} \, \tr \bigl( \hat{\mathbb{\Theta}}^4 \bigr)\,,
\ee
we find 
\begin{subequations}
\begin{align} 
    - \sqrt{- \det \! \lr \hat{\mathbb{G}}^{}_{\mu\nu} + i \, \hat{\mathbb{\Theta}}^{}_{\mu\nu} \rr}
    & = - \mathbb{E} \, \sqrt{\Omega^{3} \, \det \bigl( \mathbb{1} + i \, {\mathbb{\Theta}_0} \bigr) + \CM + O\bigr(\Omega^{-3}\bigr)}\,, \label{eq:dbi} \\[4pt]
    - \frac{1}{4} \int d^6 \sigma \, \sqrt{-\hat{\mathbb{G}}} \, \hat{\mathbb{\Theta}}^{\mu\nu} \, \hat{\mathbb{H}}^{}_{\mu\nu\rho} \, \hat{\mathbb{N}}^\rho & = - \frac{1}{4} \int d^6 \sigma \, \mathbb{E} \Bigl[ \mathbb{\Theta}^{\mu\nu} \, \mathbb{H}^{}_{\mu\nu u'} \, \mathbb{N}^{u'} - \bigl( \tilde{\Gamma}^{\mu\nu} \, \mathbb{H}^{}_{\mu\nu u} + \mathbb{\Theta}^{\mu\nu} \, \Gamma^{}_{\mu\nu u} \bigr) \, \mathbb{N}^u \Bigr]  \notag \\[4pt]
    & \quad - \frac{\Omega^3}{2} \int \mathbb{H}^{(3)} \wedge \Gamma^{(3)} \lr 1 - \Omega^{-3} \, \mathbb{N}^{}_u \, \mathbb{N}^u \rr + O\bigl(\Omega^{-3}\bigr), \label{eq:thn} \\[4pt]
    - \frac{1}{2} \lr \hat{\mathbb{C}}^{(6)} + \mathbb{F}^{(3)} \wedge \hat{\mathbb{C}}^{(3)} \rr & = \frac{\Omega^3}{2} \, \mathbb{H}^{(3)} \wedge \Gamma^{(3)} - \frac{1}{2} \bigl( \mathbb{C}^{(6)} + \mathbb{F}^{(3)} \wedge \mathbb{C}^{(3)} \bigr)\,, \label{eq:c6f3c3}
\end{align}
\end{subequations}
where
\be \label{eq:mdef}
    \CM = \tr \bigl( {\mathbb{\Theta}}_0 \, {\mathbb{\Theta}}_3 \bigr) \Bigl[ 1 + \tfrac{1}{2} \, \tr \bigl( \mathbb{\Theta}^2_0 \bigr) \Bigr] - \tr \bigl( \mathbb{\Theta}^3_0 \, \mathbb{\Theta}_3 \bigr)\,.
\ee
Note that $\mathbb{N}^u = \gamma^\mu{}_u \, \mathbb{N}^{}_\mu$ and $\mathbb{N}^{u'} = \mathbb{E}^\mu{}_{u'} \, \mathbb{N}^{}_\mu$\,, where the inverse vielbein fields $\gamma^\mu{}_u$ and $\mathbb{E}^\mu{}_{u'}$ satisfy the orthogonality conditions 
\begin{subequations}
\begin{align}
    \gamma^\mu{}_u \, \gamma_\mu{}^v & = \delta^v_u\,, 
        &%
    \gamma^\mu{}_u \, \mathbb{E}_\mu{}^{u'} = \mathbb{E}^\mu{}^{}_{u'} \, \gamma_\mu{}^u & = 0\,, \\[4pt]
    \mathbb{E}^\mu{}_{u'} \, \mathbb{E}_\mu{}^{v'} & = \delta^{v'}_{u'}\,,
        &%
    \gamma^\mu{}^{}_{u} \, \gamma^{}_\nu{}^{u} + \mathbb{E}^\mu{}^{}_{u'} \, \mathbb{E}^{}_\nu{}^{u'} & = \delta^\mu_\nu\,.
\end{align}
\end{subequations}
Using the vielbein fields $\gamma_\mu{}^u$ and $E_\mu{}^{u'}$ to project the curved indices of Eq.~\eqref{eq:theta03} gives
\be
    \mathbb{\Theta}_0 = 
    \begin{pmatrix}
        \mathbb{\Theta}^u{}^{}_v &\quad 0 \\[4pt]
        \mathbb{\Theta}^{u'}{}^{}_{v} &\quad \epsilon^{u'}{}_{v'w'} \, \mathbb{N}^{w'}
    \end{pmatrix}\,,
        \qquad%
    \mathbb{\Theta}_3 = 
    \begin{pmatrix}
        0 &\quad \mathbb{\Theta}^u{}^{}_{v'} \\[4pt]
        0 &\quad \mathbb{\Theta}^{u'}{}^{}_{v'}
    \end{pmatrix}
    - \tfrac{1}{2} \, \mathbb{N}^{}_u \, \mathbb{N}^u \, \mathbb{\Theta}_0\,.
\ee
Together with the condition $\mathbb{N}^{}_{u'} \, \mathbb{N}^{u'} = 1$\,, we obtain
\be
    \det \bigl( \mathbb{1} + i \, {\mathbb{\Theta}_0} \bigr) = \bigl( 1 - \mathbb{N}^{}_{u'} \, \mathbb{N}^{u'} \bigr) \, \det \bigl( \delta^u_v + i \, \mathbb{\Theta}^u{}_v \bigr) = 0\,.
\ee
Therefore, Eq.~\eqref{eq:dbi} is free of divergence in $\Omega$\,. Moreover, the $\Omega^3$ divergences in Eqs.~\eqref{eq:thn} and \eqref{eq:c6f3c3} cancel each other.
Assembling the expressions above, we find that the $\Omega \rightarrow \infty$ limit of the relativistic M5-brane action \eqref{eq:pstaction} is 
\begin{align} \label{eq:nrm5a}
    S^{}_\text{M5} & = - T^{}_\text{M5} \int d^3 \sigma \, \mathbb{E} \, \sqrt{\CM} - \frac{T_{\text{M}5}}{4} \int d^6 \sigma \, \mathbb{E} \Bigl[ \mathbb{\Theta}^{\mu\nu} \, \mathbb{H}^{}_{\mu\nu\rho} \, \mathbb{E}^{\rho\sigma} \, \mathbb{N}^{}_\sigma - \bigl( \tilde{\Gamma}^{\mu\nu} \, \mathbb{H}^{}_{\mu\nu\rho} + \mathbb{\Theta}^{\mu\nu} \, \Gamma^{}_{\mu\nu\rho} \bigr) \, \gamma^{\rho\sigma} \, \mathbb{N}^{}_\sigma \Bigr] \notag \\[4pt]
    & \quad - \frac{T^{}_\text{M5}}{2} \int \Bigl( \mathbb{C}^{(6)} + \mathbb{F}^{(3)} \wedge \mathbb{C}^{(3)} - \mathbb{N}^{}_u \, \mathbb{N}^u \, \mathbb{H}^{(3)} \wedge \Gamma^{(3)} \Bigr)\,.
\end{align}
Note that we have written the action~\eqref{eq:nrm5a} in a covariant way with respect to general background fields, such that the condition in Eq.~\eqref{eq:special} is not necessary anymore. Therefore $\mathbb{E}^{}_{\mu\nu} = \mathbb{E}^{}_\mu{}^{A'} \, \mathbb{E}_\nu{}^{A'}$ and  $\mathbb{E}^{\mu\nu} = \mathbb{E}^\mu{}_{A'} \, \mathbb{E}^\nu{}^{}_{A'}$\,, where the transverse vielbein $\mathbb{E}^{}_\mathbb{M}{}^{A'}$ is kept general and orthogonality conditions defined in the target space to be
\begin{subequations}
\begin{align}
    \gamma^\mathbb{M}{}_u \, \gamma^{}_\mathbb{M}{}^v & = \delta^v_u\,, 
        &%
    \gamma^\mathbb{M}{}_u \, \mathbb{E}_\mathbb{M}{}^{A'} = \mathbb{E}^\mathbb{M}{}^{}_{A'} \, \gamma^{}_\mathbb{M}{}^u & = 0\,, \\[4pt]
    \mathbb{E}^\mathbb{M}{}_{A'} \, \mathbb{E}^{}_\mathbb{M}{}^{v'} & = \delta^{v'}_{A'}\,,
        &%
    \gamma^\mathbb{M}{}^{}_{u} \, \gamma^{}_\mathbb{N}{}^{u} + \mathbb{E}^\mathbb{M}{}^{}_{A'} \, \mathbb{E}^{}_\mathbb{N}{}^{A'} & = \delta^\mathbb{M}_\mathbb{N}\,.
\end{align}
\end{subequations}
The definition for the determinant $\mathbb{E} = \sqrt{-\mathbb{G}}$ in Eq.~\eqref{eq:defe} can be covariantized to be 
\be 
    \mathbb{G} =  \frac{1}{(3!)^2} \, \epsilon^{\alpha^{}_1 \cdots \alpha^{}_6} \, \epsilon^{\beta^{}_1 \cdots \beta^{}_6} \, \gamma^{}_{\alpha^{}_1\beta^{}_1} \cdots \gamma^{}_{\alpha^{}_3\beta^{}_3} \, E^{}_{\alpha^{}_4\beta^{}_4} \cdots E^{}_{\alpha^{}_6\beta^{}_6}\,.
\ee
The action~\eqref{eq:nrm5a}, together with the definition of $\CM$ in Eqs.~\eqref{eq:theta03} and \eqref{eq:mdef}, describes a single nonrelativistic M5-brane coupled to the spacetime membrane Newton-Cartan geometry.

\subsubsection{Double-Dimensional Reduction over an Anisotropic Torus} \label{sec:ddrat}

Next, we consider the compactification of the M5-brane action~\eqref{eq:nrm5a} over an anisotropic torus. In the upcoming subsection, we will demonstrate that the resulting four-dimensional worldvolume action exhibits duality with nonrelativistic IIB string theory's D3-brane, which possesses a manifest SL($2,,\mathbb{Z}$) symmetry, a subject studied in \cite{Bergshoeff:2022iss}. This match will provide a strong crosscheck of the action principle derived in Eq.~\eqref{eq:nrm5a}. 

We consider the following dimensional reduction over an anisotropic two-torus whose cycles are along the $x^9$ and $x^{10}$ directions. The reduction ansatz for the vielbein fields $\gamma^{}_\mathbb{M}{}^u$ and $\mathbb{E}^{}_\mathbb{M}{}^{A'}$ are taken to be
\be
    \gamma^{}_\mathbb{M}{}^u = 
    \begin{pmatrix}
        \gamma^{}_M{}^A & 0\\[4pt]
        0 & e^{}_m
    \end{pmatrix}\,,
        \qquad%
    \mathbb{E}^{}_\mathbb{M}{}^{A'} = 
    \begin{pmatrix}
        E^{}_M{}^{\tilde{A}'} & 0\\[4pt]
        0 & v^{}_m
    \end{pmatrix}\,,
\ee
where
\be
    v^{}_m = 
        \sqrt{\frac{\Gamma}{\tau^{}_2}}
    \begin{pmatrix}
        1 \\[4pt]
        - \tau^{}_1
    \end{pmatrix},
        \qquad%
    e^{}_m =
    \sqrt{\Gamma \, \tau^{}_2}
    \begin{pmatrix}
        0 \\[4pt]
        1
    \end{pmatrix}
\ee
with $u = 0\,, \, 1\,, \, 9$ and $A' = 2\,, \, \cdots\,, \, 8\,, \, 10$ in eleven-dimensional membrane Newton-Cartan geometry, and $A = 0\,, \, 1$ and $\tilde{A}' = 2\,, \, \cdots\,, 8$ in nine-dimensional string Newton-Cartan geometry after dimensionally reducing over the anisotropic two-torus. 

In order to perform a double-dimensional reduction of the M5-brane over the anisotropic torus, we require that the brane wraps around the torus such that $x^9 = \sigma^4$ and $x^{10} = \sigma^5$\,. The pullback vielbein fields factorize as
\be
    \gamma^{}_\mathbb{\mu}{}^u = 
    \begin{pmatrix}
        \tau^{}_\alpha{}^A & 0 \\[4pt]
        0 & v^{}_\text{m}
    \end{pmatrix}\,,
        \qquad%
    \mathbb{E}^{}_\mathbb{\mu}{}^{A'} =
    \begin{pmatrix}
        E^{}_\alpha{}^{\tilde{A}'} & 0 \\[4pt]
        0 & e^{}_\text{m}
    \end{pmatrix}\,.
\ee
Here, $\mu = 0\,, \, \cdots\,, \, 5$ is the curved index on the six-dimensional M5-brane worldvolume and $\alpha = 0\,, \, \cdots\,, \, 3$ is the curved index on the four-dimensional D3-brane worldvolume. The pullbacks of the inverse vielbein fields on the target space torus are given by
\be
    v^\text{m} = 
    \sqrt{\frac{\tau^{}_2}{\Gamma}}
    \begin{pmatrix}
        1 \\[4pt]
        0
    \end{pmatrix}\,,
        \qquad%
    e^\text{m} = 
    \frac{1}{\sqrt{\Gamma \, \tau^{}_2}}
    \begin{pmatrix}
        \tau^{}_1 \\[4pt]
        1
    \end{pmatrix}\,. 
\ee
We truncated the Kaluza-Klein excitations as in section~\ref{sec:ddrt}. This simplification will not affect our comparison between the compactified M5-brane and D3-brane action. 

We use the same gauge-fixing of the one-form $a^{(1)}$ as in section~\ref{sec:ddrt}, \emph{i.e.}, 
\be
    a^{(1)} = q \, d\sigma^4 - p \, d\sigma^5\,,
        \qquad%
    p\,,\,q \in \mathbb{Z}\,,
\ee
and gauge fix $\mathbb{A}^{(2)}_\text{mn} = 0$\,, which implies that $\mathbb{F}^{}_{\alpha\text{mn}} = 0$\,. Moreover, $\mathbb{N}^{}_\mu$ is only non-vanishing on the torus, with\,\footnote{Note that we are \emph{not} complexifying the dilaton as in section~\ref{sec:bvbut}.}
\be \label{eq:nvne}
    \mathbb{N}^{}_\text{m} \, v^\text{m} = - \sgn \bigl( p -q \, C^{(0)} \bigr) \, \chi\,,
        \qquad%
    \mathbb{N}^{}_\text{m} \, e^\text{m} = - \sgn \bigl( p -q \, C^{(0)} \bigr)\,,
\ee
where
\be \label{eq:defchi}
    \chi = - \frac{q \, e^{-\Phi}}{p - q \, C^{(0)}}\,.
\ee
Note that $\chi$ transforms under the SL($2\,,\mathbb{Z}$) group as 
\be
    \chi \rightarrow \chi - \kappa\,.
\ee
For simplicity, the IIA fields are set to zero and using the following dimensional reduction prescriptions: 
\begin{align}
    \mathbb{C}^{(6)}_{45} = 2 \, {\CC}^{(4)}\,,
        \qquad%
    \mathbb{A}^{(2)}_\text{m} = 
    \begin{pmatrix}
        A^\text{\scalebox{0.8}{B}} \\[2pt]
        A^\text{\scalebox{0.8}{C}}
    \end{pmatrix}\,,
        \qquad%
    \mathbb{C}^{(3)}_\text{m} = 
    \begin{pmatrix}
        B^{(2)} \\[2pt]
        C^{(2)}
    \end{pmatrix}\,,
\end{align}
Based on the above ingredients, we find that the dimensionally reduced action is\,\footnote{For simplicity, we focus on the IIB sector and assume $C^{(1)} = C^{(3)} = 0$\,. Moreover, we set $R = 1$\,, where $R$ is the circle's radius over which the resulting type II superstring theory compactifies.}
\begin{align} \label{eq:sdualnrd30}
\begin{split}
    S^{}_\text{4D} = & - T^{}_\text{D3} \int d^4 \sigma \! \sqrt{- \det \!
    \begin{pmatrix}
        0 &\,\, \tau^{}_\beta \\[4pt]
        \bar{\tau}^{}_\alpha &\,\, E^{}_{\alpha\beta} + \mathscr{F}^{}_{\!\alpha\beta}    \end{pmatrix} +
        \det \!
        \begin{pmatrix}
            0 &\,\, \CJ^{}_\beta &\,\, 0 \\[4pt]
            \CJ^{}_\alpha &\,\, \tau^{}_{\alpha\beta} + i \, \tilde{\mathscr{F}}^{}_{\!\alpha\beta} &\,\, E^{}_\alpha \\[4pt]
            0 &\,\, E^*_\beta &\,\, 0
        \end{pmatrix}} + S^{}_\text{CS}\,,
\end{split}
\end{align}
where 
\begin{align} \label{eq:scsdt}
\begin{split}
    S^{}_\text{CS} = - T^{}_\text{D3} \int \bigg[ \, \CC^{(4)} + \frac{1}{2} \, \mathscr{F}^{(2)} \wedge \mathscr{F}^\text{\scalebox{0.8}{C}} & - \chi \, \mathscr{F}^{(2)} \wedge \ell^{(2)} - \frac{1}{4} \, \chi^2 \, \mathscr{F}^\text{\scalebox{0.8}{C}} \wedge \ell^{(2)} \\[4pt]
    & + \frac{1}{2} \, \Bigl( \CF^\text{\scalebox{0.8}{B}} \wedge \CC^{(2)} - \CF^\text{\scalebox{0.8}{C}} \wedge \CB^{(2)} \Bigr) \biggr]\,,
\end{split}
\end{align}
and 
\begin{subequations}
\begin{align}
    \CJ^{}_\alpha & = \frac{e^{\Phi/4}}{2 \, \sqrt{-G}} \, h^{}_{\alpha\beta} \, \varepsilon^{\beta\gamma\delta\lambda} \, \p^{}_\gamma \mathbb{A}^{}_{\delta\lambda}\,, 
        &%
    E^{}_\alpha & = E^{}_\alpha{}^2 + i \, E^{}_\alpha{}^3\,, \\[4pt]
    \tilde{\mathscr{F}}^{}_{\alpha\beta} & = \frac{1}{2 \, \sqrt{-G}} \, h^{}_{\alpha\gamma} \, h^{}_{\beta\delta} \, \varepsilon^{\gamma\delta\lambda\kappa} \, \mathscr{F}^{}_{\lambda\kappa}\,,
        &%
    E^*_\alpha & = E^{}_\alpha{}^2 - i \, E^{}_\alpha{}^3\,. 
\end{align}
\end{subequations}
Note that the cursive $\varepsilon$ denotes the Levi-Civita tensor, distinguishing it from the normal $\epsilon$ that denotes the Levi-Civita symbol. The quantity $h^{}_{\alpha\beta}$ is defined to be
\be
    h^{}_{\alpha\beta} = \tau^{}_\alpha{}^A \, \tau^{}_\beta{}^B \, \eta^{}_{AB} + E^{}_\alpha{}^{A'} \, E^{}_\beta{}^{A'}
\ee
and
the measure $G$ is given by
\be \label{eq:defg}
    G = \frac{1}{4} \, \epsilon^{\alpha^{}_1\alpha^{}_2\alpha^{}_3\alpha^{}_4} \, \epsilon^{\beta^{}_1\beta^{}_2\beta^{}_3\beta^{}_4} \, \tau^{}_{\alpha^{}_1\beta^{}_1} \, \tau^{}_{\alpha^{}_2\beta^{}_2} \, E^{}_{\alpha^{}_3\beta^{}_3} \, E^{}_{\alpha^{}_4\beta^{}_4}\,.
\ee
Moreover, $\CF^\text{\scalebox{0.8}{B}} = e^{-\Phi/2} \, d A^\text{\scalebox{0.8}{B}}$\,, $\CF^\text{\scalebox{0.8}{C}} = e^{\Phi/2} \, \Bigl( d A^\text{\scalebox{0.8}{C}} + C^{(0)} \, d A^\text{\scalebox{0.8}{B}} \Bigr)$\,, and
\bea \label{eq:ftbc}
    \mathscr{F}^{(2)} &= \mathscr{F}^\text{\scalebox{0.8}{B}} - \chi \, \mathscr{F}^\text{\scalebox{0.8}{C}} + \frac{1}{2} \, \chi^2 \, \ell^{(2)}\,,
        \qquad%
    \mathscr{F}^\text{\scalebox{0.8}{B}} = {B}^{(2)} + {\CF}^\text{\scalebox{0.8}{B}}\,,
        \qquad%
    \mathscr{F}^\text{\scalebox{0.8}{C}} = {C}^{(2)} + {\CF}^\text{\scalebox{0.8}{C}}\,.
\eea
Note that the four-dimensional action \eqref{eq:sdualnrd30} is independent of $\sgn(p - q \, C^{(0)})$ appearing in Eq.~\eqref{eq:nvne}. 
The above formalism is only valid when $p - q \, \tau^{}_1 \neq 0$\,. We will discuss the case where $p - q \, \tau^{}_1 = 0$ later in section~\ref{sec:ibdtbr}.
In the next subsection, we will show that the novel action~\eqref{eq:sdualnrd30} is indeed dual to the D3-brane action in nonrelativistic string theory.

\subsection{D3-Brane Action in Nonrelativistic IIB Superstring Theory} \label{sec:dtbantbst}

We now show that the action~\eqref{eq:sdualnrd30} is dual to the manifestly SL($2\,,\mathbb{Z}$)-invariant D3-brane action in nonrelativistic string theory. We start with the D3-brane action \eqref{eq:d3f} and use the reparametrizations of $\hat{\CB}^{(2)}$ and $\hat{\CC}^{(2)}$  as in Eqs.~\eqref{eq:mcb2c2exp}, with $\CB^{(2)}$ and $\CC^{(2)}$ defined in Eq.~\eqref{eq:hmcbcnh}. Moreover, we use the reparametrization of the RR four-form $\hat{\CC}^{(4)}$ in Eq.~\eqref{eq:hcfcf}. Note that
\begin{align} \label{eq:exphghf}
    \hat{G}^{}_{\alpha\beta} & = \omega^{3/2} \, \tau^{}_{\alpha\beta} + \omega^{-1/2} \, E^{}_{\alpha\beta}\,,
        \qquad%
    \hat{\mathscr{F}}^{(2)} = - \omega^{3/2} \, \ell^{(2)} + \omega^{-1/2} \, \mathscr{F}^{(2)}\,,
\end{align}
where $\hat{\mathscr{F}}^{(2)}$ and $\mathscr{F}^{(2)}$ are defined in Eq.~\eqref{eq:defhmsf} and \eqref{eq:ftbc}, respectively. 
In the Einstein frame, the D3-brane action \eqref{eq:d3f} in relativistic IIB superstring theory gives rise to the following D3-brane action in nonrelativistic IIB superstring theory:
\begin{align} \label{eq:d33}
\begin{split}
    S_\text{D3} = & - T^{}_\text{D3} \int d^4\sigma \sqrt{-\det \! 
    \begin{pmatrix}
        0 &\,\, \tau^{}_\beta \\[4pt]
        \bar{\tau}^{}_\alpha &\,\, E^{}_{\alpha\beta} + \mathscr{F}^{}_{\alpha\beta}
    \end{pmatrix}} + S^{}_\text{CS}\,,
\end{split}
\end{align}
where the Chern-Simons term $S^{}_\text{CS}$ is the same as in Eq.~\eqref{eq:scsdt}. Up to a boundary term, Eq~\eqref{eq:d33} is equivalent to the D3-brane action in nonrelativistic IIB theory found in \cite{Bergshoeff:2022iss}, except that the Chern-Simons terms in Eq.~\eqref{eq:d33} take a much simpler form and is manifestly independent of the auxiliary SL($2\,,\mathbb{Z}$) doublet $(\tilde{p}\,, \tilde{q}\,)^\intercal$\,. 

Next, we consider the action of a single D3-brane localized in a circle of radius $R$\,. For simplicity, we assume $R = 1$ and $\hat{C}^{(1)} = \hat{C}^{(3)} = 0$\,. The D3-brane action~\eqref{eq:d3f} in relativistic IIB superstring theory now becomes
\begin{align} \label{eq:d3f3}
    \hat{S}^{}_\text{D3} = & - T^{}_\text{D3} \int d^4\sigma \sqrt{-\det \! \lr \hat{G}^{}_{\alpha\beta} + \hat{f}^{}_\alpha \, \hat{f}^{}_\beta + \hat{\mathscr{F}}^{}_{\alpha\beta} \rr} \\[4pt]
    & - T^{}_\text{D3} \int \ls \hat{\CC}^{(4)} + \frac{\bigl( \hat{\chi} \, \hat{\mathscr{F}}^{\text{\scalebox{0.8}{B}}} + \hat{\mathscr{F}}^{\text{\scalebox{0.8}{C}}} \bigr) \wedge \hat{\mathscr{F}}}{2 \sqrt{1+\hat{\chi}^2}} + \tfrac{1}{2} \lr \hat{\CF}^{\text{\scalebox{0.8}{B}}} \wedge \hat{\CC}^{(2)} - \hat{\CF}^\text{\scalebox{0.8}{C}} \wedge \hat{\CB}^{(2)} \rr \rs, \notag
\end{align}
where $\hat{f}^{}_\alpha = e^{-\hat{\Phi}/4} \, \p^{}_\alpha \pi$ is related to the Nambu-Goldstone boson perturbing the shape of the D3-brane along the circle\,.
We dualize $\pi$ by introducing the generating functional
\be
    \hat{S}^{}_\text{g.f.} = - \frac{T^{}_\text{D3}}{3!} \int d^4 \sigma \, J^\alpha \Bigl( \pi^{}_\alpha - \p^{}_\alpha \pi \Bigr)\,.      
\ee
Integrating out $J^\alpha$ gives back the original action~\eqref{eq:d3f3}. Instead,
integrating out $\pi$ imposes $\p^{}_\alpha J^{\alpha} = 0$\,, which is solved locally by 
$J^{\alpha} = \frac{1}{2} \, \epsilon^{\alpha\beta\gamma\delta} \, \p^{}_\beta \mathbb{A}^{}_{\gamma\delta}$\,.
In terms of the quantity 
\be \label{eq:hja}
    \hat{\CJ}^\alpha \equiv \frac{e^{\hat{\Phi}/4}}{\sqrt{-\hat{G}}} \, J^\alpha\,,
\ee
we find the dual action 
\begin{align} \label{eq:sdualdt}
\begin{split}
    \hat{S}^{}_\text{dual} = & - T^{}_\text{D3} \int d^4 \sigma \, \sqrt{-\det \Bigl[ \hat{G}^{}_{\alpha\beta} + i \, \bigl( \star \hat{\mathscr{F}} \bigr)^{}_{\alpha\beta} - \hat{\CJ}^{}_\alpha \, \hat{\CJ}^{}_\beta \Bigr]} \\[4pt]
    & - T^{}_\text{D3} \int \ls \hat{\CC}^{(4)} + \frac{\bigl( \hat{\chi} \, \hat{\mathscr{F}}^{\text{\scalebox{0.8}{B}}} + \hat{\mathscr{F}}^{\text{\scalebox{0.8}{C}}} \bigr) \wedge \hat{\mathscr{F}}}{2 \sqrt{1+\hat{\chi}^2}} + \tfrac{1}{2} \lr \hat{\CF}^{\text{\scalebox{0.8}{B}}} \wedge \hat{\CC}^{(2)} - \hat{\CF}^\text{\scalebox{0.8}{C}} \wedge \hat{\CB}^{(2)} \rr \rs.
\end{split}
\end{align}
In the $\omega \rightarrow \infty$ limit, the action~\eqref{eq:sdualdt} gives rise to Eq.~\eqref{eq:sdualnrd30}, where the latter is derived from dimensionally reducing nonrelativistic M5-brane described by Eq.~\eqref{eq:nrm5a} over the anisotropic two-torus. See Appendix~\ref{app:D3} for a derivation.
Since the nonrelativistic D3-brane action~\eqref{eq:d33} arises from the same $\omega \rightarrow \infty$ limit of the relativistic D3-brane action~\eqref{eq:d3f3}, it implies that the exotic action~\eqref{eq:sdualnrd30} comes from dualizing the Nambu-Goldstone boson perturbing the shape of the D3-brane described by the action~\eqref{eq:d33} along a circle in nonrelativistic string theory. In this sense, compactifying the nonrelativistic M5-brane action~\eqref{eq:nrm5a} over an anisotropic torus gives rise to the nonrelativistic D3-brane action~\eqref{eq:d33}. 

\subsection{Inter-Branched D3-Brane Revisited} \label{sec:ibdtbr}

Throughout this section, the case $p - q \, \tau^{}_1 \neq 0$ was of interest. In this subsection, we focus on the case where $p - q \, \tau^{}_1 = 0$\,, \emph{i.e.}, $\chi \rightarrow \infty$\,.  

We have learned in section~\ref{sec:onebranelimit} that the SL($2\,,\mathbb{Z}$) transformations satisfying $\gamma \, \tau^{}_1 + \delta = 0$ map nonrelativistic string theory to the critical RR two-form limit of relativistic IIB string theory. Note that, under the SL($2\,,\mathbb{Z}$) transformation, we have
\be \label{eq:pqt}
    p' - q' \, \tau'_1 = \bigl( \gamma \, \tau^{}_1 + \delta \bigr) \, \bigl( p - q \, \tau^{}_1 \bigr)\,. 
\ee
When $\gamma \, \tau^{}_1 + \delta = 0$\,, the dual of $p - q \, \tau^{}_1$ vanishes in the critical RR two-form limit of relativistic IIB string theory. Therefore, the $p - q \, \tau^{}_1 \rightarrow 0$ limit, \emph{i.e.}, the $\chi \rightarrow \infty$ limit of Eq.~\eqref{eq:d33} defines the associated D3-brane action, which takes the following form: 
\begin{align} \label{eq:inbdt}
\begin{split}
    S^{}_\text{D3} & \rightarrow \frac{T^{}_\text{D3}}{2} \int d^4 \sigma \,  \sqrt{-G} \, \frac{1 + \bigl( \star\mathscr{F}^{\text{\scalebox{0.8}{C}}} \bigr)^{\alpha\beta} \, \tau^{}_{\beta\gamma} \, \bigl( \star\mathscr{F}^{\text{\scalebox{0.8}{C}}} \bigr)^{\gamma\delta} \, E^{}_{\delta\alpha} - \frac{1}{16} \,  \Bigl[ \bigl( \star\mathscr{F}^{\text{\scalebox{0.8}{C}}} \bigr)^{\alpha\beta} \, \mathscr{F}^{\text{\scalebox{0.8}{C}}}_{\alpha\beta} \Bigr]^2}{\bigl( \star \ell \bigr)^{\alpha\beta}\, \mathscr{F}^{\text{\scalebox{0.8}{C}}}_{\alpha\beta}} \\[4pt] 
    & \quad \, -  T^{}_\text{D3} \int \bigg[ \CC^{(4)} - \tfrac{1}{2} \, \mathscr{F}^{\text{\scalebox{0.8}{B}}} \wedge \mathscr{F}^{\text{\scalebox{0.8}{C}}} + \tfrac{1}{2} \, \Bigl( \CF^{\text{\scalebox{0.8}{B}}} \wedge \CC^{(2)} - \CF^{\text{\scalebox{0.8}{C}}} \wedge \CB^{(2)} \Bigr) \biggr]\,.
\end{split}
\end{align}
Here, $G$ is the measure defined in Eq.~\eqref{eq:defg}. Note that cancelling the large $\chi$ divergences in Eq.~\eqref{eq:d33} to derive Eq.~\eqref{eq:inbdt} requires $\bigl( \star \ell \bigr)^{\alpha\beta} \mathscr{F}^{\text{\scalebox{0.8}{C}}}_{\alpha\beta} < 0$\,. This theory is equivalent to the inter-branched D3-brane action in \cite{Bergshoeff:2022iss}, which is closely related to noncommutative Yang-Mills (NCYM) theory \cite{Gopakumar:2000na}. Under the Seiberg-Witten map \cite{Seiberg:1999vs}, the quantity $\bigl|\bigl( \star \ell \bigr)^{\alpha\beta} \mathscr{F}^{\text{\scalebox{0.8}{C}}}_{\alpha\beta}\bigr|$ in Eq.~\eqref{eq:inbdt} plays the role of the NCYM gauge coupling. 

%%%%%%%%%%%%%%%%%%%%%%%
\section{Conclusions} \label{sec:concl}
%%%%%%%%%%%%%%%%%%%%%%%

We studied the anisotropic toroidal compactification of nonrelativistic M-theory and its application to a single M5-brane. We established a geometrical interpretation of the polynomial realization of the SL($2\,,\mathbb{Z}$) duality in nonrelativistic type IIB superstring theory, where the IIB background fields transform as polynomials of the parameter $\kappa$ defined in Eq.~\eqref{eq:trnsfeaaomega}.\,\footnote{As a side remark, it is also natural to construct the F-theory \cite{Vafa:1996xn} associated with nonrelativistic type IIB superstring theory. This line of inquiry could offer insights for understanding the compactifications of nonrelativistic IIB string theory to lower dimensions.} We have shown that $\kappa$ receives a physical interpretation as an effective Galilean boost velocity on the anisotropic two-torus over which nonrelativistic M-theory compactifies. We then reviewed the relativistic M5-brane over a torus \cite{Berman:1998va}, where we found a new expression for the manifestly SL($2\,,\mathbb{Z}$)-invariant D3-brane action, without introducing the auxiliary SL($2\,,\mathbb{Z}$) doublet $(\tilde{p}\,, \tilde{q})$ as in \cite{Bergshoeff:2006gs}. Next, we constructed the covariant M5-brane action~\eqref{eq:nrm5a} in nonrelativistic M-theory. Compactifying this M5-brane over the anisotropic torus gives rise to a manifestly SL$(2, \mathbb{Z})$-invariant D3-brane action~\eqref{eq:d33} in nonrelativistic IIB string theory, which significantly simplifies the previous result in \cite{Bergshoeff:2022iss}. 

This paper is a first step towards understanding the U-duality unifying nonrelativistic and DLCQ M-theory, which bears intriguing connections to the Matrix theory description of M-theory \cite{uduality}. We conclude with an outlook for future research directions.

\vspace{3mm}
\noindent \textbf{U-duality and polynomial realizations.} The moduli space of M-theory compactified on $\mathbb{R}^{11-n} \times T^n$ is invariant under an infinite discrete U-duality group. The U-duality group is the exceptional group $ E_{n(n)}(\mathbb{Z})$ generated by $\text{SL}(n, \mathbb{Z})$ and $ \text{SO}(n-1, n-1, \mathbb{Z})$. In this paper, we studied $E_{2(2)}(\mathbb{Z})$ in nonrelativistic M-theory. It would be interesting to explore the realization of these exceptional groups in nonrelativistic M-theory. We expect an interesting interplay between the exceptional groups and polynomial realizations.\,\footnote{The exceptional groups $E^{}_{3(3)} (\mathbb{Z})$ and $E^{}_{4(4)} (\mathbb{Z})$ have been considered in \cite{Blair:2021waq}, which provides useful ingredients for future studies of their interplay with polynomial realizations.} Moreover, these novel U-duality relations reveal a fascinating duality web that unifies different decoupling couplings of string/M-theory, containing Matrix gauge theories, DLCQ, and nonrelativistic string/M-theory and beyond. See \cite{uduality, wfmt} for detailed constructions of this duality web.

\vspace{3mm}
\noindent \textbf{Self-duality condition.} Understanding the action principle of (non)relativistic quantum field theories with self-dual field strength and general covariance is a crucial yet subtle facet of string theory and supergravity. For example, a fundamental property of the M5-brane effective action is the self-duality of the field strength $\mathbb{F}^{(3)} = \star \mathbb{F}^{(3)}$ on the six-dimensional worldvolume. The PST formalism provides an elegant way to incorporate this self-duality condition at the level of the action by introducing an auxiliary scalar. However, the analog of this self-duality condition on the M5-brane worldvolume in nonrelativistic M-theory appears elusive. Even at the quadratic order of the field strength $\mathbb{F}^{(3)}$\,, the nonrelativistic M5-brane action still takes a rather complicated form, which makes it arduous to identify the self-duality condition in any simple form. We will present this quadratic action in Appendix~\ref{app:quadM5}, which still requires further studies to reveal the underlying structure. A proper understanding of the self-duality condition on the nonrelativistic M5-brane will, for example, facilitate the study of nonrelativistic one-brane solitons carrying self-dual charge, where it will be interesting to compute the associated tension as in \cite{Perry:1996mk}.      

\vspace{3mm}
\noindent \textbf{Anisotropic Calabi-Yau manifolds.} We focused on anisotropic toroidal compactifications in this paper. It would also be natural to consider compactifications over general Calabi-Yau manifolds with similar anisotropic twists in the future, which plays a crucial role in further advancing the string/M-theory duality web. For example, in \cite{Cherkis:1997bx}, wrapping the relativistic M5-brane on a Calabi-Yau K3 surface leads to heterotic string theory in a seven-dimensional target spacetime. This computation does not explicitly rely on the existence of the K3 metric but instead uses the 22 harmonic representatives of K3's integral second cohomology classes $H^2(\text{K}3\,, \mathbb{Z})$\,. This construction has been supersymmetrized in \cite{Park:2009me}. Studying the compactification of nonrelativistic M5-brane over an anisotropic K3 manifold would provide a powerful tool for understanding nonrelativistic heterotic string theory, which remains unexplored, and possibly also the DLCQ of relativistic heterotic string theory \cite{Danielsson:1996es,Kachru:1996nd,Motl:1997tb}.

\vspace{3mm}
\noindent \textbf{Membrane scattering amplitudes.} Another interesting future direction would be to construct brane amplitudes in nonrelativistic M-theory, using the techniques developed in \cite{Heydeman:2017yww}, which computed tree-level scattering amplitudes for n-particles involving D3-, D5-, and M5-branes using twistor-like spinor-helicity coordinates. Regarding the equivalence between nonrelativistic and DLCQ string scattering amplitudes studied in \cite{Danielsson:2000gi, Yan:2021hte}, the study of the brane amplitudes could potentially provide insights into resolving some of the long-standing ambiguities and discrepancies between Matrix theory and supergravity computation as discussed in previous works \cite{Dine:1997sz, Douglas:1997uy}.

%%%%%%%%%%%%%%%%%%%%%%%%%%%

\acknowledgments

We would like to thank Eric Bergshoeff, Chris Blair, Eric D'Hoker, Thomas Dumitrescu, Kevin Grosvenor, Johannes Lahnsteiner, Per Kraus, Florian Niedermann, Niels Obers, and Utku Zorba for their useful discussions and comments on the paper. The main results of this paper were presented on February 8, 2023, at the \href{https://indico.ph.ed.ac.uk/event/233/timetable/?view=standard}{\color{black}\emph{Beyond Lorentzian Geometry II}} workshop held at ICMS in Edinburgh and on May 9, 2023, at the \href{https://indico.fysik.su.se/event/8008/}{\color{black}\emph{Non-Relativistic Strings and Beyond}} workshop held at Nordita in Stockholm. S.E. is supported by the Bhaumik Institute. Z.Y. is supported by the European Union’s Horizon 2020 research and innovation programme under the Marie Sk\l{}odowska-Curie grant agreement No 31003710. Nordita is supported in part by NordForsk.

%%%%%%%%%%%%%%%%%%%%%%%%%%%

\newpage

\vfill

\appendix

\section{Iwasawa Decomposition of Local \texorpdfstring{SL($2\,,\mathbb{R}$)}{SLR}}
\label{sec:prid}

In section~\ref{sec:atbsld}, we discussed the realization of SL($2\,,\mathbb{Z}$) in terms of the variables $\kappa$ in relativistic M-theory and $\hat{\kappa}$ in nonrelativistic M-theory on a two-torus. In this appendix, we reveal a natural relation between this unconventional SL($2\,,\mathbb{Z}$) realization and the Iwasawa decomposition of $\textbf{G} = \text{SL}(2\,,\mathbb{R})$\,, which takes the following form \cite{e543455c-2183-31cf-aa50-d7a5d5d9f85a}:\,\footnote{See \cite{Obers:1998fb} for a review of applications of the Iwasawa decomposition in U-duality. We thank Niels Obers for pointing out the connection to the Iwasawa decomposition. Here, we choose to represent $\mathbf{N}$ as a lower triangular matrix, aligning with the notation used in the paper, as opposed to the more common practice of representing it as an upper triangular matrix.} 
\be
    \textbf{G} = \textbf{K} \cdot \textbf{A} \cdot \textbf{N}
\ee 
where
\begin{subequations}
\begin{align}
    & \hspace{2cm} \textbf{K} = \Biggl\{ 
    \begin{pmatrix}
        \cos \theta &\quad - \sin \theta \\[4pt]
        \sin \theta &\quad \cos \theta
    \end{pmatrix}
    : \, \theta \in \mathbb{R} \Biggr\}\,, \\[6pt]
    & \textbf{A} = \Biggl\{
    \begin{pmatrix}
        r &\quad 0 \\[4pt]
        0 &\quad r^{-1}
    \end{pmatrix}
    : \, r > 0
    \Biggr\}\,,
        \qquad%
    \textbf{N} = \Biggl\{
    \begin{pmatrix}
        1 &\quad 0 \\[4pt]
        x &\quad 1
    \end{pmatrix}
    : \, x \in \mathbb{R}
    \Biggr\}\,.
\end{align}
\end{subequations}
Here, \textbf{K} are the SO($2\,, \mathbb{R}$) matrices associated (rotations), \textbf{A} are positive diagonal matrices of determinant one (dilatations), and \textbf{N} are unipotent matrices (translations).   

In relativistic M-theory, the $\hat{\kappa}$ representation induces an embedding of SL($2\,,\mathbb{Z}$) within SL($2\,,\mathbb{R}$). In Eq.~\eqref{eq:trnsfe}, we showed that SL($2\,,\mathbb{Z}$) acts on the zweibein one-form $\hat{e}^a = \hat{e}^{}_m{}^a \, d\hat{x}^m$ on the two-torus as
\be \label{eq:ktheta}
    \hat{e}^a \rightarrow \hat{k}^{a}{}_b \, \hat{e}^b\,,
        \qquad%
    \hat{k} = 
    \begin{pmatrix}
        \cos \theta &\quad - \sin \theta \\[4pt]
        \sin \theta &\quad \cos \theta
    \end{pmatrix} 
    \in \textbf{K}\,,
\ee
where $\theta$ is defined in Eq.~\eqref{eq:rdhk}. This expression is the same as a SO$(2\,, \mathbb{R})$ transformation, if one regards $\theta$ in Eq.~\eqref{eq:ktheta} as the group parameter. This group action maps to the subgroup \textbf{K} in the Iwasawa decomposition. Moreover, the SL($2\,,\mathbb{Z}$) transformations of the component $\hat{e}_m{}^a$ in the one-form field $\hat{e}^a$ are given by
\be
    \hat{e}_m{}^a
        \rightarrow%  
    \bigl(\hat{a} \cdot \hat{n}\bigr)_{m}{}^{n} \, \hat{e}^{}_n{}^a\,,
\ee
where
\be
    \hat{a} = 
    \begin{pmatrix}
        \hat{r} &\quad 0 \\[4pt]
        0 &\quad \hat{r}^{-1}
    \end{pmatrix} 
        \in \textbf{A}\,,
        \qquad%
    \hat{n} = 
    \begin{pmatrix}
        1 &\quad 0 \\[4pt]
        \hat{x} &\quad 1
    \end{pmatrix} 
        \in \textbf{N}\,,
\ee
with $\hat{r} = \bigl|\gamma \, \hat{\tau}^{}_1 + \delta \bigr| \sqrt{1 + \hat{\kappa}^2}$ and $\hat{x} = - \bigl( \beta \, \hat{r}^2 + \gamma \, \bigl| \hat{\tau} \bigr|^2 \bigr) / \delta$\,.
Note that the matrices in \textbf{K} act on the frame index $a$ of the zweibein $\hat{e}^{}_m{}^a$\,, while \textbf{A} and \textbf{N} act on the curved index $m$\,. 

Next, we investigate the embedding of SL($2\,,\mathbb{Z}$) within SL($2\,,\mathbb{R}$) on the anisotropic torus in nonrelativistic M-theory by performing an $\omega \rightarrow \infty$ limit of the above results. Complexifying the dilaton field as in section~\ref{sec:bvbut}, we find the following embedding of SL($2\,,\mathbb{Z}$) within SL($2\,, \mathbb{R}$)\,:
\vspace{-5mm}
\begin{subequations}
\begin{align}
    e^a & \rightarrow \tilde{n}^a{}_b \, e^b\,,
        &%
    \tilde{n} & = 
    \begin{pmatrix}
          1 &\quad 0 \\[4pt]
          \kappa &\quad 1
    \end{pmatrix} \in \mathbf{N}\,, \\[4pt]
    e^{}_m{}^a & \rightarrow \bigl( a \cdot n \bigr){}_{m}{}^{n} \, e^{}_{n}{}^a\,, 
        &%
    a & = 
    \begin{pmatrix}
        {r} &\quad 0 \\[4pt]
        0 &\quad {r}^{-1}
    \end{pmatrix} 
        \in \textbf{A}\,,
        \qquad%
    n = 
    \begin{pmatrix}
        1 &\quad 0 \\[4pt]
        {x} &\quad 1
    \end{pmatrix} 
        \in \textbf{N}\,, 
\end{align}
\end{subequations}
where $r = |\gamma \, \tau^{}_1 + \delta|$ and $x = - \bigl( \beta \, r^2 + \gamma \, \tau^2_1 \bigr) / \delta$\,.
Note that the realization of SL($2\,,\mathbb{Z}$) on the anisotropic torus embeds SL($2\,,\mathbb{Z}$) within $\mathbf{A} \cdot \mathbf{N}$ but \emph{not} within the SO(2) subgroup $\mathbf{K}$\,.

\section{A Quadratic Gauge Nonrelativistic M5-Brane Action}
\label{app:quadM5}

This appendix presents the quadratic part of nonrelativistic M5-brane action~\eqref{eq:nrm5a} expanded with respect to the three-form field strength $\mathbb{F}^{(3)}$ on the worldvolume. In analog with Eq.~\eqref{eq:self-dual}, we also work in a flat spacetime with the M5-brane extending in the three-dimensional longitudinal sector and $\mathbb{C}^{(3)} = \mathbb{C}^{(6)} = 0$\,. Furthermore, we make the extra assumption that the norm $n \equiv \sqrt{\mathbb{N}_u \, \mathbb{N}^u}$ is nonzero. Keeping only the quadratic $\mathbb{F}^{(3)}$ terms from the nonrelativistic M5-brane action~\eqref{eq:nrm5a} gives rise to the following free gauge theory in six-dimensions:
\begin{align}
\label{eq:quadDBIPST1}
\begin{split}
    S^{}_{\text{quad.}} \! = & - \frac{T^{}_{\text{M}5}}{12} \! \int \!\! \frac{d^6\sigma}{n^{3}} \, \bigg\{ \Bigl[ \mathbb{F}_{uvw} \, \mathbb{F}^{uvw}  - 6 \, \mathbb{N}^u \, \mathbb{N}^{}_{u'} \, \bigl( \mathbb{F}^{}_{uvw} \, \mathbb{F}^{vwu'} + \mathbb{F}^{}_u{}^{wu'} \, \mathbb{F}^{}_{vwv'} \, \mathbb{N}^v \, \mathbb{N}^{v'} \bigr) \Bigr] \\[4pt]
    & \hspace{1.1cm} + \, 3 \, n^2 \Bigl[ \mathbb{F}^{}_{uvu'}\mathbb{F}^{uvu'} - 2 \, \mathbb{N}^{u} \, \mathbb{N}^{v'} \, \bigl( 2 \, \mathbb{F}^{}_{uvu'} \, \mathbb{F}^{vu'}{}_{v'} + \mathbb{F}^{}_{uu'w'} \, \mathbb{F}_{vv'}{}^{w'} \, \mathbb{N}^{u'} \, \mathbb{N}^v \bigr) \Bigr] \\[6pt]
    & \hspace{2.4cm} + 3 \, n^4 \, \mathbb{F}^{}_{uu'v'} \, \bigl( \mathbb{F}^{uu'v'} - 2 \, \mathbb{F}^{u'v'w'} \, \mathbb{N}^u \, \mathbb{N}^{}_{w'} \bigr) + n^6 \, \mathbb{F}_{u'v'w'}  \mathbb{F}^{u'v'w'} \bigg\} \\[4pt]
    & - \frac{T_{\text{M}5}}{24} \int d^6 \sigma \, \epsilon^{uvw} \, \epsilon^{u'v'w'} \, \mathbb{N}^{s'} \, \Bigl[ 6 \, \mathbb{N}_w \, \bigl( \mathbb{F}_{uu's'} \, \mathbb{F}_{vv'w'} - \mathbb{F}_{uvw'} \mathbb{F}_{u'v's'} \bigr) \\ 
    & \hspace{3.49cm} + \mathbb{N}_{w'} \, \bigl(  \mathbb{F}_{uvw} \, \mathbb{F}_{u'v's'} + 6 \, \mathbb{F}_{vwv'} \mathbb{F}_{uu's'} + 3 \, \mathbb{F}_{wu'v'} \mathbb{F}_{uvs'} \bigr) \Big]\,.
\end{split}
\end{align}
The action up to quadratic order also contains a constant and linear piece in $\mathbb{F}^{(3)}$\,, namely, 
$$
    -T_{\text{M}5} \! \int \! d^6\sigma \Bigl[ n + \frac{1}{6 \, n} \, \epsilon^{uvw} \, \bigl( \mathbb{F}_{uvw} - 3 \, \mathbb{F}_{uvw'} \, \mathbb{N}^{w'} \, \mathbb{N}_{w} \bigr) + \frac{n^{2}}{6} \, \epsilon^{u'v'w'} \, \mathbb{F}_{u'v'w'} - \frac{1}{2} \, \epsilon^{u'v'w'}  \mathbb{F}_{uv'w'} \mathbb{N}^u \mathbb{N}_{u'}  \Bigr].
$$
Note that we have split the Levi-Civita symbol $\epsilon^{\alpha\beta\gamma\delta\kappa\lambda}$ into a product of a longitudinal piece $\epsilon^{uvw}$ defined via $\epsilon^{012} = 1$ and a transverse piece $\epsilon^{u'v'w'}$ defined via $\epsilon^{345} = 1$\,.
In principle, this action should help us understand the analogs of the PST gauge symmetries and self-duality constraint of $\mathbb{F}^{(3)}$ that we have reviewed in section~\ref{sec:pstfmb} for relativistic M5-brane. Unfortunately, in its current form, extracting useful insights directly from the rather involved expression~\eqref{eq:quadDBIPST1} is challenging. Further studies of the free theory still await. For example, it might be useful to consider the expansion with respect to a small $\mathbb{F}^{(3)}$ around a different background field configuration. To derive Eq.~\eqref{eq:quadDBIPST1}, we had to assume that $n \neq 0$\,. This condition can be relaxed if a nonzero longitudinal $\mathbb{C}^{(3)}$ is turned on, with $\mathbb{C}^{}_{uvw} \neq 0$\,.  

\section{Dual D3-Brane in Nonrelativistic IIB Superstring Theory}
\label{app:D3}

In this appendix, we derive the dual D3-brane action~\eqref{eq:sdualnrd30} by taking the nonrelativistic string limit of Eq.~\eqref{eq:sdualdt}. We focus on the Dirac-Born-Infeld (DBI) Lagrangian in the first line of Eq.~\eqref{eq:sdualdt}, \emph{i.e.},
\be
    \hat{\CL} = - \sqrt{-\det \! \ls \hat{G}^{}_{\alpha\beta} + i \, \bigl( \star \hat{\mathscr{F}} \bigr)^{}_{\alpha\beta} - \hat{\CJ}^{}_\alpha \, \hat{\CJ}^{}_\beta \rs}\,.
\ee
For simplicity, we focus on flat spacetime with $\tau^{}_\text{M}{}^A = \delta_\text{M}^A$ and $E^{}_\text{M}{}^{A'} = \delta_\text{M}^{A'}$ and covariantize the resulting action at the end of the derivation. Plugging Eqs.~\eqref{eq:exphghf} and \eqref{eq:hja}, we find
\be \label{eq:hcl}
    \hat{\CL} = - \omega^{-1} \sqrt{-\det \Bigl( \hat{\CO} - \hat{\CJ} \, \hat{\CJ}^\intercal \Bigr)} = - \omega^{-1} \sqrt{- \det \hat{\CO} - 
    \det
    \begin{pmatrix}
        0 & \quad \hat{\CJ}^\intercal \\[4pt]
        \hat{\CJ} &\quad \hat{\CO}
    \end{pmatrix}}\,,
\ee
where
\begin{align}
    \hat{\CO} = 
    \begin{pmatrix}
        \omega^2 \, \bigl( \eta^{}_{AB} + i \, \tilde{\mathscr{F}}^{}_{AB} \bigr) &\quad i \, \tilde{\mathscr{F}}^{}_{Aj} \\[4pt]
        i \, \tilde{\mathscr{F}}^{}_{iB} &\quad \delta^{}_{ij} - i \, \epsilon^{}_{ij} + i \, \omega^{-2} \, \tilde{\mathscr{F}}^{}_{ij} 
    \end{pmatrix},
        \qquad%
    \hat{\CJ} = 
    \begin{pmatrix}
        \omega \, \CJ^{}_{A} \\[4pt]
        \omega^{-1} \, \CJ^{}_{i}
    \end{pmatrix},
\end{align}
with $A = 0\,, 1$ and $i = 2\,, 3$\,. 
Note that
\begin{subequations} \label{eq:identities}
\begin{align}
    \det \hat{\CO} 
    & =
    \omega^2 \, \det 
    \begin{pmatrix}
        0 &\quad \tau^{}_\beta \\[4pt]
        \bar{\tau}^{}_\alpha &\quad E^{}_{\alpha\beta} + \mathscr{F}^{}_{\alpha\beta} 
    \end{pmatrix} + O(\omega^0)\,, \label{eq:detorw} \\[4pt] 
    \det
    \begin{pmatrix}
        0 & \quad \hat{\CJ}^\intercal \\[4pt]
        \hat{\CJ} &\quad \hat{\CO}
    \end{pmatrix} 
    & = \det \Bigl( \omega^2 \, \mathcal{\CE} \, \mathcal{\CE}^\dagger + \CN \Bigr) = - \omega^2 \, \det 
    \begin{pmatrix}
        \CN &\quad \CE \\[4pt]
        \CE^\dagger &\quad 0
    \end{pmatrix}+ O(\omega^0)\,.
\end{align}
\end{subequations}
We have covariantized the determinant in Eq.~\eqref{eq:detorw}. Moreover, $\CE = (0\,, 0\,, 0\,, 1\,, i)^\intercal$ and 
\be
    \CN = 
    \begin{pmatrix}
        0 &\quad \CJ^{}_B &\quad \CJ^{}_{j} \\[4pt]
        \CJ^{}_A &\quad \eta^{}_{AB} + i \, \tilde{\mathscr{F}}^{}_{AB} &\quad i \, \tilde{\mathscr{F}}^{}_{Aj} \\[4pt] 
        \CJ^{}_{i} &\quad i \, \tilde{\mathscr{F}}^{}_{iB} &\quad i \, \tilde{\mathscr{F}}^{}_{ij}
    \end{pmatrix}.
\ee
These ingredients covariantized are
\be
    \CE = 
    \begin{pmatrix}
        0 \\[4pt]
        E^{}_\alpha
    \end{pmatrix},
        \qquad%
    \CN = 
    \begin{pmatrix}
        0 &\quad \CJ^{}_\beta \\[4pt]
        \CJ^{}_\alpha &\quad \tau^{}_{\alpha\beta} + i \, \tilde{\mathscr{F}}^{}_{\alpha\beta} 
    \end{pmatrix},
\ee
where $E_\alpha = E_\alpha{}^2 + i \, E_\alpha{}^3$ and $\CJ_\alpha = ( \tau^{}_{\alpha\beta} + E^{}_{\alpha\beta} ) \, \CJ^\beta$\,. 
Plugging Eq.~\eqref{eq:identities} into Eq.~\eqref{eq:hcl} and taking the $\omega \rightarrow \infty$ limit derives the DBI part of the desired action~\eqref{eq:sdualnrd30}. The $\omega \rightarrow \infty$ limit of the Chern-Simons terms in \eqref{eq:sdualdt} reproduces the Chern-Simons terms in the action~\eqref{eq:sdualnrd30} in a straightforward way, which we do not elaborate further here.

\newpage

\bibliographystyle{JHEP}
\bibliography{acnmt}
\end{document}